\documentclass{elsart}
\usepackage{epsfig}

\def\lsim{\mathrel{\rlap{\lower4pt\hbox{\hskip1pt$\sim$}}
    \raise1pt\hbox{$<$}}} 
\def\gsim{\mathrel{\rlap{\lower4pt\hbox{\hskip1pt$\sim$}}
    \raise1pt\hbox{$>$}}} 

\def\beas{\begin{eqnarray*}}
\def\eeas{\end{eqnarray*}}
\def\bq{\begin{equation}}
\def\eq{\end{equation}}
\def\ben{\begin{enumerate}}\def\een{\end{enumerate}}
\def\nn{\nonumber}
\def\del{\partial}

\def\r{\rho}

\def\la{\langle}
\def\ra{\rangle}

\def\Re{{\rm Re}\, }

\def\tr{{\rm tr}\, }
\def\Tr{{\rm Tr}\, }

\def\s{\sigma}

\def\ve{\varepsilon}
\def\vs{\varsigma}

\def\to{\rightarrow}
\def\bra{\langle}
\def\ket{\rangle}

\def\slash#1{{\rlap{$#1$} \thinspace /}}

\def\be{\begin{eqnarray}}
\def\ee{\end{eqnarray}}
\def\Fc{{\mathcal F}}
\def\L{{\mathcal L}}
\def\V{{\mathcal V}}
\def\Del{{\mathcal D}}
\def\E{{\mathcal E}}
\def\O{{\mathcal O}}
\def\R{{\mathcal R}}
\def\ln{{\rm ln}}
\def\C{\tilde{C}}

\def\kF{k_F}

\def\F1{F_1}
\def\ft1{{\tilde{F}_1}}
\def\ft1omega{\tilde{F}_1^\omega}

\def\A0{A_0}
\def\Jvec{\vec{J}}
\def\Jb{\mbox{\boldmath$ J$}}\def\Kb{\mbox{\boldmath$ K$}}

\def\Vb{\mbox{\boldmath$ V$}}
\def\Ab{\mbox{\boldmath$ A$}}
\def\jb{\mbox{\boldmath$ j$}}
\def\pb{\mbox{\boldmath$ p$}}\def\qb{\mbox{\boldmath $q$}}
\def\tb{\mbox{\boldmath$ t$}}
\def\ub{\mbox{\boldmath$ u$}} \def\vb{\mbox{\boldmath$ v$}}
\def\xb{\mbox{\boldmath$ x$}}
\def\Bp{\mbox{\boldmath$ p$}}\def\Bq{\mbox{\boldmath $q$}}
\def\Bk{\mbox{\boldmath$ k$}}\def\Br{\mbox{\boldmath $r$}}
\def\Bs{\mbox{\boldmath $s$}}
\def\qbhat{\mbox{\boldmath $\hat{q}$}}\def\taub{\mbox{\boldmath $\tau$}}
\def\pbhat{\mbox{\boldmath $\hat{p}$}}
\def\pbbhat{\mbox{\boldmath $\hat{\bar{p}}$}}
\def\pib{\mbox{\boldmath $\pi$}}
\def\kbhat{\mbox{\boldmath $\hat{k}$}}
\def\nubhat{\mbox{\boldmath $\hat{\nu}$}}
\def\ubhat{\mbox{\boldmath $\hat{u}$}}
\def\gammab{\mbox{\boldmath $\gamma$}}
\def\kappab{\mbox{\boldmath $\kappa$}}
\def\rhob{\mbox{\boldmath $\rho$}}
\def\sigmab{\mbox{\boldmath $\sigma$}}
\def\thetab{\mbox{\boldmath $\theta$}}
\def\Bnabla{\mbox{\boldmath $\nabla$}}
\def\Bsigma{\mbox{\boldmath $\sigma$}}

\def\Beta{\mbox{\boldmath $\eta$}}
\def\roughly#1{\mathrel{\raise.3ex\hbox{$#1$\kern-.75em%
\lower1ex\hbox{$\sim$}}}}
\def\bm#1{\mbox{\boldmath$#1$\/}}

\def\lsim{\roughly<}
\def\gsim{\roughly>}

\def\prl {Phys. Rev. Lett.}\def\pr{Phys. Rev.}
\def\np{Nucl. Phys.}\def\pl{Phys. Lett.}
\def\PL #1 #2 #3 {Phys.~Lett.~{\bf #1} (#2) #3}
\def\NP #1 #2 #3 {Nucl.~Phys.~{\bf #1} (#2) #3}
\def\PR #1 #2 #3 {Phys.~Rev.~{\bf #1} (#2) #3}
\def\PRL #1 #2 #3 {Phys.~Rev.~Lett.~{\bf #1} (#2) #3}
\def\AP #1 #2 #3 {Ann.~Phys.~{\bf #1} (#2) #3}

\begin{document}
\runauthor{Chaejun Song}

\begin{frontmatter}
\title{Dense Nuclear Matter:\\
Landau Fermi-Liquid Theory\\ and\\ Chiral Lagrangian with Scaling
\thanksref{thesis}}
\author[suny]{Chaejun Song\thanksref{chaejun}}

\address[suny]{Department of Physics \& Astronomy,
        State University of New York,\\
        Stony Brook, New York 11794-3800, USA}

\thanks[thesis]{In part based on the Ph.D. thesis (February 99) 
of Seoul National University.}
\thanks[chaejun]{E-mail: song@silver.physics.sunysb.edu}

\begin{abstract}
The relation between the effective chiral Lagrangian whose
parameters scale according to Brown and Rho scaling(``BR scaling")
and Landau Fermi-liquid theory for hadronic matter is discussed in
order to make a basis to describe the fluctuations under the
extreme condition relevant to neutron stars. It is suggested that
BR scaling gives the background around which the fluctuations are
weak. A simple model with BR-scaled parameters is constructed and
reproduces the properties of the nuclear ground state at normal
nuclear matter density successfully. It shows that the tree level
in the model Lagrangian is enough to describe the fluctuations
around BR-scaled background. The model Lagrangian is consistent
thermodynamically and reproduces relativistic Landau Fermi-liquid
properties. Such  points are  important for dealing  with hadronic
matter under extreme condition. On the other hand it is shown that
the vector current obtained from the chiral Lagrangian is the same
as that obtained from Landau-Migdal approach. We can determine the
Landau parameter in terms of BR-scaled parameter. However these
two approaches provide different results, when applied to the
axial charge. The numerical difference is small. It shows that the
axial response is not included properly in the Landau-Migdal
approach.

\end{abstract}

\end{frontmatter}

\newpage
\tableofcontents

\newpage
\listoffigures
\listoftables

\newpage
\section{Introduction}
\label{intro}

Although QCD which deals with quarks and gluons is believed to be
the fundamental theory for strong interactions, it is generally
accepted  that the appropriate theory at very low energy is the
effective quantum field theory which incorporates the observed
degrees of freedom in low-energy nuclear process, i.e., pions,
nucleons and other low-mass
hadrons~\cite{w67,weinberg,pmr93,kolck,tsp,pmr,PKMR}. The
effective Lagrangian in matter-free or dilute space is governed by
QCD symmetries with its parameters to be determined from
experiments in free space. The energy scale of the experiments in
which one is interested determines which hadrons play an important
role in the theory. For example, it is shown~\cite{PKMR} that one
can integrate out even the pions for the two nucleon systems at
very low energy. According to the results of \cite{PKMR}, the
deuteron and low-energy nucleon-nucleon scattering properties can
be described very accurately by an effective theory given in terms
of the nuclonic degrees of freedom only with a cutoff around the
natural scale of the theory which for low energy is the pion mass.

Since there is currently growing interplay between the physics of
hadrons and the physics of compact objects in astrophysics through
the properties of hot and dense environments, we want to extend
such successful strategy of effective field theories to a dense
medium. First of all, our main goal is to understand the
properties of dense hadronic medium which can be tested in various
heavy ion collisions. Recent dilepton experiments (CERES and
HELIOS-3) gave us very important information on the properties of
hadrons in dense medium, which we will detail later.
Understanding the properties of dense hadronic medium through
heavy ion collision experiments is essential for understanding the
properties of neutron stars which are believed to be formed in the
center of supernovae at the time of explosions. Especially the
determination of the maximum neutron star mass is one of the most
important issues in astrophysics in explaining controversies
between the observations and theoretical estimations, and it will
give some hints for the detectibility of the gravitational wave
detectors.
A dense matter makes the probed energy scale larger than that in
free space. Therefore we have to introduce more massive degrees of
freedom which are usually vector meson and/or higher order
operators in the nucleon fields. We must also consider a new
energy scale, Fermi energy of nucleons in bound system.

The standard strategy to attack the dense hadronic matter is to
obtain the ground state of matter and compute the excitations
around it, based on an effective Lagrangian whose parameters are
obtained in free space. Although such an approach can give
satisfactory results with a sufficient number of parameters, it is
not obvious whether we can extend the results for the different
density. When the extension does not work, we usually face with
very complicated loop diagrams which may lead to the impasse.

Another strategy is to start from the in-medium Lagrangian which
is built on the reasonable assumptions, instead of deriving the
hadronic matter properties from the matter-free effective
Lagrangian defined in free space. In this approach one regards its
mean field solution as a solution of the Wilsonian effective
action in which the high energy modes are integrated out into the
coefficient. We can compare it with the well-known Landau
Fermi-liquid theory, which works at low energy excitation in
strongly correlated Fermi system. Landau Fermi-liquid theory is
described by quasiparticles, which are the low energy excitations
in Fermi liquid, and their interactions under the assumption of
one-to-one correspondence between the quasiparticle in the liquid
and the particle in a non-interacting gas. It is a fixed point
theory~\cite{shankar,froehlich} under Wilsonian renormalization to
the Fermi surface~\cite{bg90} with $\Lambda /k_F\rightarrow 0$ if
Bardeen-Cooper-Schrieffer(BCS) instability does not exist, where
$\Lambda$ is the cutoff of the theory relative to the Fermi
surface and $k_F$ is Fermi momentum. Since the result after the
repeated renormalization does not depend on $k_F$, the argument
for Fermi liquid holds as long as there is no phase transition.
A famous example of such a Lagrangian is Walecka model. Its
extension and justification were studied recently by Furnstahl
{\it et al}.\ \cite{tang,tang2,tang3}. Their Lagrangian is
constrainted by QCD symmetry and by Georgi's naturalness condition
and the coupling constants in it are tuned in order to describe
nuclear ground state. Bulk properties of nuclei are described by
it very successfully. But it is somewhat unclear to know how to
approach the fluctuation on the ground state.

In this review we will approach the nuclear matter in a different
way. We wish to apply the strategy of the effective field theory
to the in-medium theory. We assume that the in-medium effective
Lagrangian has the same structure as in free space according to
the symmetry constraint of the fundamental theory QCD but that its
parameters are modified in medium. It means that the effect of
embedding a hadron in matter appears mainly in the change of the
``vacuum," i.e., quark and gluon condensate in QCD variables and
the parameters in an effective theory. In this strategy the
density-dependent parameters include many-body correlations.

Brown-Rho(BR) scaling~\cite{BR91} is one specific way to define
such in-medium parameters. Brown-Rho scaling is the scaling of the
dynamically generated masses of hadrons which consist of chiral
quarks, i.e., $u$ and $d$ quarks. Brown and Rho phrased the
scaling with the large $N_c$ Lagrangian, i.e., Skyrmion, under the
assumption that the chiral symmetry and the scale symmetry of QCD
are relevant. Their Lagrangian is implemented with scale anomaly
of QCD, too. The masses and pion decay constant of this QCD
effective theory scale universally:
\be
\Phi (\rho) \approx \frac{f^\star_\pi (\r )}{f_\pi}
\approx\frac{m_v^\star (\r )}{m_v} \approx \frac{m^\star_\s (\r
)}{m_\s} \approx \frac{M^\star (\r )}{M}. \ee The star represents
in-medium quantities here. $v$ is vector meson degree of freedom
and $s$ an isoscalar scalar meson which has a mass $\sim$ 500 MeV
in nuclear matter. $M$ represents a free nucleon mass and
$M^\star$ a scaled nucleon mass which is somewhat different from
the Landau effective mass discussed in Section \ref{vector}.

It is known that BR scaling describes that the light-quark vector
meson property in the extreme condition very successfully. One can
make the extreme condition through relativistic heavy ion
collisions. Specially the dileptons provide a good probe of the
earlier dense and hot stage of relativistic heavy ion collision
since the interaction of leptons is not subject to the strong
interactions of the final state. CERES(Cherenkov Ring Electron
Spectrometer) collaboration observed in heavy ion collision
(S$+$Au) that the dilepton production with invariant masses from
250 MeV to about 500 MeV is enhanced much more than the predicted
from the superposition of $pp$ collision~\cite{CERES}. And
HELIOS-3(CERN Super Proton Synchrotron detector)~\cite{helios3}
also observed the dilepton enhancement in S$+$W. It is shown by
Li, Ko, and Brown~\cite{LKB} that a chiral Lagrangian with
BR-scaled meson masses describes most economically and beautifully
the enhancement, which is found to come from the dropping vector
meson masses in dense matter. Furthermore the excitation into the
kaonic direction above the given ground state seems to describe
the properties of kaons in medium~\cite{kaos,fopi} with scaled
parameters~\cite{gebgsi}.

Since BR scaling gives the universal scaling mass relation among
hadrons, it must work for the baryon properties in medium. In
recent works~\cite{FR96,sbmr,smr,frs,rmdb97} it has been discussed
how the BR scaling, which is applied to meson properties
successfully, can be applied to the baryon property in dense
medium and how it can be extrapolated to a hadronic matter under
extreme conditions from the known normal nuclear matter. The
construction of such bridges will be necessary to understand the
various phenomena in relativistic heavy ion collisions and in
compact stars in the universe. For those purposes we develop
arguments for mapping the effective chiral Lagrangian whose
parameters are governed by BR scaling to Landau Fermi-liquid
theory. We will relate the meson mass scaling with baryon mass
scaling and also relate matter properties with BR scaling
parameter, which implies the vacuum structure characterized by
quark condensate, via Landau parameter. Fermi liquid theory may
work up to chiral phase transition, though the relevant degrees of
freedom are changed to quasiquarks from quasihadrons.

In this review we approach the aim in two ways. The first is to
obtain the ground state with BR scaling. How the Fermi surface is
obtained in BR scaling framework is not yet understood clearly. So
people usually assume that the ground state of hadronic matter is
determined by the conventional matter from a standard many-body
theory and use density-dependent effective chiral Lagrangians to
compute mesonic fluctuations above the ground state. Though
various fluctuation phenomena can be described successfully in
this way, such a treatment gives no constraints for consistency
between the excitations and the ground state. Though BR scaling is
applied very successfully to describe meson properties in medium
\cite{LKB,LBLK96}, the way to deal with the matter properties is
disconnected from BR scaling. Such a procedure is not satisfactory
in going to the higher density region from the normal nuclear
matter density. We can take the kaon condensation in neutron star
as an example. In dealing with it, we come across a change of the
ground state from that of a non-strange matter to a strange
matter. The works up to date~\cite{kl86,lbmr95} treated $KN$
interaction and the ground state separately. This is not
satisfactory. Ground state properties might effect the
condensation. For example, the effect of the four-Fermi
interactions which play an important role in determining ground
state, suppresses a pion condensation~\cite{mpw92}. So we must
deal with the whole bulk involving the ground state and
excitations on top of it on the same footing. We assume that the
ground state is given by the same effective Lagrangian that is
supposed to include higher order corrections as the mean field of
the BR-scaled chiral Lagrangian. We want to make an initial step
for dealing with the ground state and the fluctuation on top of it
on the same basis. We bridge these two properties
 by constructing a simple model whose parameters scale
in the manner of BR and which describes nuclear matter properties
well. It is shown  that the model can be mapped to Landau
Fermi-liquid theory.

The next step to achieve our aim is to identify the parameters of
the BR-scaled effective Lagrangian with the fixed point quantities
in Landau Fermi-liquid theory, given a hadronic matter with a
Fermi surface. With this identification, certain mean field
quantities of heavy meson (e.g.\ $\rho$, $\omega$) can be related
to BR scaling through the Landau parameters. We show how such
arguments work for the electromagnetic currents in nuclear matter.
We can link a set of BR-scaled parameters at nuclear matter
density with the orbital gyromagnetic ratio in terms of the Landau
parameter $F_1^\omega$ which comes from integrated-out isoscalar
vector degree of freedom in the effective Lagrangian. Then we will
try to derive the corresponding formulas for the axial current in
a similar way.

This review is organized as follows. In Section 2 Landau
Fermi-liquid theory and its interpretation in terms of
renormalization group language are summarized briefly. And we show
how thermodynamic observables are related to the relativistic
Landau parameters. In Section 3 the strategy of an effective field
theory and how it is applied to a dense matter are explained. We
proceed to discuss
how nuclear matter described by an in-medium effective Lagrangian
can be identified with Landau Fermi liquid. The model of
Furnstahl, Serot and Tang~\cite{tang} (referred to FTS1) which
imposes the symmetries of QCD is examined as an example of the
application of a general strategy of an effective chiral theory to
a medium. In Section 4 Brown-Rho scaling is derived with
QCD-oriented effective Lagrangian. A model where BR scaling
governs the parameters of a chiral Lagrangian and determines the
background at a given density is constructed in Section 5 in order
to describe in weak coupling the same physics as FTS1 which has
strong coupling in the form of a large anomalous dimension of a
dilaton. It describes well normal nuclear matter properties and
has thermodynamic consistency and Fermi-liquid structure needed
for the extrapolation to higher density region. In Section 6
vector current and axial charge transition matrix elements for a
nucleon above the given Fermi sea in Landau-Migdal theory and in
chiral effective Lagrangian are calculated and compared. A summary
and comments on some unsolved problems are given in Section 7 and
Section 8 respectively. Appendix A shows how sensitive the
equation of state is to the many-body correlation parameters for
$\rho >\rho_0$. Appendix B shows how to compute relativistically
the pionic contribution to Landau parameter $F_1$ by means of
Fierz transformation. And the vector-mesonic contribution to $F_1$
and to electromagnetic convection current is calculated in
Appendix C relativistically with random phase approximation.
\section{Landau Fermi-liquid theory}

We discuss briefly Landau Fermi-liquid theory in this section,
before presenting the relation between the chiral effective theory
for nuclear matter and Landau Fermi-liquid theory. A mini-primer
on Landau Fermi-liquid theory is given in the first subsection to
define the quantities involved. In the second subsection, it is
discussed that Landau Fermi-liquid theory is considered as an
effective theory and is shown to be a fixed point theory in
renormalization group(RG) language. And we will show how the
thermodynamic quantities are related to relativistic Landau
parameters in Section \ref{relation}.
\subsection{A mini-primer}\label{concept}

Landau's Fermi-liquid theory is a semi-phenomenological approach
to strongly interacting normal Fermi systems at small excitation
energies. The elementary excitations of the Fermi-liquid, which
correspond to single particle degrees of freedom of the Fermi gas,
are called quasiparticles in Landau Fermi-liquid theory. It is
assumed that a one-to-one correspondence exists between the
low-energy excitations of the Fermi liquid near Fermi surface,
i.e., quasiparticles, and those of a non-interacting Fermi gas. A
quasiparticle state of the interacting liquid is obtained by
turning on the interaction adiabatically at the corresponding
state of non-interacting Fermi gas. The quasiparticle properties,
e.g.\ the mass, in general differ from those of free particles due
to interaction effects. In addition there is a residual
quasiparticle interaction, which is parameterized in terms of the
so called Landau parameters.

The adiabatic process described above is possible in the vicinity
of the Fermi surface only. Let us see the Fermi liquid at $T=0$.
Since Pauli exclusion principle makes the states below the Fermi
surface filled, quasiparticle with energy $\ve$ loses energy less
than $\ve -\ve_F$ when colliding the background particles. It
means that the quasiparticles which can interact with the
quasiparticle are those with an energy within $|\ve -\ve_F |$ of
the Fermi surface. And the final state momenta are also restricted
by $\ve^\prime <\ve$. Pauli exclusion principle and the
corresponding rarity of final states make the quasiparticle life
time proportional to $|\ve -\ve_F|^{-2}$ at $T=0$ case.

Fermi-liquid theory is a prototype effective theory, which works
because there is a separation of scales. The theory is applicable
to low-energy phenomena, while the parameters of the theory are
determined by interactions at higher energies. The separation of
scales is due to the Pauli principle and the finite range of the
interaction. Pauli principle makes the low energy quasiparticle
physics possible near the Fermi surface and the finite range of
interaction makes a few quasiparticles around Fermi surface, who
appear by the small change of the energy in low energy physics,
form a gas. Fermi-liquid theory has proven very useful
\cite{baym-pethick} for describing the properties of e.g.\ liquid
$^3$He and provides a theoretical foundation for the nuclear shell
model~\cite{migdal} as well as nuclear dynamics of low-energy
excitations~\cite{speth,BWBS}.

The interaction between two quasiparticles ${\pb_1}$ and ${\pb_2}$
at the Fermi surface of symmetric nuclear matter can be written in
terms of a few spin and isospin invariants \cite{BSJ}
\be\label{qpint} { f}_{\pb_1\sigma_1\tau_1,\pb_2\sigma_2\tau_2}&=&
\frac{1}{N(0)}\left[F(\cos \theta_{12})+F^\prime(\cos
\theta_{12})\taub_1\cdot \taub_2+G(\cos \theta_{12})\sigmab_1\cdot
\sigmab_2\phantom{\frac{\qb^{\,
2}}{k_F^2}}\right.\nonumber\\&+&\left. G^\prime(\cos
\theta_{12})\sigmab_1\cdot \sigmab_2\taub_1\cdot\taub_2
+\frac{\qb^{\, 2}}{k_F^2}H(\cos
\theta_{12})S_{12}(\qbhat)\right.\nonumber\\&+&\left.
\frac{\qb^{\, 2}}{k_F^2}H^\prime(\cos \theta_{12})S_{12}
(\qbhat)\taub_1\cdot\taub_2\right] \ee where $\theta_{12}$ is the
angle between ${\pb_1}$ and ${\pb_2}$ and $N(0)=\frac{\gamma
k_F^2}{(2\pi^2)}\left(\frac{dp}{d\ve}\right)_F$ is the density of
states at the Fermi surface. In this review natural units where
$\hbar=1$ are used. The spin and isospin degeneracy factor
$\gamma$ is equal to 4 in symmetric nuclear matter. Furthermore,
$\qb=\pb_1-\pb_2$ and \bq\label{tensor} S_{12}(\qbhat) = 3
\sigmab_1\cdot\qbhat\sigmab_2\cdot\qbhat -
\sigmab_1\cdot\sigmab_2, \eq where $\qbhat = \qb/|\qb |$. The
tensor interactions $H$ and $H^\prime$ turn out to be important
for the axial charge~\cite{BSJ}
The functions $F, F^\prime, \dots$ are expanded in Legendre
polynomials, \bq F(\cos \theta_{12})=\sum_\ell F_\ell P_\ell(\cos
\theta_{12}), \eq with analogous expansion for the spin- and
isospin-dependent interactions. The energy of a quasiparticle with
momentum $p=|{\pb}|$, spin $\sigma$ and isospin $\tau$ is denoted
by $\epsilon_{p,\sigma,\tau}$ and the corresponding quasiparticle
number distribution by $n_{p,\sigma,\tau}$. From now on the spin
and isospin indices $\sigma$ and $\tau$ will be omitted from the
formulas to avoid overcrowding, except where needed to avoid
ambiguities.
\be
n_p(\Br ,t)&=&n_p^0(\ve^0_p)+\delta n_P(\Br ,t)\label{denfl}\\
\ve_p(\Br ,t)&=&\ve_p^0+ \sum_{\sigma^\prime ,\tau^\prime}
\int\frac{d^3p^\prime}{(2\pi)^3} f_{pp^\prime}\delta
n_{p^\prime}(\Br ,t)\label{enefl} \ee where $\delta n_p(\Br ,t)$
is the long wave length excitations from the ground state $n_p^0
(\ve^0)$ in the vicinity of Fermi surface. The space and time
dependence of the quantities will also be omitted, e.g.,
$\ve\equiv\ve ({\Br },t)$. The Landau effective mass and velocity
of a quasiparticle on the Fermi surface is defined by \bq
\left.\frac{d\ve_{p}}{dp}\right|_{p=\kF} = \frac{\kF}{m_L^\star}
\equiv v_F^\star .\label{Landaumass} \eq

We must note that the total current is not
\be
\Jb_{locQP} =\sum_{\sigma ,\tau}\int \frac{d^3p}{(2\pi )^3}
\vb_F^\star \delta n_p \ee in Landau Fermi-liquid theory. The
quasiparticle distribution $n_p(\Br ,t)$ obeys
\be
\frac{\del n_p}{\del t}+\frac{d\Br}{dt}\cdot\Bnabla_rn_p
+\frac{d\pb}{dt}\cdot\Bnabla_p n_p=0. \ee The key assumption of
Landau Fermi-liquid kinetic theory is that $\ve_p(\Br ,t)$ plays
the role of the quasiparticle Hamiltonian;
\be
\frac{d\pb}{dt}&=&-\Bnabla_r\ve_p\\
\frac{d\Br}{dt}&=&\Bnabla_p\ve_p . \ee The kinetic equation is
\be
\frac{\del n_p}{\del t}+\Bnabla_r n_p\cdot\Bnabla_p\ve_p
-\Bnabla_pn_p\cdot\Bnabla_r\ve_p=I[n_{p^\prime}]\label{balance}
\ee where $I[n_{p^\prime}]$ is an internal collision integral
which represents the sudden change of quasiparticle momenta. We
consider the system without external forces. Integrating over
$\pb$, that cancels the effect of the internal collision under the
assumption that the quasiparticle energy, momentum and number are
locally conserved, (\ref{balance}) becomes (to order $\delta
n_p(\Br ,t)$), by integration by part,
\be
\sum_{\sigma ,\tau}\int \frac{d^3p}{(2\pi )^3} \frac{\del\delta
n_p}{\del t}+ \Bnabla_r \cdot (n_p\Bnabla_p\ve_p)
&=&\frac{\del\delta\rho}{\del t}+ \sum_{\sigma ,\tau}\Bnabla_r
\cdot\int \frac{d^3p}{(2\pi )^3} \delta n_p\Bnabla_p \ve_p
+\delta\ve_p\Bnabla_p n_p^0\nonumber\\ &=&0 \ee where $\delta\rho$
is the total particle (or quasiparticle) density fluctuation. We
can easily see that the conserved current is
\be
\Jb &=&\sum_{\sigma ,\tau}\int \frac{d^3p}{(2\pi
)^3}n_p\Bnabla_p\ve_p \nonumber\\ &=&\sum_{\sigma ,\tau}\int
\frac{d^3p}{(2\pi )^3}\vb_F^\star\delta n_p^{loc} \ee with the
excitation from the local equilibrium $n_p^0(\ve (\Br ,t))$;
\be
n_p(\Br ,t)=n_p^0(\ve (\Br ,t))+\delta n_p^{loc}(\Br
,t)\label{loceq}. \ee Comparing (\ref{loceq}) with (\ref{denfl})
and (\ref{enefl}), we obtain
\be
\delta n_p^{loc} =\delta n_p+\frac{\del n_p^0}{\del\ve}
\sum_{\sigma^\prime ,\tau^\prime} \frac{d^3p^\prime}{(2\pi
)^3}f_{pp^\prime}\delta n_p^\prime. \ee The modification
$\Jb_{locQP}-\Jb$ can be interpreted as the effect of the return
flow of the surrounding matter to the localized wave packet
carrying $\Jb$ and is called back-flow current.
\subsection{Renormlization group approach}\label{rgliq}

Landau Fermi-liquid theory was based on Landau's reasonable
intuition. After his work, his theory was derived and justified
microscopically~\cite{abri}. Recent development of Wilsonian RG
method in medium~\cite{bg90} provides us a new understanding of
Fermi liquid theory. Landau Fermi-liquid theory is a fixed point
theory described by marginal coupling, i.e. $\Fc$,
$m_L^\star$~\cite{shankar}. In this section we will review the RG
arguments for Fermi liquid theory.

\begin{figure}
\setlength{\epsfysize}{2.0in} \centerline{\epsffile{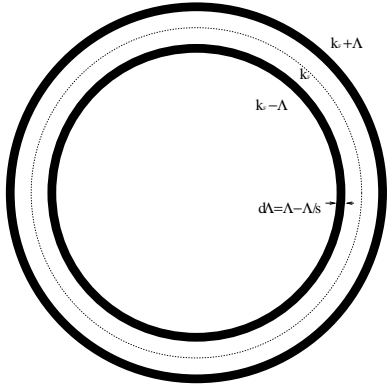}}
\caption{Our interest lies in the spherical shell which is of the
thickness of $\Lambda$ on either side of Fermi sphere. In RG
transformation we eliminate the mode within the two thick-lined
shells whose thickness is $d\Lambda$.} \label{fsphere}
\end{figure}

To do this we consider a nonrelativistic system of spinless
fermions whose Fermi surface is spherical characterized by $k_F$
for simplicity. Then non-interacting one particle Hamiltonian near
Fermi surface is
\be
H=\frac{\Kb^2}{2m}-\frac{k_F^2}{2m}\approx \frac{k}{m}k_F \equiv
v_Fk \ee with $k=|\Kb |-k_F$. The free fermion field action
\be
S_0=\int_\Lambda \bar{\psi}(\omega k\Omega)(i\omega -vk) \psi
(\omega k\Omega )\label{free} \ee in momentum space where
\be
\int_\Lambda:=\int \frac{d\Omega}{(2\pi)^2}\int_{-\Lambda}^\Lambda
\frac{dk}{(2\pi)}\int_{-\infty}^\infty \frac{d\omega}{(2\pi)}. \ee
 $\bar{\psi}$ and $\psi$ are Grassmannian eigenvalue
with fermion operator $\hat\Psi$; $\hat\Psi |\psi\ket =\psi
|\psi\ket$ and $\bra\bar{\psi}|\hat\Psi^\dagger =\bra\bar\psi
|\bar\psi$. Note that a shell of thickness $\Lambda$ on either
side of the Fermi surface is taken for low energy physics as seen
in Fig.\ \ref{fsphere}.
Pauli exclusion principle lets only small deviations from the
Fermi surface, not from the origin, be important in low energy
physics of fermion matter. This defines the starting point of an
{\it in-medium} renormalization group procedure.

The first step for the renormalization is decimation; to integrate
out the high energy mode whose momentum is larger than $\Lambda
/s$ and to reduce the cutoff from $\Lambda$ to $\Lambda /s$. For
example, the free action (\ref{free}) becomes
\be
S_0^{eff} =\int \frac{d\Omega}{(2\pi)^2} \int_{-\Lambda
/s}^{\Lambda /s} \frac{dk}{(2\pi)}\int_{-\infty}^\infty
\frac{d\omega}{(2\pi)} \bar{\psi}(\omega k\Omega)(i\omega -vk)
\psi (\omega k\Omega ). \ee Next we rescale the momenta in order
to compare the old and the new;
\be
(\omega ,\Bk )\to (s\omega ,s\Bk ). \ee The last step is to absorb
the uninteresting multiplicative constant by
\be
\psi \to s^{-3/2}\psi . \ee The RG transformation consists of
these three steps. After such RG transformation, the free action
(\ref{free}) returns to the old. When a coupling is turned on, the
coupling is called relevant if it increases after RG
transformation. If it decreases, it is called irrelevant and if it
remains fixed like the free action, it is called marginal.

Now let us turn on four-Fermi interaction. The appropriate action
is
\be
S&=&\int_\Lambda \bar{\psi}[i\omega-v_F^\star k]\psi
+\delta\mu^\star \int_\Lambda \bar{\psi}\psi\nonumber\\ &
&+\frac{1}{2!2!} \int_{\Lambda_4}
u(4,3,2,1)\bar{\psi}(4)\bar{\psi}(3)\psi (2)\psi (1) \label{Flag}
\ee where $(i)$ represents $(\omega_i,k_i,\Omega_i)$ and
\be
\int_{\Lambda_4}:=\prod_i\int\frac{d\Omega}{(2\pi)^2}
\int_{-\Lambda}^\Lambda\frac{dk}{(2\pi )^2}
\int_{-\infty}^{\infty}\frac{d\omega}{(2\pi)} \Theta (\Lambda
-|\Bk_4|).\label{4-int} \ee with a cutoff function for $k_4$,
$\Theta (\Lambda -|\Bk_4|)$, which is needed to make all the
momenta lie within the band of width $2\Lambda$ around the Fermi
surface. Here $v_F^\star=k_F/m^\star$ and \be m^\star =
\frac{1}{Z(1+\frac{m}{k_F}\frac{\del\Sigma}{\del k})} \ee is the
effective mass of the nucleon which will be equal to the Landau
mass $m_L^\star$ as will be elaborated on later. $\Sigma (\omega
,k)$ is self-energy and $1/Z =1+i\frac{\del\Sigma}{\del\omega}$.
The effective mass arises because the eliminated mode contributes
to $i\omega \bar{\psi}\psi$ and $k\bar{\psi}\psi$ differently. By
defining $\psi^\prime =s^{-3/2}Z^{-1/2}\psi$ we fix the
coefficient of $i\omega \bar{\psi}\psi$ and define the effective
mass. The term with $\delta \mu^\star$ is a counter term added to
assure that the Fermi momentum is fixed (that is, the density is
fixed). What this term does is to cancel loop contributions
involving the four-Fermi interaction to the nucleon self-energy
(i.e., the tadpole) which contributes marginally so that the
$v_F^\star$ is at the fixed point. Since other contributions which
can be written as $-\Sigma\bar\psi\psi$ are irrelevant, $m$ moves
to $m^\star$ in earlier stage of renormalization but becomes a
fixed point characterized by some $m^\star$. This means that the
counter term essentially assures that the effective mass $m^\star$
be at the fixed point. Without this procedure, the term quadratic
in the fermion field would be ``relevant'' and hence would be
unnatural~\cite{shankar}.

\begin{figure}
\setlength{\epsfysize}{1.8in} \centerline{\epsffile{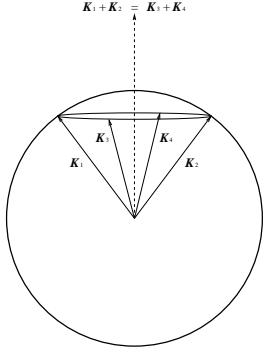}}
\caption{Momentum conservation and the physics near Fermi surface
force the angle between $\Kb_3$ and $\Kb_4$ to be the same as that
between $\Kb_1$ and $\Kb_2$ unless $\Kb_2=-\Kb_1$.} \label{2sum}
\end{figure}

Let us see the quartic coupling $u$ at tree level. The cutoff
function $\Theta (\Lambda -|\Bk_4|)$ in (\ref{4-int}) makes the
coupling depend on angles on the Fermi surface. Since all the
momenta are on the thin spherical shell near $k_F$, momentum
conservation makes the new quartic coupling $u^\prime
(\omega^\prime ,k^\prime , \Omega^\prime )$ decay after the RG
transformation at tree level except for the two cases. One is that
$\Kb_3$ and $\Kb_4$ are the rotation of $\Kb_1$ and $\Kb_2$ around
the $\Kb_1+\Kb_2$ as seen in Fig.\ \ref{2sum}.
In that case, the opening angle $\cos^{-1} (\ubhat_1\cdot\ubhat_2
)$ is fixed where $\ubhat_i$ is a unit vector in the direction of
$\Kb_i$. The other, so-called BCS coupling, is that $\ubhat_1
=-\ubhat_2$ and $\ubhat_3=-\ubhat_4$. These cases are represented
by functions;
\be
u(\theta_{12}=\theta_{34})&=&\Fc (\theta_{12},\vartheta)\\
u(\theta_{13}=\theta_{24})&=&\V(\theta_{13})\nonumber \ee where
$\theta_{ij}=\ubhat_i\cdot\ubhat_j$ and $\vartheta$ is the angle
between the planes containing $(\Kb_1,\Kb_2)$ and $(\Kb_3,\Kb_4)$
respectively. $\Fc$ and $\V$ are marginal at tree level.

\begin{figure}
\setlength{\epsfysize}{1.8in} \centerline{\epsffile{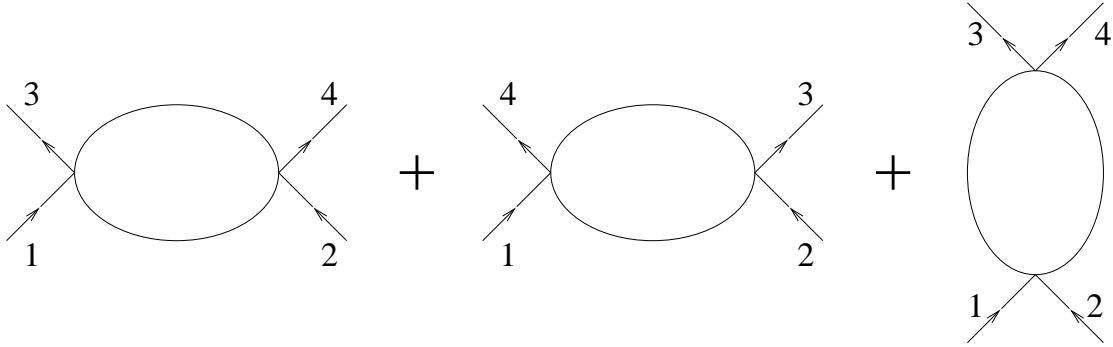}}
\caption{One loop diagrams for the renormalization of the
marginal quartic couplings $\Fc$ and $\V$. } \label{loop1}
\end{figure}

Then the next question is how $\Fc$ and $\V$ evolve at one loop
level. Figure \ref{loop1} shows the one loop diagrams for the
evolution.
Integrating out the momentum shell of thickness $d\Lambda$ at
$k=\pm\Lambda$ and sending $\Lambda /k_F\to 0$, all the diagrams
in Fig.\ \ref{loop1} do not contribute to $\Fc$. So $\Fc$ is left
marginal to the one loop order. In the case of $\V$, the third
diagram in Fig.\ \ref{loop1} makes a flow. When we expand $\V$ in
terms of angular momentum eigenfunction, $\V$ becomes irrelevant
only if all the $\V_l$'s are repulsive. However, if any $\V_l$ is
attractive, it becomes relevant and causes BCS instability. We
call it BCS channel. Since Landau theory assures that there is no
phase transition for one-to-one correspondence between particles
and quasiparticles, BCS channel destabilizes in Landau
Fermi-liquid theory.

If we divide the angular part of the shell integration in the
action (\ref{free}) into the cells of size
$\Delta\Omega\sim\Lambda /k_F$, the shell is split into $\sim
k_F/\Lambda$ cells. By analogy with large $N$ theory each cell
corresponds to one species, i.e., $N \sim k_F/\Lambda$. $1/N$
expansion tells that all higher loop corrections except for the
bubbles of the first diagram in Fig.\ \ref{loop1} are down by
powers of $1/N=\Lambda /k_F$. And the contributions of bubbles of
the first diagram survive but are irrelevant to the $\beta$
function. Only one loop contributions to $\beta$ function survive
in the large $N$ limit. So the four-Fermi interactions in the
phonon channel $\Fc$ are also at the fixed points in addition to
the Fermi surface fixed point with the effective mass $m^\star$.
Note that only forward scattering $\Fc (\vartheta =0)$ is
important for responses to soft probes, since nonforward
amplitudes in loop calculations are also suppressed by $1/N$.
Six-Fermi and higher-Fermi interactions are irrelevant and
contribute at most screening of the fixed-point constants. The
Landau parameter $F$ can be identified with forward scattering
$\gamma\frac{m^\star}{2\pi^2 k_F}\Fc (\vartheta =0)$
easily~\cite{shankar}. We arrive at the Fermi-liquid fixed point
theory in the absence of BCS interactions.

Chen, Fr\"ohlich, and Seifert~\cite{froehlich} obtain the same
result in the $1/N$ expansion where their $N$ is taken to be
$N\sim \lambda$ with $1/\lambda$ being the width of the effective
wave vector space around the Fermi sea which can be considered as
the ratio of the microscopic scale to the mesoscopic scale. More
specifically if one rescales the four-Fermi interaction such that
one defines the dimensionless constant $g \sim u_0/k_F^2$ where
$u_0$ is the leading term (i.e., constant term) in the Taylor
series of the quantity $u$ in (\ref{Flag}), then the fermion wave
function renormalization $Z$, the Fermi velocity $v_F$  and the
constant $g$ are found not to flow up to order ${\O}(g^2/N)$. Thus
in the large $N$ limit, the system flows to Landau fixed point
theory to all orders of loop corrections. This result is correct
provided there are no long-range interactions and if the BCS
channel is turned off.
\subsection{Relations for relativistic Fermi-liquid}\label{relation}

In this section we briefly summarize the relations of physical
properties of the relativistic Landau Fermi-liquid. The extension
of Landau Fermi-liquid theory to relativistic region for high
density matter is found in \cite{baymchin}. It should be pointed
out that one can use all the standard Landau Fermi-liquid
relations established below in our calculations once fixed point
quantities are identified in the chiral Lagrangian.
\subsubsection{Compression modulus and $F_0$}

The density of states at Fermi surface is
\be
N(0)&=&\left( \frac{\del \rho}{\del\ve}\right)_{k_F} =\frac{\gamma
k_Fm_L^\star}{2\pi^2}.\label{no} \ee Note that $m_L^\star
=\sqrt{m_N^2+k_F^2}$ for relativistic non-interacting Fermi gas.
$m_L^\star$ defined by (\ref{Landaumass}) includes the kinetic
energy for relativistic Fermi liquid. The chemical potential is
defined by
\be
\mu\equiv\frac{\del\E}{\del\rho}=\ve_F\label{chemi} \ee where $\E$
represents the energy per volume. Using (\ref{no}) and
(\ref{chemi}) one can derive the relativistic relation between
compression modulus which represents the change of volume with
pressure and $F_0$ in the same way as the nonrelativistic one:
\be
K&=&9\rho\frac{\del\mu}{\del \rho} =9\rho\frac{\del}{\del
\rho}\left(\ve^0_p+ \int d\tau^\prime
f_{pp^\prime}n_{p^\prime}\right)\nonumber\\
&=&\frac{3k_F^2}{m_L^\star}(1+F_0)\label{commodul} \ee Here $\rho$
is the baryon number density and $n_p$ is the Fermi distribution
function for the state of momentum $p$.
\subsubsection{Landau effective mass}

In deriving the Landau mass formula, we shall compare the rest
frame and the boosted frame with very small velocity $\ub$ and
check Lorentz symmetry. Since $u\equiv |\ub |$ is small we neglect
the order of $u^2$ in the process of the derivation. (Note that
$\gamma_u =(1-u^2)^{-1/2}\approx 1$ in that order. )

Let us first derive the relativistic relation between the Landau
effective mass and the velocity dependence of the quasiparticle
interaction. When we add a particle (or equivalently
quasiparticle) of momentum $\pb$ in the rest frame to the system,
the energy and the momentum of the system increases by $\pb$ and
$\ve_p(0)$ respectively. In a moving frame with velocity $-\ub$,
the momentum and energy increase by
\be
\pb^\prime &=&\pb-\ubhat (\ubhat\cdot\pb )(1-\gamma_u )
+\ve_p(0)\ub\gamma_u\nonumber\\ &\approx
&\pb+\ve_p(0)\ub\label{pboost}\\
\ve_{p^\prime}(u)&=&(\ve_p(0)+\pb\cdot\ub)\gamma_u\nonumber\\
&\approx &\ve_p(0)+\pb\cdot\ub .\label{eboost} \ee From
$\ve_{p}(u)$. (\ref{pboost}) and (\ref{eboost}), we have
\be
\ve_p(u)=\ve_{p-\ve_p(0)u}(0)+\pb\cdot\ub =
\ve_p(0)-\ve_p(0)\ub\cdot\Bnabla_p\ve_p(0)+\pb\cdot\ub .\label{eb}
\ee and
\be
\ve_{p}(u)=\ve_{p}(0) +\int
d\bar{\tau}f_{p\bar{p}}(n_{\bar{p}}(u)-n_{\bar{p}}(0)).\label{ep}
\ee Since $n_{p^\prime}(u)=n_p(0)$, we can obtain
\be
n_{\bar{p}}(u)\approx n_{\bar{p}-\ve_{\bar{p}}u}(0)
=n_{\bar{p}}(0)-e_{\bar{p}}(0)\ub\cdot\Bnabla_{\bar{p}}
n_{\bar{p}}(0) \ee using (\ref{pboost}) and (\ref{eboost}).
Then (\ref{ep}) becomes
\be
\ve_p(u)=\ve_p(0)
-\int d\bar{\tau}f_{p\bar{p}}\ve_{\bar{p}}(0)
\ub\cdot\Bnabla_{\bar{p}}n_{\bar{p}}(0). \ee Comparing it with
(\ref{eb})
\be
\pb=\ve_p(0)\Bnabla_p\ve_p(0) -\int
d\bar{\tau}f_{p\bar{p}}\ve_{\bar{p}}(0)
\Bnabla_{\bar{p}}n_{\bar{p}}(0).\label{co} \ee In the ground state
\be \Bnabla_{\bar{p}}n_{\bar{p}}(0) =-\delta (\ve_{\bar{p}}-\mu )
\frac{\del \ve_{\bar{p}}}{\del\bar{p}}\pbbhat \ee gives
\be
p=\ve_p\frac{\del \ve_p}{\del p}+ \int d\bar\tau\delta
(\ve_{\bar{p}}-\mu )\ve_{\bar{p}}f_{p\bar{p}}\pbhat\cdot
\pbbhat\frac{\del \ve_{\bar{p}}}{\del\bar{p}}. \ee The chemical
potential $\mu$ is $\ve_{p_F}$. Thus (\ref{co}) becomes on the
Fermi surface
\be
\mu\left( \frac{\del \ve_p}{\del p}\right)_{p_F}
=p_F-\mu\frac{\gamma p_F^2}{2\pi^2}\frac{f_1}{3}. \ee Using
(\ref{no}), one obtains
\be
\frac{m_L^\star}{\mu} =1+\frac{F_1}{3}. \label{mass}\ee { This is
the relativistically extended relation of the famous Landau mass
formula \bq \label{eff-mass} \frac{m_L^\star}{M}=1 +
\frac{F_1}{3}, \eq that follows from Galilean invariance in the
same way~\cite{baym-pethick}. }
\subsubsection{First sound velocity}

The first sound is a density oscillation mode under the
circumstances where there are many quasiparticle collisions during
the time of interest. The sufficiently often collisions produce
the required local equilibrium in the time scale of the period of
motion.

When first sound wave gives small change in density of a static
homogeneous relativistic fluid in a comoving frame without
changing entropy, the continuity equation requires
\be
\frac{\del}{\del t}\delta\rho =-\rho\Bnabla\cdot\vb
.\label{numberc} \ee Under the condition of the fixed entropy
\be
kds =\frac{1}{T}(p\delta v +\delta \ve
)=\frac{1}{\rho^2T}\{\rho\delta\ve -(p+\E )\delta\rho\} =0 \ee
with the entropy per particle $ks$, the volume per particle
$v=1/\rho$, and the energy per particle $\ve =\E /\rho$, the
relatvistic equation of motion becomes
\be
\frac{\del\vb}{\del t}=-\frac{1}{p+\E}\Bnabla\delta p
=-\frac{\delta p}{\rho\delta\E} \Bnabla\delta\rho .\label{reom}
\ee Applying (\ref{reom}) to (\ref{numberc}), we obtain
\be
\frac{\del^2}{\del t^2}\delta\rho =\left( \frac{\del
p}{\del\E}\right)\Bnabla^2\delta\rho . \ee Thus the first sound
velocity of the relativistic Landau Fermi-liquid is
\be
c_1^2&\equiv&\frac{\del p}{\del\E}\nonumber\\
&=&\frac{\del\rho}{\del\E}\frac{\del }{\del \rho}\left(
\rho^2\frac{\del\E /\rho}{\del\rho}\right)\nonumber\\
&=&\frac{K}{9\mu}\label{firstsound}\\
&=&\frac{k_F^2}{3\mu^2}\frac{1+F_0}{1+F_1/3}\nonumber \ee from the
above results.
\section{Chiral effective Lagrangian for nuclei}

In this section we come to hadron/nuclear phenomena. We simply
explain the basic theory of strong interaction, QCD, and how
successfully low energy phenomena can be described by its
effective theory. Then it is discussed how the effective theory
can be applied to the nuclear/hadronic matter ground state. FTS1
model will be studied as an example.
\subsection{QCD: basis of strong interactions}\label{qcd}

Quantum Chromodynamics is believed to be the fundamental theory
for the strong interaction. It is a nonabelian gauge description
of the strong interaction and its building blocks are quarks and
gluons. The QCD Lagrangian is
\be
\L_{QCD}=-\frac12 \Tr G_{\mu\nu}G^{\mu\nu} +\bar{q}(i\slash{\del}
-g\slash{G} -m_q)q \ee with $q=(u,d,...)^T$ and
$m_q=$diag$[m_u,m_d,...]$ in flavor space. In this section $\Tr$
represents the trace over flavor indices. The gauge field strength
tensor is
\be
G_{\mu\nu}=\del_\mu G_\nu-\del_\nu G_\mu -ig[G_\mu ,G_\nu ] \ee
where $G_\mu =G_\mu^a \frac{\lambda^a}{2}$. $\lambda^a$ is the
well-known Gell-Mann's $3\times 3$ traceless hermitian matrix with
color indices $a =1,2,...,8$.

Quantum Chromodynamics conserves its relevant symmetries; Lorentz
symmetry, parity, charge conjugation, time reversal and color
SU(3) gauge symmetry. In addition, it has approximate symmetries.
Isospin symmetry is a good one because of small difference between
$u$ and $d$ masses.

There is another well-known and important QCD approximate
symmetry, chiral symmetry. It is based on the fact that the mass
of light quarks $u$ and $d$ are much smaller than chiral symmetry
breaking scale $\Lambda_\chi\sim 4\pi f_\pi\sim 1$ GeV. We can
consider the massless limit, i.e. $m_q\rightarrow 0$ when dealing
with light quark physics. It means that chiral symmetry becomes
relevant. With projection operator
\be
P_{\pm}=\frac{1\mp \gamma_5}{2} \ee we can decompose the spinor
into the eigenstates of helicity $\sigmab\cdot \kbhat$,
\be
\sigmab\cdot\kbhat q=\pm q_\pm. \ee with $q_{\pm}\equiv P_{\pm}q$.
Without quark mass term, $q_+$ and $q_-$ are decoupled each other
\be
\L_{QCD}\to -\frac12 \Tr G_{\mu\nu}G^{\mu\nu}
+\bar{q}_+(i\slash{\del} -g\slash{G})q_+ -\bar{q}_-(i\slash{\del}
-g\slash{G})q_- \ee So the separate transformations
\be
q_\pm \to e^{-i\epsilon_\pm }q_\pm \ee and
\be
q_\pm \to e^{-i\thetab_\pm\cdot\tb}q_\pm \ee with real parameters
$\epsilon_\pm$ and $\thetab_pm$ leave the massless QCD Lagrangian
invariant, where $\tb$ is the general SU(2) generators. We can
construct SU(2) and U(1) vector transformations as
\be
q\to e^{-i\thetab_V\cdot\tb}q,\ \ \ q\to e^{-i\epsilon_V}q \ee and
axial ones as
\be
q\to e^{-i\thetab_A\cdot\tb\gamma_5}q,\ \ \ q\to
e^{-i\epsilon_A\gamma_5}q \ee with $\thetab_\pm
=\thetab_V\pm\thetab_A$ and $\epsilon_\pm
=\epsilon_V\pm\epsilon_A$. The corresponding currents are
\be
\Vb^\mu &=&\bar{q}\gamma^\mu\tb q,\ \ \ V_s^\mu =\bar{q}\gamma^\mu
q,
\\
\Ab^\mu &=&\bar{q}\gamma^\mu\gamma_5\tb q,\ \ \ A_s^\mu
=\bar{q}\gamma^\mu\gamma_5 q. \ee

Although massless QCD Lagrangian is invariant under these
transformations, its U(1) axial current of QCD is not conserved
due to anomaly even if the quark mass $m_q\to 0$;
\be
\del_\mu A_s^\mu =\frac{N_Fg^2}{16\pi^2}\Tr
G_{\mu\nu}\tilde{G}^{\mu\nu} +2i\bar{q}m_q\gamma_5 q \ee with
$\tilde{G}^{\mu\nu}=\varepsilon^{\mu\nu\alpha\beta}G_{\alpha\beta}$.
So the massless QCD has chiral symmetry (SU(2)$_R\times
$SU(2)$_L$) and fermion number symmetry (U(1)$_V$). Chiral
symmetry does not appear in hadron spectrum. So it is generally
assumed that the chiral symmetry is dynamically broken into
SU(2)$_V$ in hadron physics and the light pseudoscalar mesons are
regarded as Goldstone bosons.

Since the massless QCD Lagrangian contains no dimensional
parameters, massless QCD action is invariant under scale
transformation;
\be
x^\prime =ax,\ q(x^\prime )=a^{-3/2}q(x), \ G_\mu (x^\prime
)=a^{-1}G_\mu (x). \ee Thus massless QCD seems to have another
approximate symmetry, scale symmetry. However, the renormalization
prescriptions which have the running coupling constant break the
scale symmetry. It means that we need to introduce a dimensional
scale in order to specify the value of running coupling constant.
Such specification of a scale breaks the scale invariance of QCD;
\be
\del_\mu D^\mu =\theta^\mu_\mu =\frac{\beta}{g}\Tr
G_{\mu\nu}G^{\mu\nu}+\bar{q}(1+\gamma_q)m_qq \label{tranomaly} \ee
with the anomalous dimension of quarks $\gamma_q
=$diag$[\gamma_u,\gamma_d,...]$ even if quark masses are zero.
$D^\mu$ is dilatation current, $\theta^\mu_\nu$ is the improved
energy-momentum tensor~\cite{coleman} and
\be
\beta\equiv \mu\frac{\del g}{\del\mu}=(\frac23
N_F-11)\frac{g^3}{16\pi^2} +\O (g^5) \ee with the scale of
renormalization $\mu$. Contrary to the axial anomaly which
includes only one-loop contributions, scale anomaly includes
multi-loop contributions in $\beta$.
Scale anomaly which plays a major role in breaking scale
invariance of light quark QCD gives a basis of BR scaling argument
which will appear in Section \ref{scaling}.
\subsection{Effective field theories}\label{eft}

We know that the particles which appear in nuclear physics, e.\
g.\ pions, nucleons, etc., are not elementary particles. In the
framework of QCD, they are known to consist of quarks and gluons
which give non-perturbative contributions in low energy processes.
In addition, QCD
is a part of the Standard Model which is also an effective theory,
not a fundamental theory.

In order to deal with low energy hadrons, one can construct an
effective field theory that is appropriate to the probed energy
scale $Q$. The relevant degrees of freedom in such effective field
theory are the low-energy particles which appear at the energy
scale of the observed experiments. The particles with energies
higher than the relevant energy scale are integrated out and
absorbed into the couplings among the relevant degrees of freedom;
\be
& &\int [\Del q][\Del\bar{q}][\Del g] e^{i\int d^4x
\L_{fund}[q,\bar{q},g,\eta ,\bar{\eta},j]}\nonumber\\ &=&\int
[\Del B][\Del\bar{B}][\Del M] e^{i\int d^4x
\L_{eff}[B,\bar{B},M,\eta ,\bar{\eta},j]} \label{integout}\ee
where $\eta$, $\bar{\eta}$, and $j$ are the external sources of
elementary fermions $q$(quarks), anti-fermions
$\bar{q}$(anti-quarks), and bosons $g$(gluons) respectively in
fundamental theory(QCD) and $B$, $\bar{B}$, and $M$ represent
fermions(baryons), anti-fermions(anti-baryons), and bosons(mesons)
respectively in effective field theory. Although one can in
principle calculate the couplings in effective theory for strong
interactions from the fundamental QCD from equation
(\ref{integout}), it seems an impossible mission because we do not
know how to perform such a highly non-perturbative calculation.
However, we do not have to deal with, nor to know exactly, the
fundamental theory.

The strategy to build an effective theory is simple.  It consists
of writing down the most general Lagrangian which conserves all
the relevant symmetries that figure at a given energy scale, and
satisfy the basic principles(e.g.\ quantum mechanics, cluster
decomposition) of the theory. Then any theory under the same
constraints looks like his/hers at sufficiently low energy scale
though he/she cannot insist that the right theory necessarily
leads only to his/hers. (This is called ``folk theorem" by
Weinberg~\cite{wein97} though he used it to explain the usefulness
of quantum field theory.) By imposing the relevant symmetries of
the QCD or of a more fundamental theory, e.g., chiral symmetry,
and the basic principles we can build an effective field theory
for low-energy nuclear processes where the composite hadrons are
taken as elementary.

The number of terms in the effective Lagrangian which are
consistent with fundamental or assumed symmetries may be infinite,
but one can manage to describe the probed physical processes by
expanding the terms in $Q/\Lambda$, where $\Lambda$ is s suitable
cut-off scale and $Q$ the scale probed ($Q << \Lambda$). So only a
few leading terms are relevant for low-energy processes. And the
couplings can be determined from available experiments.

Such effective field theories with chiral symmetry so designed
work very well in matter-free or dilute space. They are used to
calculate soft-pion processes firstly~\cite{w67}, and recently
extended to the process in light nuclei though nucleons are not
soft~\cite{weinberg,pmr93,kolck,tsp,pmr,PKMR}. Park {\it et al.}'s
work for two-nucleon systems at very low energies~\cite{PKMR} is
one of the best and newest examples of the success. Setting the
cutoff near one pion mass, they integrate out all mesonic degrees
of freedom, even the pions. Since the deuteron bound state is
dilute, the parameters of the theory can be determined by free
space experiments. The results in \cite{PKMR} confirm that the
strategy of the effective field theory works remarkably well. When
the pion field is included in addition, it provides a new degree
of freedom and improves the theory even further allowing one to go
higher in energy scale~\cite{PKMR}.

But in heavier nuclei, the energy scales of the system will be
higher since the interactions between nucleons in such systems
sample all length scales and hence other degrees of freedom than
nucleonic and pionic need be introduced. It means that the
irrelevant and neglected terms in the computation for the light
nuclei become more relevant. We need to consider more and more
terms as the density of the system becomes higher and higher. The
power of effective field theory whose parameters are determined
from the matter-free experiments gets weakened in dense system.
How can we proceed as density increases beyond the ordinary matter
density for which there are practically no experimental data?
\subsection{Effective Lagrangian in medium and Landau theory}\label{integ}

Recently Lynn made a progress to give a good hint to answer the
above questions~\cite{lynn}. He proposed that the ground-state
matter is ``chiral liquid"  which arises as a non-topological
soliton. Fluctuation around this ground state should give an
accurate description of the observables that we are dealing with.
We shall here extend this argument further and make contact with
Landau's Fermi-liquid theory of nuclear matter. This will allow us
to understand the nuclear/hadron matter description in terms of
chiral Lagrangians and Fermi-liquid fixed point theory thereby
giving a unified picture of ordinary nuclear matter and extreme
state of matter probed in heavy-ion collisions. This is the
attempt to connect the physics of the two vastly different
regimes.

The basic assumption we start with is that the chiral liquid
arises from a quantum effective action resulting from integrating
out the degrees of freedom lying above the chiral scale
$\Lambda_\chi \sim 4\pi f_\pi \sim 1$ GeV.
\be
\int [d\phi_<]e^{iS_\chi (\phi_< )} =\int
[d\phi_<][d\phi_>]e^{iS(\phi_< ,\phi_> )} \ee where the subscript
$<$($>$) represents the sector $\omega <\Lambda_\chi$($\omega
>\Lambda_\chi$) of the given set of fields $\phi$. As explained in
Sec.\ \ref{eft},
\be
S_\chi =\sum_i g_i\hat{O}_i\label{mfea} \ee is the sum of all
possible terms consistent with symmetries of QCD. This corresponds
to the first stage of ``decimation''~\cite{froehlich} in our
scheme. The mean field solution of this action is then supposed to
yield the ground state of nuclear matter with the Fermi surface
characterized by the Fermi momentum $k_F$. The effective
Lagrangian was given in terms of the baryon, pion, quarkonium
scalar and vector fields. The gluonium scalars are integrated out.
Instead of treating the scalar and vector fields explicitly, we
will consider here integrating them out further from the effective
Lagrangian. This will lead to four-Fermi, six-Fermi, etc.\ ,
interactions in the Lagrangian with various powers of derivatives
acting on the Fermi field. The resulting effective Lagrangian will
then consist of the baryons and pions coupled bilinearly in the
baryon field and four-Fermi and higher-Fermi interactions with
various powers of derivatives, all consistent with chiral
symmetry. A minimum version of such Lagrangian in mean field can
be shown to lead to the original (naive) Walecka model~\cite{tsp}.

The next step is to decimate successively the degrees of freedom
present in the excitations with the scale
$E<\Lambda_\chi$~\cite{froehlich}. To do this, we consider
excitations near the Fermi surface, which we take to be spherical
for convenience characterized by $k_F$. First we integrate out the
excitations with momentum $p\geq \pm \Lambda$ (where $p=|\pb|$ and
$\Lambda <\Lambda_\chi$) measured relative to the Fermi surface
(corresponding to the particle-hole excitations with momentum
greater than $2\Lambda$). We are thus restricting ourselves to the
physics of excitations whose momenta lie below $2\Lambda$ as in
Section \ref{rgliq}. Leaving out the pion for the moment and
formulated non-relativistically, the appropriate action to
consider can be written in a simplified and schematic form as Eq.\
(\ref{Flag}).\footnote{The pion will be introduced in the Section
\ref{vector} in terms of a non-local four-Fermi interaction that
enters in the ground state property and gives the nucleon Landau
mass formula in terms of BR scaling and pionic Fock term.}

In nuclear matter, the spin and isospin degrees of freedom need to
be taken into account into the four-Fermi interaction. All these
can be written symbolically in the action (\ref{Flag}). The
function $u$ in the four-Fermi interaction term therefore contains
spin and isospin factors as well as space dependence that takes
into account non-locality and derivatives. For simplicity we will
consider it to be a constant depending in general on spin and
isospin factors. Non-constant terms will be ``irrelevant.'' In our
discussions, the BCS channel that corresponds to a
particle-particle channel does not figure and hence will not be
considered explicitly.

The successive mode elimination
\be
\int [d\phi_<^l]e^{iS^\star (\phi_<^l)}= \int
[d\phi_<^l][d\phi_<^h]e^{iS_\chi (\phi_<^l ,d\phi_<^h )}, \ee
which satisfies RG equation explained in Sec.\ \ref{rgliq}, will
give
\be
S^\star =\sum_i g_i^\star \hat{O}^\star_i. \ee The starred
coupling constant $g^\star_i$ and operators $\hat{O}^\star_i$
depend on density through $s$ and have the same structure of
(\ref{mfea}). $l(h)$ represents the components $p<\Lambda
/s$($p>\Lambda /s$). The upshot of the analysis in Section
\ref{rgliq} is that the resulting theory is the Fermi-liquid fixed
point theory with the limit $\Lambda /k_F\to 0$. In sum, we arrive
at the picture where the chiral liquid solution of the quantum
effective chiral action gives the Fermi-liquid fixed point theory.
The parameters of the four-Fermi interactions in the phonon
channel are then identified with the fixed-point Landau
parameters.

There are two steps to apply such scheme to nuclear/hadron
phenomenology. The first is to derive the in-medium effective
Lagrangian directly from QCD or from the matter-free effective
Lagrangian and the second is to solve the built in-medium
effective Lagranian. The first is very difficult. So we usually go
to the second step, after building the in-medium effective
Lagrangian by reasonable guesses. We assume that the effective
Lagrangian satisfies
\be
S^{eff} =\int d^4x\L^{eff} \ee where $S^{eff}$ is a Wilsonian
effective action arrived at after integrating out high-frequency
modes of the nucleon and other heavy degrees of freedom. This
action is then given in terms of sum of terms organized in chiral
order in the sense of effective theories at low energy.

One way to build the chiral effective Lagrangian for nuclear
matter has been studied by Furnstahl, Serot and
Tang~\cite{tang,tang2}\footnote{The model in \cite{tang} is
referred to as FTS1. That in \cite{tang2} shall as FTS2.} They
formulated their theory in terms of a chiral Lagrangian
constructed by using the terms which are governed by QCD symmetry
and applying the ``naturalness" condition for all relevant fields.
In doing this, they introduced in the FTS1 a quarkonium field that
is associated with the trace anomaly with its potential
constrained by Vainshtein {\it et al.}'s low energy
theorem~\cite{vzns}. And Georgi's ``naive dimensional
analysis"~\cite{NDA}\footnote{Since the strong interactions have
two relevant scales $f_\pi$ and $\Lambda_\chi$ and $\Lambda_\chi$
is much larger than $f_\pi$, we can apply Georgi's naive
dimensional analysis to the low energy hadron physics.} was used
in the FTS2 instead of the trace anomaly. It was argued in
\cite{tang2} that a Lagrangian so constructed contains in
principle arbitrarily higher-order many-body effects
including those loop corrections that can be expressed as
counter terms involving matter fields (e.g., baryons). This is
essentially equivalent to Lynn's chiral effective
action~\cite{lynn} that purports to include all orders of quantum
loops in chiral expansion supplemented with counter terms
consistent with the order to which loops are calculated. Though it
is a little hard to define the fluctuation on its ground state,
their models are very successful in describing the bulk properties
of nuclei.

Another way is to apply the strategy of the effective theory to
the in-medium Lagrangian. The parameters of the effective
Lagrangian are related to the vacuum state at a given density so
depend on the density. One famous example is BR
scaling~\cite{BR91}. Brown and Rho point \cite{BR96} that the mean
field solution of the chiral effective Lagrangian given by BR
scaling approximates
\be
\delta S^{eff}=0.\label{extreme} \ee The detail of BR scaling is
reviewed in the next section. The aim of this review is to cast BR
scaling in a suitable form starting with a chiral Lagrangian
description of the ground state as specified above around which
fluctuations in a various flavour sectors are to be made. To do
this we study phenomenologically successful FTS1 model here.
\subsection{A chiral effective model: FTS1 model}\label{matter}

Furnstahl, Tang and Serot~\cite{tang} constructed an effective
nonlinear chiral model that will be referred to as FTS1 model in
this review. The FTS1 model incorporates the scale anomaly of QCD
in terms of a light (``quarkonium") scalar field $S$ and a heavy
(``gluonium") scalar field $\vs$ and gives a good description of
basic nuclear properties in mean field. One can also build a model
appealing to general notions of effective field theories such as
``naturalness condition" as in \cite{tang2}
This avoids the use of the scale anomaly of QCD. The FTS2 model is
also an effective mean field theory which gives an equally
satisfactory phenomenology as the FTS1. However the FTS1 is found
to be more convenient for studying the role of the light scalar
field in the scale anomaly we are interested in since the FTS1
Lagrangian includes the scale anomaly term explicitly. In
addition, Li, Brown, Lee, and Ko found that the FTS1 model
reproduces quite successfully nucleon flow in heavy ion
collisions~\cite{LBLK96}.

The underlying assumption in the FTS1 is that the light scalar
field transforms under scale transformation as
\be
S(a^{-1}x)=a^d S(x) \ee with a parameter $d$ that can be different
from its canonical scale dimension, i.e.\ unity, while the scale
dimension of the heavy gluonium, which is integrated out in the
effective Lagrangian for normal nuclear matter, is taken to be
unity. This assumption imposes that quantum fluctuations in the
scalar channel be incorporated into an anomalous dimension
$d_{an}=d-1\neq 0$. A RG flow argument in Section \ref{integ}
justifies this assumption heuristically. One further assumption of
the FTS1
is that there is no mixing between the light scalar $S(x)$ and the
heavy scalar $\vs$ in the trace anomaly.
The FTS1 Lagrangian has the form
\be
\L^{eff}=\L_s-H_g\frac{\vs^4}{\vs_0^4}\left(
\ln\frac{\vs}{\vs_0}-\frac{1}{4} \right) -H_q\left(
\frac{S^2}{S_0^2}\right)^{\frac{2}{d}}\left(
\frac{1}{2d}\ln\frac{S^2}{S_0^2} -\frac{1}{4}\right) \ee where
$\L_s$ is the chiral- and scale-invariant Lagrangian containing
$\vs ,S,N,\pi , \omega$, etc. Here $\vs_0$ and $S_0$ are the
vacuum expectation values(VEV) with the vacuum $|0\ra$ defined in
the matter-free space:
\be
\vs_0\equiv\la 0|\vs |0\ra, \ \ \ \ S_0\equiv\la 0|S|0\ra. \ee The
trace of the improved energy-momentum tensor~\cite{coleman}, i.\
e.\ the divergence of the dilatation current $D^{\mu}$, from the
Lagrangian is;
\be
\del_{\mu} D^{\mu}=\theta_\mu^\mu = -H_g\frac{\vs^4}{\vs_0^4}
-H_q\left( \frac{S^2}{S_0^2}\right)^{2/d}. \ee The mass scale
associated with the gluonium degree of freedom is higher than that
of chiral symmetry, $\Lambda_\chi\sim 1$ GeV. For example the mass
of the scalar glueball calculated by Weingarten~\cite{w94} is
1.6$\sim$1.8 GeV. This invites us to integrate out the gluonium.
The resulting FTS1 effective Lagrangian takes the form
\be
\L&=&\bar{N}[i\gamma_{\mu}(\del^{\mu}+iv^{\mu}
+ig_v\omega^{\mu}+g_A\gamma_5 a^{\mu}) -M+g_s\phi ]N \nonumber\\ &
&-\frac{1}{4}F_{\mu\nu}F^{\mu\nu}+ \frac{1}{4!}\zeta g_v^4
(\omega_{\mu}\omega^{\mu})^2 \nonumber\\ & &+\frac{1}{2}\left(
1+\eta \frac{\phi}{S_0}\right)\left\{
\frac{f_{\pi}^2}{2}\tr(\del_{\mu}U\del^{\mu}U^{\dagger})+ m_v^2
\omega_{\mu}\omega^{\mu}\right\}\nonumber\\ &
&+\frac{1}{2}\del_{\mu}\phi\del^{\mu}\phi
-\frac{m_s^2}{4}S_0^2d^2\left[ \left(
1-\frac{\phi}{S_0}\right)^{4/d} \left\{\frac{1}{d} \ln\left(
1-\frac{\phi}{S_0}\right)-\frac{1}{4}\right\} +\frac{1}{4}\right]
\label{leff} \ee where $S=S_0-\phi$, $\eta$ and $\zeta$ are
unknown parameters to be fixed and
\be
\xi^2 &=& U =e^{i\pib\cdot\taub /f_\pi}\nonumber\\ v_\mu &=&
-\frac{i}{2}(\xi^{\dagger}\del_\mu\xi +\xi\del_\mu\xi^\dagger
)\nonumber\\ a_\mu &=& -\frac{i}{2}(\xi^\dagger\del_\mu\xi
-\xi\del_\mu\xi^\dagger ).\nonumber \ee Note that a given VEV of
the $\phi$ field scales down the pion decay constant and the
$\omega$ mass in the same way at the lowest chiral order as in
\cite{BR91}. The static mean field equations of motion for FTS1
are
\be
g_s\sum_i\bar{N}_iN_i&=&\nabla^2\phi_0 -m_s^2\left(
1-\frac{\phi_0}{S_0}\right)^{\frac{4-d}{d}} \ln
\left(1-\frac{\phi_0}{S_0}\right)
-\frac{\eta}{2S_0}m_v^2\omega_0^2\\ g_v\sum_i\bar{N}_i^\dagger
N_i&=&\nabla^2\omega_0 +m_v^2\left(
1+\eta\frac{\phi_0}{S_0}\right) \omega_0
+\frac{\zeta}{6}g_v^4\omega_0^3\label{ftsv} \ee with the static
mean field solutions $\phi_0$ and $\omega_0$. Equation
(\ref{ftsv}) is a constraint because $\omega_0$ is not a dynamical
degree of freedom.

It is important to note that the FTS1 Lagrangian is an effective
Lagrangian in the sense explained in Section \ref{eft}. The effect
of high frequency modes of the nucleon field and other massive
degrees of freedom appears in the parameters of the Lagrangian and
in the counter terms that render the expansion meaningful. It
presumably includes also vacuum fluctuations in the Dirac sea of
the nucleons \cite{tang,tang3}. In general, it could be much more
complicated. Indeed,
 one does not yet know how to implement this strategy in
full rigor given that one does not know what the matching
conditions are. In \cite{tang,tang2}, the major work is, however,
done by choosing the relevant parameters of the FTS1 Lagrangian to
fit the empirical informations.

The energy density for uniform nuclear matter obtained from
(\ref{leff}) is
\be
\varepsilon
 &=& \frac{\gamma}{(2\pi)^3}\int^{k_F}d^3k \sqrt{\Bk^2+(M-g_s\phi_0 )^2}
\nonumber\\ & & -\frac{m_v^2}{2}\left( 1+\eta\frac{\phi_0}{S_0}
\right) \omega_0^2 +g_v\rho_B
\omega_0-\frac{\zeta}{4!}g_v^4\omega_0^4\nonumber\\ &
&+\frac{m_s^2}{4}S_0^2d^2\left\{ \left( 1-\frac{\phi_0}{S_0}
\right)^{\frac{4}{d}} \left(
\frac{1}{d}\ln(1-\frac{\phi_0}{S_0})-\frac{1}{4}\right)
+\frac{1}{4} \right\} . \label{energydensity} \ee Here $\gamma$ is
the degeneracy factor.
\subsection{Anomalous dimension in FTS1 model}\label{anomal}

The best fit to the properties of nuclear matter and finite nuclei
is obtained with the parameter set T1\footnote{Explicitly the T1
parameters are: $g_s^2=99.3$, $g_v^2=154.5$, $\eta=-0.496$, and
$\zeta =0.0402$.} when the scale dimension of the scalar $S$ is
near $d=2.7$ in the FTS1~\cite{tang}. The large anomalous
dimension means that one is fluctuating around a wrong ground
state. Brown-Rho scaling is meant to avoid this. In this section,
we analyze how this comes out and present what we understand on
the role of the large anomalous dimension $d_{an}=d-1\approx 1.7$
in nuclear dynamics. In what follows, the parameter T1 with this
anomalous dimension will be taken as a canonical parameter set.

\subsubsection{Scale anomaly}

Following Coleman and Jackiw~\cite{coleman}, the scale anomaly
can be discussed in terms of an anomalous Ward identity.  Define
$\Gamma_{\mu\nu}(p,q)$ and $\Gamma(p,q)$ by
\be
G(p)\Gamma_{\mu\nu}(p,q)G(p+q) &=&\int d^4xd^4ye^{iq\cdot
x}e^{ip\cdot y} \langle 0\mid T^\ast \theta_{\mu\nu}(x)\varphi
(y)\varphi (0) \mid 0\rangle\\ G(p)\Gamma G(p+q) &=&\int
d^4xd^4ye^{iq\cdot x}e^{ip\cdot y} \langle 0\mid T^\ast[\del^\mu
D_{\mu}](x)\varphi (y)\varphi (0) \mid 0\rangle\nn \ee with the
renormalized propagator $G(p)$ and the renormalized fields
$\varphi (x)$. Here $T^\ast$ represents the covariant T-product,
$D_{\mu}(x)$ the dilatation current, and $\theta_{\mu\nu}$ the
improved energy momentum tensor with
$D_{\mu}=x^\nu\theta_{\mu\nu}$. A naive consideration on Ward
identities would give
\be
g_{\mu\nu}\Gamma^{\mu\nu}(p,q) =\Gamma
(p,q)-idG^{-1}(p)-idG^{-1}(p+q) \ee with $d$ the scale dimension
of $\varphi (x)$. However $\Gamma$ is ill-defined due to
singularity and so has to be regularized. With the regularization,
the Ward identity reads
\be
g_{\mu\nu}\Gamma^{\mu\nu}(p,q) &=&\Gamma
(p,q)-idG^{-1}(p)-idG^{-1}(p+q)+A(p,q)\\ A(p,q)&\equiv
&\lim_{\Lambda\rightarrow\infty} \Gamma (p,q,\Lambda )-\Gamma
(p,q)\label{Anomaly} \ee where the additional term, $A$, is the
anomaly. This anomaly corresponds to a shift in the dimension of
the field involved at the lowest loop order but at higher orders
there are vertex corrections. One obtains however a simple result
when the beta functions vanish at zero momentum
transfer~\cite{coleman}. Indeed in this case, the only effect of
the anomaly will appear as an anomalous dimension.  In general
this simplification does not occur. However it can take place when
there are nontrivial fixed points in the low energy theory. Under
the reasoning developed in condensed matter physics~\cite{shankar}
it is argued later that nuclear matter is given, in the absence of
BCS channel, by a Landau Fermi-liquid fixed point theory with
vanishing beta functions of the four-Fermi interactions and that
all quantum fluctuation effects would therefore appear in the
anomalous dimension of the scalar field $S$. That nuclear matter
is a Fermi-liquid fixed point seems to be well verified at least
phenomenologically as seen in Sect. 5. However that fluctuations
into the scalar channel can be summarized into an anomalous
dimension is a conjecture that remains to be proved. We conjecture
here that this is one way we can understand the success of the
FTS1 model.
\subsubsection{Nuclear matter properties at $d_{an}\approx 5/3$}

The FTS1 theory has some remarkable features associated with the
large anomalous dimension. Particularly striking is the dependence
on the anomalous dimension of the compression modulus $K$ and
many-body forces.

\begin{figure}
\setlength{\epsfysize}{3in} \centerline{\epsffile{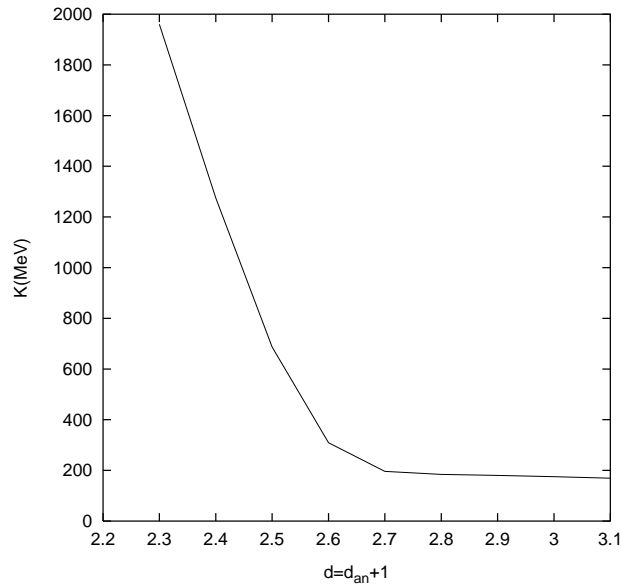}}
\caption{Compression modulus vs. anomalous dimension. The
parameter set used here is the T1 in FTS1. This shows the
sensitivity of the compression modulus to the anomalous
dimension.}\label{Kd}
\end{figure}

In Table \ref{p} are listed the compression modulus $K$ and the
equilibrium Fermi momentum $k_{eq}$ vs. the scale dimension $d$ of
the scalar field $\phi$. As the $d$ increases, the $K$ drops very
rapidly and stabilizes at $K\sim 200$ MeV consistent with
experiments for $d\approx 2.6$ and stays nearly constant for
$d>2.6$. This can be seen in Fig.\ \ref{Kd}. The equilibrium Fermi
momentum on the other hand slowly decreases uniformly as the $d$
increases.

Unfortunately, we have no simple understanding of the mechanism
that makes the compression modulus $K$ stabilize at the particular
value $d_{an}\approx 5/3$. We believe there is a non-trivial
correlation between this behavior of $K$ and the observation made
below that the scalar logarithmic interaction brought in by the
scale anomaly is entirely given at the saturation point by the
quadratic term at the same $d_{an}$ with the higher polynomial
terms (i.e.\ many-body interactions) contributing more repulsion
for increasing anomalous dimension. At present our understanding
is purely numerical and hence incomplete.
\begin{table}
\caption{Equilibrium Fermi momentum $k_{eq}$ and binding energy
$B=M-E/A$ as a function of $d$ for Fig.\ \ref{Kd}}\label{p} \vskip
.3cm
\begin{center}
\begin{tabular}{cccc}\hline\hline
$d$&$K$(MeV)&$k_{eq}$(MeV)&$B$(MeV)\\ \hline 2.3&1960&313&50.4\\
2.4&1275&308&37.0\\ 2.5&687&297&27.1\\ 2.6&309&279&20.4\\
2.7&196&257&16.4\\ 2.8&184&241&14.0\\ 2.9&180&231&12.4\\
3.0&175&223&11.2\\ 3.1&169&217&10.3\\ \hline\hline
\end{tabular}
\end{center}
\end{table}

\begin{figure}[bthp]
\setlength{\epsfysize}{8in} \centerline{\epsffile{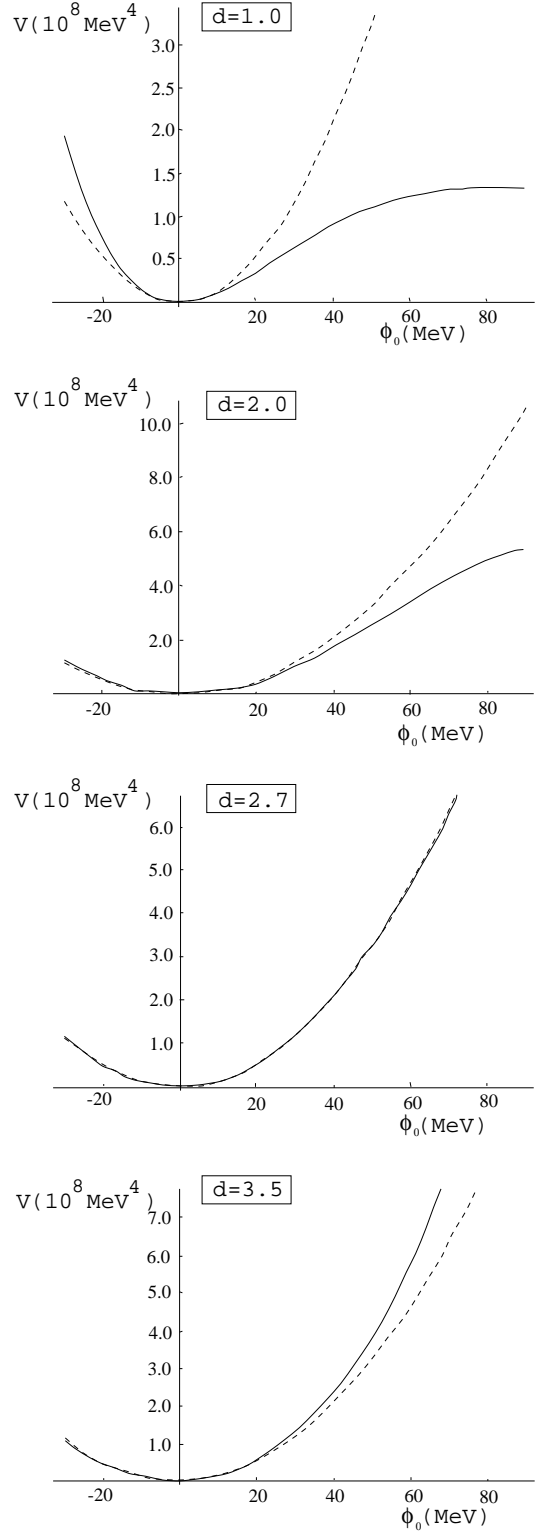}}
\caption{\small Comparison between the $\phi^2$ interaction and
the logarithmic self-interaction of the scalar field { with the
FTS1 parameters.} The dashed lines represent $
\frac{m_s^2}{2}\phi^2$ and the solid lines
$\frac{m_s^2}{4}S_0^2d^2[(1-\frac{\phi}{S_0})^{4/d}[\frac{1}{d}
\ln (1-\frac{\phi}{S_0})-\frac{1}{4}]+\frac{1}{4}]$ for (from top
to bottom) $d=1,2,2.7,3.5$  respectively.}\label{pot}
\end{figure}

In mean field, the logarithmic potential in Eqs. (\ref{leff}) and
(\ref{energydensity}) contains n-body-force (for $n\geq 2$)
contributions to the energy density. For the FTS1 parameters,
these n-body terms are strongly suppressed for $d\gsim 2.6$. This
is shown in Fig.\ \ref{pot} where it is seen that the entire
potential term is accurately reproduced by the quadratic term
$\frac 12 m_s^2\phi^2$ for $d_{an}\sim 5/3$. Furthermore a close
examination of the results reveals that each of the n-body terms
are separately suppressed. This phenomenon is in some sense
consistent with chiral symmetry~\cite{weinberg} and is observed in
the spectroscopy of light nuclei~\cite{friar}.

\begin{figure}
\setlength{\epsfysize}{3in} \setlength{\epsfxsize}{4in}
\centerline{\epsffile{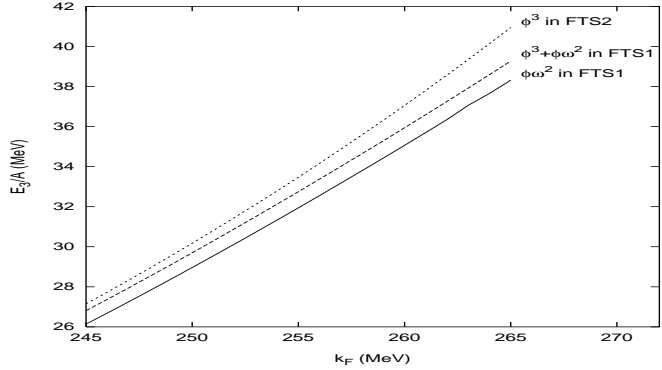}} \caption{\small The 3-body
contributions to the energy per nucleon vs. Fermi momentum in the
FTS models. The short-dashed line represents the contribution of
the $\phi^3$ term in the FTS2 with the Q1 parameters. The
long-dashed and the solid lines represent the contributions of the
cubic terms ($\phi \omega^2$ and $\phi^3$) in the FTS1 with the T1
parameters for $d=2.7$.}\label{three}
\end{figure}

Since there are additional polynomial terms in vector fields
(i.e.\ terms like $\phi\omega^2$), the near complete suppression
of the scalar polynomials does not mean the same for the total
many-body forces. In fact we should not expect it. To explain why
this is so, we plot in Fig.\ \ref{three} the three-body
contributions of the $\phi^3$ and $\phi\omega^2$ forms. We also
compare the FTS1 results with the FTS2 results that are based on
the naturalness condition. In FTS1 the $\phi^3$ term that turns to
repulsion from attraction for $d>8/3$ contributes little, so the
main repulsion arises from the $\phi\omega^2$-type term. This,
together with an attraction from a $\omega^4$ term, is needed for
the saturation of nuclear matter at the right density. This raises
the question as to how one can understand the result obtained by
Brown, Buballa, and Rho~\cite{BBR} where it is argued that the
chiral phase transition in dense medium is of mean field with the
bag constant given entirely by the quadratic term $\sim \frac 12
m_s^2\phi^2$. The answer to this question is as follows. First we
expect that the anomalous dimension will stay locked at
$d_{an}=d-1\sim 5/3$ near the phase transition (this is because
the anomalous dimension associated with the trace anomaly -- a
consequence of ultraviolet regularization -- is not expected to
depend upon density), so the $\phi^n$ terms for $n>2$ will
continue to be suppressed as density approaches the critical
value. Secondly near the chiral phase transition the gauge
coupling of the vector meson will go to zero in accordance with
the Georgi vector limit~\cite{vector}, where chiral symmetry is
restored by $m_\rho\to 0$, and vector meson decoupling takes place
as argued in \cite{BRPR96}. So the many-body forces associated
with the vector mesons will also be suppressed as density
increases to the critical density.
\subsubsection{Anomalous dimension and the scalar meson mass}\label{dim}

We would like to understand how the large anomalous dimension
needed here could arise in the theory and its role in the scalar
sector. As suggested in \cite{FR96} and elaborated more in Section
\ref{integ}, one appealing way of understanding the FTS1 mean
field theory is to consider all channels to be at Fermi-liquid
fixed points except that because of scale anomaly, the scalar
field develops an anomalous dimension, thereby affecting
four-Fermi interaction in the scalar channel resulting when the
scalar field is integrated out. If the anomalous dimension were
sufficiently negative so that marginal terms became marginally
relevant, then the system would become unstable as in the case of
the NJL model or superconductivity, with the resulting spontaneous
symmetry breaking. However if the anomalous dimension is positive,
then the resulting effect will instead be a screening. The
positive anomalous dimension we need here belongs to the latter
case. We can see this as follows. Consider the potential given
with the low-lying scalar $S$ (with the gluonium component
integrated out):
\be
V(S,\cdots)=\frac 14 m_S^2 d^2 S_0^2
\left(\frac{S}{S_0}\right)^{\frac 2d} \left(\frac 1d
\ln\frac{S}{S_0} -\frac 14\right) +\cdots\label{V} \ee where $m_S$
is the light-quarkonium mass in the free space ($\sim 700$ MeV)
and the ellipses stand for other contributions such as pions,
quark masses etc. that we are not interested in. The scalar
excitation on a given background $S^\star$ is given by the double
derivative of $V$ with respect to $S$ at $S=S^\star$
\be
{m_S^\star}^2=m_S^2 \left(\frac{S^\star}{S_0}\right)^{ \frac
4d-2}\left[1+\left(\frac 4d
-1\right)\ln\frac{S^\star}{S_0}\right]. \ee We may identify the
ratio $S^\star/S_0$ with the BR scaling factor $\Phi$~\cite{FR96}:
\be
\frac{S^\star}{S_0}=\Phi=\frac{f_\pi^\star}{f_\pi}=\frac{m_v^\star}{m_v}.
\ee Then we have
\be
\frac{m_S^\star}{m_S}=\Phi (\rho) \kappa_d (\rho) \ee with
\be
\kappa_d (\rho)= \Phi^{\frac 2d -2} \left[1+\left( \frac 4d
-1\right) \ln\Phi\right]^{\frac 12}. \ee One can see that for
$d=1$ which would correspond to the canonical dimension of a
scalar field the scalar mass falls much faster, for a $\Phi
(\rho)$ that decreases as a function of density, than what would
be given by BR scaling. Increasing the $d$ (and hence the
anomalous dimension) makes the scalar mass fall less rapidly. With
$d\approx 2$, $\kappa_d \approx 1$ and we recover the BR scaling.
Since the dropping scalar mass is associated with an increasing
attraction, we see that the anomalous dimension plays the role of
bringing in an effective repulsion. One may therefore interpret
this as a screening effect of the scalar attraction. In
particular, that $d-2\approx .7 >0$ means that in FTS1, an
additional screening of the BR-scaled scalar exchange (or an
effective repulsion) is present.
\section{Brown-Rho scaling}
Brown and Rho develop a strategy of density-dependent effective
field theory for an in-medium effective theory in \cite{BR91}.
They assume that the effective Lagrangian even for the hadrons in
matter also keeps the symmetries of QCD (e.g., chiral symmetry)
and that the parameters of the effective theory are determined at
a given density. The change of vacuum in the presence of medium is
assumed to be expressible by the change of parameters of the
theory. Brown-Rho scaling is the relation among the
density-dependent parameters in medium. The quasiparticle picture
associated with BR scaling is a successful way to describe hadrons
in medium. In this section BR scaling is derived from QCD-oriented
effective Lagrangian and discussed.
\subsection{Freund-Nambu model}\label{fn}

Before discussing the effective theory for dense matter and BR
scaling, we study in this section the Freund-Nambu (FN) model, in
order to describe the idea of BR scaling. The FN model is a simple
model where scale symmetry is realized in Goldstone mode. The FN
model Lagrangian is
\be
\L^{FN}&=&\L_{inv}^{FN}+\L_{sb}^{FN}\label{fnmodel}\\
\L_{inv}^{FN} &=&\frac12 \del^\mu\phi\del_\mu\phi +\frac12
\del^\mu\chi\del_\mu\chi -\frac12 f^2\chi^2\phi^2\nonumber\\
\L_{sb}^{FN}&=&-\frac{c}{8}(\chi^2-v^2)^2\nonumber \ee with matter
field $\phi$ and dilaton field $\chi$ both of which have scale
dimension 1;
\be
\delta\phi =\epsilon (1+x\cdot\del )\phi\ \ \ \delta\chi =\epsilon
(1+x\cdot\del )\chi .\label{scaletr} \ee Equation (\ref{scaletr})
comes from $\Phi (ax)=a^{-1}\Phi (x)$ with $a=1+\epsilon$ where
$\Phi$ represents fields of scale dimension 1. The equations of
motion for the FN model
\be
\del^2 \phi -f^2\chi^2\phi &=&0\\
\del^2\chi-f^2\chi\phi^2-\frac{c}{2}\chi (\chi^2-v^2)&=&0 \ee have
non-trivial constant solutions;
\be
\phi =0, \ \ \chi=\pm v. \ee

Let $\la\chi\ra =v$ and shift the dilation field as
\be
\chi =\chi^\prime +v. \ee The scale transformation of the newly
defined $\chi^\prime$ does not show a definite scale dimension,
but show a symptom of Goldstone mode;
\be
\delta\chi^\prime =\epsilon (1+x\cdot\del )\chi^\prime +\epsilon
v. \ee

With the field redefinition
\be
\sigma\equiv\frac{v}{2}(\frac{\chi^2}{v^2}-1) \ee Eq.
(\ref{fnmodel}) becomes
\be
\L_{inv}^{FN}&=&\frac12\del^\mu\phi\del_\mu\phi
-\frac{m_\phi^2}{2}\phi^2
-\frac{m_\phi^2}{v}\sigma\phi^2+\frac12\del^\mu\sigma\del_\mu\sigma
\frac{1}{1+2\chi /v}\nonumber\\
\L_{sb}^{FN}&=&-\frac{m_\sigma^2}{2}\sigma^2 \ee with
\be
m_\phi&=&f\la\chi\ra\\ m_\sigma &=&\sqrt{c}\la\chi\ra.\nonumber
\ee Scale symmetry in $\L_{inv}$ becomes invisible.

We can see some interesting features in the FN model. Both the
matter and dilaton field masses depend on $\la\chi\ra$. It means
that the both masses move in the same way if the vacuum
expectation value of $\chi$ is changed. In addition, the $\sigma$
mass can be left massive even if $c\to 0$. We must also note that
the explicit scale symmetry breaking term is necessary for the FN
model picture. It makes the spontaneously scale symmetry breaking
of $\L_{inv}$ possible.
\subsection{Brown-Rho scaling in in-medium effective Lagrangian}\label{scaling}

We can introduce the elementary scalars that QCD has. As Schechter
did~\cite{schechter80}, the scalar fields represented by the trace
anomaly (\ref{tranomaly}) can be introduced. We divide the trace
anomaly into a hard part, which is associated with the gluonium
$\chi_g$, and a soft part, which comes from the quarkonium $\chi$.
\be
\theta^\mu_\mu=(\theta^\mu_\mu )_{hard}+(\theta^\mu_\mu )_{soft}.
\ee The gluon condensate in $(\theta^\mu_\mu )_{hard}$ determines
the gluonium mass $m_{\chi_g}\sim 1.6-1.8$ GeV~\cite{w94} and does
not vanish~\cite{kb93}. Since gluonium mass is much larger than
the chiral scale $\Lambda_\chi\sim 1$ GeV, gluonium fields are
integrated out in low energy physics. And Adami and
Brown~\cite{ab} show by QCD sum rule that the gluon condensate is
less important for the masses of light-quark hadrons. So we focus
on the quarkonium scalar which has mostly to do with
$(\theta^\mu_\mu )_{soft}$.

In matter-free space, there seems to be no scalar whose mass is
small enough, i.e.,$\ll\Lambda_\chi$. However, according to
Weinberg's mended symmetry~\cite{mendedweinberg} there must be a
scalar to form a quartet with pions near the chiral phase
transition. In addition, lattice simulations~\cite{ikky97} near
the chiral phase transition shows that two light flavor QCD
transition reproduces a scaling relation with O(4) exponents as
argued first by Pisarski and Wilczek~\cite{pw84}. Beane and van
Kolck~\cite{beane} suggest that the Goldstone boson from
spontaneous scale symmetry breaking plays the role of the chiral
partner of the pion, i.e., the chiral singlet scalar in the scale
anomaly approaches the pions and makes up the quartet of O(4)
symmetry in medium as density increases to the critical density of
the chiral restoration. And we must note that the quark
condensate, which contributes to the quarkonium mass, goes to zero
as chiral symmetry is restored. So we assume that the quarkonim
mass in the scale anomaly becomes $m_\chi^\star \ll\Lambda_\chi$
as density increases, though $m_\chi\sim\Lambda_\chi$ in free
space.

Integrating out the hard part of the scalar field $\chi_g$, we
introduce into a Skyrme Lagrangian - which represents the low
energy QCD for the infinite number of colors - the soft part of
scalar, $\chi$, in order to make our effective theory consistent
with QCD scale property.
\be
\L &=&\L_{inv}+\L_{SB}\\ \L_{inv}&=&\frac{f_\pi^2}{4}\left(
\frac{\chi}{\chi_0}\right)^2 \Tr (\del_\mu U\del^\mu U^\dagger
)+\frac12\del_\mu\chi\del^\mu\chi\nn\\ &
&+\frac{1}{32e^2}\Tr[U^\dagger\del_\mu U,U^\dagger\del_\nu
U]^2+\cdots \label{skyrme}\\ \L_{SB}&=&-V(\chi )+\mbox{pion mass
term}+ \cdots \ee with
\be
U &=& e^{\frac{i\taub\cdot\pib}{f_\pi}}\label{ufield}\\ \chi_0 &=&
\la 0|\chi |0\ra . \ee We make $\L_{inv}$ scale correctly by
multiplying the proper powers of $\chi$. Since $\L_{inv}$ do not
contribute to $\theta_\mu^\mu$, we add the scale breaking
potential term $V(\chi )$ due to scale anomaly. $V(\chi )$
includes radiative corrections of high order and gives the trace
anomaly of QCD in terms of $\chi$.

Let us break the scale symmetry spontaneously following the
strategy of the FN model. If we define $\chi_\star$ as the mean
field value of $\chi$ in dense matter
\be
\chi_\star\equiv \bra\chi\ket_\rho , \ee we can expand $\chi$ as
\be
\chi =\chi_\star +\chi^\prime . \ee Stars will be affixed to all
the quantities that appear in nuclear matter from now on.
$\L_{inv}$ becomes in terms of $\chi^\prime$,
\be
\L_{inv}&=&\frac{f_\pi^{\star 2}}{4} \Tr (\del_\mu U\del^\mu
U^\dagger )+ \frac12\del_\mu\chi^\prime\del^\mu\chi^\prime\nn\\ &
&+\frac{1}{32e^2}\Tr[U^\dagger\del_\mu U,U^\dagger\del_\nu
U]^2+\cdots \ee with the effective pion decay constant defined at
mean field level as
\be
f^\star_\pi = f_\pi\frac{\chi_\star}{\chi_0}.\label{fpi} \ee Note
that $U^\star =e^{i\taub\cdot\pib^\star /f_\pi^\star}$ with
$\pib^\star\equiv \pi\chi_\star /\chi_0$ is the same as $U$ in
(\ref{ufield}) by the definition (\ref{fpi}). In our effective
field theory both scale symmetry and chiral symmetry are realized
in the Goldstone mode and their Goldstone bosons are $\chi$ and
$\pi$'s, respectively.

According to Gell-Mann-Oaks-Renner (GMOR) relation
\be
f_\pi^2 m_\pi^2 =-\frac{m_u+m_d}{2}\bra 0|\bar{u}u+\bar{d}d|0\ket
,\label{gmor} \ee The pion mass is proportional to the quark mass.
Since the quark masses comes from explicit chiral symmetry
breaking which has something to do with the electroweak symmetry
breaking scale $\Lambda_{EW}\gg \Lambda_\chi$, the pion mass
problem is out of our interesting range. In the prediction of
chiral perturbation theory in medium~\cite{thorsson} the pole mass
of the pion does not decrease up to nuclear matter density. In
fact a recent analysis of deeply bound pionic states in heavy
nuclei~\cite{waas} shows that the pole mass of the pion could be
even a few per cents higher than the free space value at nuclear
matter density. The $m_\pi^\star$ in our in-medium effective
chiral Lagrangian is not necessarily the pole mass. Thus it is not
clear how to incorporate this empirical information into our
theory. We will assume here that $m_\pi^\star$ does not scale.
This assumption may be justified by using the fact that the pion
is an almost perfect Goldstone boson. Under the assumption
$m_\pi^\star =m_\pi$, Eq. (\ref{gmor}) may give
\be
\frac{f^\star_\pi}{f_\pi}\sim \left( \frac{\bra
\bar{q}q\ket^\star}{\bra\bar{q}q\ket}\right)^{1/2}. \ee It is an
example to show the relation between the BR scaling factor and the
change of the vacuum in medium.

Since the hedgehog solution of the Skyrme Lagrangian
(\ref{skyrme}) gives the Skyrmion mass proportional to
$\sqrt{g_A}f_\pi$, the nucleon mass must scale in the matter as
\be
\frac{m_N^\star}{M}=\sqrt{\frac{g_A^\star}{g_A}}\frac{f_\pi^\star}{f_\pi}.
\ee Here $M$ represents the nucleon mass in free space and
$m_N^\star$ is the in-medium effective mass of nucleon (later
identified with the Landau effective mass). In the simple Skyrme
model $g_A$ is inversely proportional to $e^2$ which does not
scale since the quartic Skyrme term is classically
scale-invariant. So $g_A$ does not change in our simple model and
the nucleon mass scales in the same way as pion decay constant in
medium.

The successful low energy results of the tree level in the chiral
effective theory implemented with hidden local symmetry, i.e.,
Kawarabayashi-Suzuki-Riazuddin-Fayyazudin(KSRF) relation, $\rho$
coupling universality, $\rho$-meson dominance, etc., are shown to
remain valid with the loop effects at low energy~\cite{hy92}.
Though it is proven in free space, it might hold in medium. If we
assume KSRF relation
\be
m_v^2=2e^2f_\pi^2 \ee is satisfied in nuclear matter, we obtain
\be
\frac{m_v^\star}{m_v}=\frac{f_\pi^\star}{f_\pi} \ee with the
subscript $v$ standing for light-quark vector mesons $\rho$ and
$\omega$.

The fluctuating component $\chi^\prime$ in the soft $\chi$ is
expected to represent multi-pion excitations in scalar channel and
to give a scalar effective field $\sigma$ in dense medium by
interpolation. It is known that the correlated 2$\pi$ exchange can
be approximated by a scalar field with a broad mass
distribution~\cite{djv80}. Durso, Kim, and Wambach's recent
calculation of $N\bar{N}\to \pi\pi$ helicity amplitude in the
scalar channel $f^{J=0}_+$ for $\rho\sim\rho_0$ with BR scaling of
vector meson mass shows that the resonance of the scalar mass
becomes very sharper and that its value shifts downward $\approx$
500 MeV~\cite{dkw93}. $\rho_0$ represents the normal nuclear
matter density (0.16/fm$^3$). The light ($\ll \Lambda_\chi$) and
decreasing mass of the scalar field suggests that it is the
expected Beane and van Kolck's dilaton~\cite{beane}. Since the
main $\pi -\pi$ rescattering comes from $\rho$-meson
exchange~\cite{djv80}, the shift of the scalar mass in medium is
affected by the density-dependence of the $\rho$ mass. It is clear
for low densities, where the linear approximation works well, that
\be
\frac{m_\sigma^\star}{m_\sigma}\approx\frac{m_v^\star}{m_v}. \ee
Now we find that the hadron masses and pion decay constant
decrease in the similar way.
\be
\Phi (\rho) \approx \frac{f^\star_\pi (\r )}{f_\pi} \approx
\frac{m^\star_\s (\r )}{m_\s}\approx\frac{m_v^\star (\r )}{m_v}
\approx \frac{m^\star (\r )}{M}. \label{BRscaling} \ee For the
moment, we are ignoring the scaling of $g_A$ to which we will
return later. $M^\star$ represents the effective nucleon mass
obtained with a fixed $g_A^\star$. The universal factor $\Phi (\r
)$ can be determined by experiments and/or QCD sum rules.

Note that the scalar field that governs BR scaling is the
quarkonium component of the scale anomaly, not the hard gluonic
component. The latter gives dominant contribution to the scale
anomaly in QCD but is integrated out in the effective Lagrangian.
This structure imposes the hadron scaling relation
(\ref{BRscaling}) by a Nambu-Jona-Lasinio(NJL) mechanism as
described in \cite{BBR}. Recently Liu {\it et al}.'s detailed
lattice analysis~\cite{liu} for the source of the mass of a
constituent quark supports this structure. They show that
dynamical symmetry breaking contribution gives most of the mass of
the chiral constituent quark. It means that the change of the
vacuum in (\ref{BRscaling}) can be related only in a subtle way to
the light quark hadron mass.

When considering BR scaling, one must note that the BR-scaled
masses and decay constants are the parameters in an effective
theory. An effective parameter defined in one theory can be
different from the parameter defined in another theory, even
though two theories have the same physics in common. So the
connection between the BR-scaled parameters and the parameters in
the other theories needs be established via observed quantities,
i.e., experiments.

One must also note that BR scaling is the result of mean field
approximation. So when one considers the cases where higher order
modifications are important, that is, the mean-field approximation
is not reliable, BR scaling cannot be applied without further
corrections. Let us take Goldberger-Treiman relation
\be
g_{\pi NN}f_\pi = g_AM \ee as an example. In the real world $g_A$
decreases from $\sim$ 1.26 in free space to $\sim $ 1 at normal
nuclear matter density since it is affected by the short range
interaction between baryons. So the naive BR scaling, which would
have $g_{\pi NN}$ remain constant, can not explain
Goldberger-Treiman relation in the low density region. It means
higher order modifications spoil the tree-order BR scaling at low
densities. At high densities above the normal nuclear matter
density, however, $g_A^\star$ remains at 1 while $f_\pi^\star$
continues to drop and hence the coupling constant ratio increases.

\begin{figure}
\setlength{\epsfysize}{4in} \centerline{\epsffile{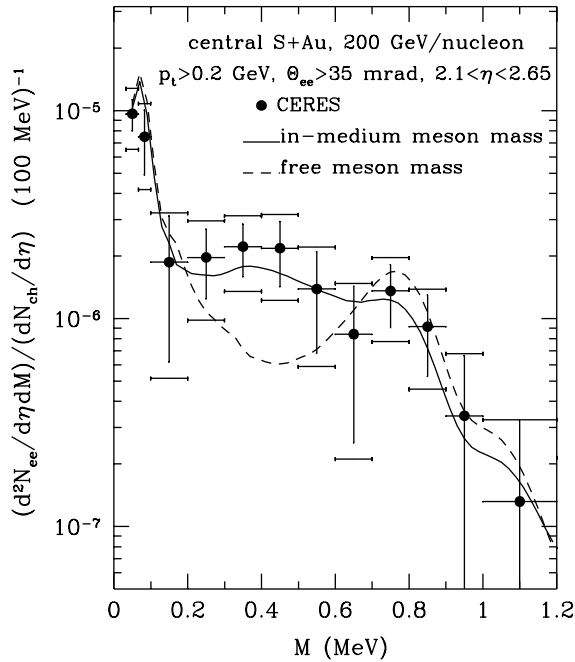}}
\caption{The comparison of CERES dilepton experiments and the
theoretical predictions with the free-mass mesons and with scaled
mass mesons. The figure comes from \cite{kl96}.}\label{dilep}
\end{figure}

Although Brown and Rho's arguments about the scaling relation of
effective parameters may seem a bit drastic, the experiments
(e.g., CERES~\cite{CERES}, HELIOS-3~\cite{helios3}) provide 
support for this scheme. The explanation of CERES data is one of
the good examples. As seen in Fig.\ \ref{dilep}, the scaling of
effective masses of hadrons in medium reproduces the CERES results
very simply at the mean field level~\cite{LKB}.\footnote{In Ref.\
\cite{LKB} BR scaling is realized on the basis of the constituent
quark model~\cite{st94}. Li, Ko, and Brown used Walecka theory to
implement the density dependence of the constituent quarks.}
\subsection{Duality}
Brown-Rho scaling describes the low-mass dilepton enhancement
on the basis of
the quasiquark picture. But the increase of dilepton yields can
also be described by a hadronic theory with free masses,
introducing the appropriate variables, e.g.\ nucleonic
excitations. Rapp, Chanfray and Wambach~\cite{rcw97} described the
CERES results successfully using conventional many-body theory.
To evaluate the in-medium rho meson propagator they renormalized the
intermediate two-pion states, which dress rho mesons and interact
with the surrounding nucleons and deltas, and considered the
direct interaction of rho mesons with surrounding hadrons,
especially $\rho$-like baryon-hole excitations (``rhosobar"). They
found that such medium effects broaden the spectral function of rho
meson and the dilepton production at $\sim m_\rho /2$
is enhanced.

Rapp {\it et al.}'s hadronic rescattering approach and Brown-Rho's
quasiparticle approach have merits and demerits, compared with
each other. Since BR scaling is approximate at mean-field
level, Rapp {\it et al.}'s theory is more reliable at low
densities than BR scaling. On the other hand at higher densities
many diagrams have to be considered in the hadronic rescattering
approach. BR scaling gives a medium-modified vacuum around which
the weak fluctuations can be dealt with.

If both descriptions are correct, the effective variables are
expected to change smoothly from hadrons to quasiquarks subject to
BR scaling and both descriptions must show duality around the
hadron-quark changeover densities\footnote{Rapp and Wambach
suggested~/cite{rw99} recently interpreting the strong broadening 
of the $\rho$-meson resonance as a reminiscence of hadron-quark 
changeover.} Such duality was suggested by
Brown {\it et al}.\ \cite{BLRRW} and Y. Kim {\it et al}.\
\cite{krbr99} studied it more precisely. For rho mesons in medium
they studied a two-level model, which consists of the collective
rhosobar [$N^*(1520)N^{-1}$]\footnote{$N^*(1520)$ gives the most
important contribution to the photoabsorption cross sections in
the dileption analysis~\cite{rubw}.} and the `elementary' $\rho$.

The in-medium $\rho$-meson propagator is
\be
D_\rho = \left[ q_0^2 -\qb^2 -(m_\rho^0)^2-\Sigma_{\rho\pi\pi}
-\Sigma_{\rho BN} \right]^{-1} \ee where the $\Sigma$ indicates
self-energies and $m_\rho^0$ is the bare mass of $\rho$. Taking the
free rho meson mass $m_\rho = (m_\rho^0)^2+\Re\Sigma_{\rho\pi\pi}$, we
obtain the dispersion relation in the $\qb =0$ limit
\be
q_0^2=m_\rho^2+\Re\Sigma_{\rho N^*N}.\label{dis} \ee The
phenomeological Lagrangian for the s-wave interaction\footnote{For
p-wave coupling, the Lagrangian is
\be
\L=\frac{f_{\rho BN}}{m_\rho}B^\dagger (\Bs\times\rhob
)\cdot\rhob_a t_aN + \mbox{h.c.}
\nonumber\ee} between the elmentary $\rho$ and
the $\rho$-like baryon-hole excitation is
\be
\L_{\rho BN}=\frac{f_{\rho BN}}{m_\rho}B^{\dagger}
(q_0\Bs\cdot\rhob_a-\rho^0_a\Bs\cdot\qb )t_aN + \mbox{h.c.}
\label{rbn}\ee with appropriate spin operator $\Bs$ and isospin
operators $t_a$~\cite{ppllm}. The self-energy from the interaction
(\ref{rbn}) for $B=N^*(1520)$ is
\be
\Sigma_{\rho N^*(1520)N}(q_0) \approx\frac83 f_{\rho N^*N}^2
\frac{q_0^2}{m_\rho^2}\frac{\rho_n}{4}\left(\frac{2\Delta
E}{(q_0+i\Gamma_{t}/2)^2-(\Delta E)^2}\right) \label{rnn}\ee
where
$\rho_n$ is the nuclear density and the total width $\Gamma_t$
includes the the free width of $M_{N^*}(1520)$ and its modification in
medium. Neglecting nuclear Fermi motion ($\qb = 0$ limit), 
$\Delta E=M_{N^*}-M_N$.

\begin{figure}
\centerline{\epsfig{file=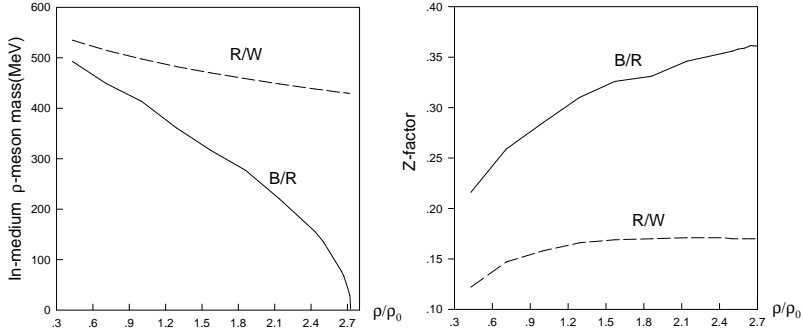,width=11cm}} 
\vskip -1.4cm
\caption{The in-medium $\rho$-meson mass and $Z$ factor obtained in
\cite{krbr99} for $\Gamma_t=0$.}\label{rwo1}
\end{figure}
Kim {\it et al.}~\cite{krbr99} showed that the dispersion relation
(\ref{dis}) gives the two 
states, which have $\rho$-meson quantum numbers, with the spectral weight
\be
Z=\left( 1-\frac{\del}{\del q_0} \Re \Sigma_{\rho
N^*N}\right)^{-1} \ee 
and that one of them can be interpreted as
an in-medium vector meson whose mass decreases. Figure
\ref{rwo1} shows the results with $\Gamma_t =0$. R/W indicates
that $m_\rho$ in (\ref{rnn}) is the free mass and B/R indicates
that $q_0/m_\rho$ in (\ref{rnn}) is replaced by 1, i.e., replacing
$m_\rho$ by $m_\rho^\star$. Note that $q_0$ i.e., the in-medium
$\rho$-meson mass, cannot go to zero at any density as seen in R/W
of Fig. \ref{rwo1} if $m_\rho$ in the denominator of (\ref{rnn})
is fixed. It matches neither BR scaling~\cite{BR91} nor the
prediction that $m_\rho^\star\rightarrow 0$ at the chiral
transition point~\cite{BBR}. Brown {\it et al}.~\cite{BLRRW}
suggested the replacement of $m_\rho$ in (\ref{rnn}) by
$m_\rho^\star$ in order to go from Rapp {\it et al.}'s hadronic
theory to BR scaling which predicts zero vector meson mass at some
high density. B/R of Fig. \ref{rwo1} shows that in-medium
$\rho$-meson mass goes to zero near $\rho\sim 2.75 \rho_0$ as
predicted in \cite{BBR}. The study of the schematic
model~\cite{BLRRW,krbr99} provides how BR scaling can be
interpolated from a hadronic rescattering description.
\section{Effective Lagrangian with BR scaling}

If the large anomalous dimension of the scalar field in FTS1 is a
symptom of a strong-coupling regime, it suggests that one should
redefine the vacuum in such a way that the fluctuations around the
new vacuum become weak-coupling. This is the basis of the BR
scaling introduced in \cite{BR91}. The basic idea is to fluctuate
around the ``vacuum" defined at $\rho\approx \rho_0$ characterized
by the quark condensate $\la \bar{q}q\ra_{\rho} \equiv
\la\bar{q}q\ra^\star$. This theory was developed with a chiral
Lagrangian implemented with the trace anomaly of QCD as seen in
the last section. The Lagrangian used was the one valid in the
large $N_c$-limit of QCD and hence given entirely in terms of
boson fields from which baryons arise as solitons (skyrmions):
Baryon properties are therefore dictated by the structure of the
bosonic Lagrangian, thereby leading to a sort of universal scaling
between mesons and baryons. One can see, using a dilated chiral
quark model, that BR scaling is a generic feature also at high
temperature in the large $N_c$ limit~\cite{hkl}.\footnote{If one
calculate $f_\pi^\star (T)$ and $m_N^\star (T)$ for zero density
in chiral perturbation theory, the temperature-dependence deviates
from BR scaling at low $T$~\cite{gls}. In \cite{hkl} Y. Kim, H.K.
Lee and M. Rho discussed the point that BR scaling holds at
non-zero density.}

In this description, one is approximating the complicated strong
interaction process at a given nuclear matter density in terms of
``quasiparticle" excitations for both baryons and bosons in
medium. This means that the properties of fermions and bosons in
medium at $\rho\approx \rho_0$ are given in terms of tree diagrams
with the properties defined in terms of the masses and coupling
constants universally determined by the quark condensates at that
density.

The question then is: How can one marry the FTS1 Lagrangian with
the BR-scaled Lagrangian? The next question is how to identify
BR-scaled parameters with the Landau parameters. In this section
we will provide some answers to these two questions.
\subsection{A Hybrid BR-scaled model}\label{hymodel}

As a first attempt to answer these questions, we consider the
hybrid model in which the ground state is given by the mean field
of the FTS1 Lagrangian ${\L}_{FTS1}$ and the fluctuation around
the ground state is described by the tree diagrams of the
BR-scaled Lagrangian $\Delta {\L}$,
\be
{\L}^{eff}={\L}_{FTS1} + \Delta {\L}. \ee
This model with the canonical parameters (T1) for the ground state
and a BR-scaled fluctuation Lagrangian in the non-strange flavor
sector was recently used by Li, Brown, Lee, and Ko~\cite{LBLK96}
for describing simultaneously nucleon flow and dilepton production
in heavy ion collisions. The nucleon flow is sensitive to the
parameters of the baryon sector, in particular, the repulsive
nucleon vector potential at high density whereas the dilepton
production probes the parameters of the meson sector. With a
suitable momentum dependence implemented to the FTS1 mean field
equation of state, the nucleon flow comes out in good agreement
with experiments. Furthermore the scaling of the nucleon mass in
the FTS1 theory in dense medium, say, at $\rho\sim 3\rho_0$, is
found to be essentially the same as that given by the NJL model.
Therefore we can conclude that the nucleon in FTS1 scales in the
same way as BR scaling.

The dilepton production involves both baryon and meson properties,
the former in the scaling of the nucleon mass and the latter in
the scaling of the vector meson ($\rho$) mass. The equation of
state correctly describing the nucleon flow and the BR-scaled
vector meson mass is found to fit the dilepton data equally well,
comparable to the fit obtained in \cite{LKB} using  Walecka mean
field. What is important in this process is the scalar mean field
which governs the BR scaling and hence the production rate comes
out essentially the same for FTS1 and Walecka mean fields. The
delicate interplay between the attraction and the repulsion that
figures importantly  in the compression modulus does not play an
important role in the dilepton process.

Let us see how the particles behave in the background of the FTS1
ground state given by ${\L}_{FTS1}$. The nucleon of course scales
{\it\`a la} BR as mentioned above. We can say nothing on the pion
and the $\rho$ meson with the FTS1 theory. However there is
nothing which would preclude the $\rho$ scaling {\it\`a la} BR and
the pion non-scaling within the scheme. What is encoded in the
FTS1 theory is the behavior of the $\omega$ and the scalar $S$
which figure importantly in Walecka mean fields. Let us therefore
focus on these two fields in medium near normal nuclear matter
density.

We have already shown in subsection \ref{dim} that the mass of the
scalar field $S$ drops less rapidly than BR scaling for $d>2$. One
can think of this as a screening of the four-Fermi interaction in
the scalar channel that arises when the scalar meson with the
BR-scaled mass is integrated out.  This feature and the property
of the $\omega$ field can be seen by the toy model of the FTS1
Lagrangian (that includes terms corresponding up to three-body
forces)
\be
\L_{toyFTS1}=\L_{BR}
 +\frac{m_\omega^2}{2}(2+\eta )\frac{\phi}{S_0}\omega^2 -\frac{m_s^2\phi^3}
{3S_0}\label{toy} \ee where
\be
\L_{BR} &=& \bar{N}(i\gamma_{\mu}(\del^\mu+ig_v\omega^\mu
)-M+g_s\phi )N \nonumber\\ & &+\frac{m_\omega^2}{2}\omega^2
(1-\frac{2\phi}{S_0}) -\frac{m_s^2}{2}\phi^2
(1-\frac{2\phi}{3S_0}).\label{toyBR} \ee We have written
${\L_{BR}}$ such that the BR scaling is incorporated {\it at mean
field level} as
\be
\Phi (\rho) = \frac{M^\star}{M}=\frac{m_s^\star}{m_s}
=\frac{m_\omega^\star}{m_\omega}\approx 1-\frac{\phi}{S_0} \ee
with
\be
S_0=\la 0|S|0\ra = M/g_s. \ee Here we are ignoring the deviation
of the scaling of the effective nucleon mass (denoted later as
$m_N^\star$) from the universal scaling $\Phi (\rho)$~\cite{FR96}.
This will be incorporated in the next subsection. We can see from
(\ref{toy}) that the FTS1 theory brings in an additional repulsive
three-body force  coming from a cubic scalar field term while if
one takes $\eta=-2$, the $\omega$ field will have a BR scaling
mass in nuclear matter. Fit to experiments favors $\eta\approx
-1/2$ instead of $-2$, thus indicating that the FTS1 theory
requires a many-body suppression of the repulsion due to the
$\omega$ exchange two-body force. (In the simple model with BR
scaling that we will construct below, we shall use this feature by
introducing a ``running'' vector coupling $g_v^\star$ that drops
as a function of density.) The effective $\omega$ mass may be
written as
\be
m_\omega^{\star 2}\approx [1+\eta\frac{\phi_0}{S_0}]m_\omega^2.
\ee For $\eta <0$, we have a falling $\omega$ mass corresponding
to BR scaling (modulo, of course, the numerical value of $\eta$).
In FTS1, there is a quartic term $\sim \omega^4$ which is
attractive and hence {\it increases} the $\omega$ mass. In fact,
because of the attractive quartic $\omega$ term, we have
\be
\frac{m_\omega^\star}{m_\omega}\approx1.12\label{omegaeff} \ee at
the saturation density with T1 parameter set.
This would seem to suggest that due to higher polynomial
(many-body) effects, the $\omega$ mass does not follow BR scaling
in medium. Furthermore the $\omega$ effective mass
increases slowly around this equilibrium value. On the other hand
Klingl, Waas, and Weise's recent sum rule analysis on current
correlation function~\cite{KKW,kkw98} follows BR scaling. It shows
that effective $\omega$ meson mass scales down by about 15$\%$ at
normal nuclear matter density from its free value. We will discuss
the shift of vector meson mass in medium in detail in Section
\ref{mesons}.
\subsection{Model with BR scaling}\label{model}

The above hybrid model suggests how to construct an effective
Lagrangian model that implements BR scaling and contains the same
physics as the FTS1 theory. The crucial point is that such a
Lagrangian is to give {in the mean field} the chiral liquid
soliton solution. This can be done by making the following
replacements in (\ref{toyBR}):
\be
M-g_s\phi_0&\rightarrow& M^\star,\nonumber\\ m_\omega^2
(1-\frac{2\phi_0}{S_0})&\rightarrow&
{m_\omega^\star}^2,\nonumber\\ m_s^2
(1-\frac{2\phi_0}{S_0})&\rightarrow& {m_s^\star}^2 \ee and write
\be
\L_{BR} &=& \bar{N}(i\gamma_{\mu}(\del^\mu+ig_v\omega^\mu
)-M^\star+h\phi )N \nonumber\\ & &-\frac 14 F_{\mu\nu}^2 +\frac 12
(\partial_\mu \phi)^2 +\frac{{m^\star_\omega}^2}{2}\omega^2
-\frac{{m^\star_s}^2}{2}\phi^2\label{toyBRP} \ee with
\be
\frac{M^\star}{M}=\frac{m_\omega^\star}{m_\omega}=\frac{m_s^\star}{m_s}
=\Phi (\rho).\label{BRscaled} \ee The additional term
$\bar{N}h\phi N$ is put in to account for the difference between
the Landau mass $m_L^\star$ to be given later and the BR scaling
mass $M^\star$. In the chiral Lagrangian approach with BR scaling,
the difference comes through the Fock term involving non-local
pion exchange. This will be discussed further in the next chapter.
For simplicity we will take the scaling in the form
\be
\Phi (\rho)=\frac{1}{1+y\rho/\rho_0} \ee with $y=0.28$ so as to
give $\Phi (\rho_0)=0.78$ (corresponding to $k_F=260$ MeV) found
in QCD sum rule calculations~\cite{jin}, as well as from the {\it
in-medium} GMOR relation~\cite{BRPR96}.

Note that the Lagrangian (\ref{toyBRP}) treated at the mean field
level would give a Walecka-type model with the meson masses
replaced by the BR scaling mass.\\ \indent Now in order to
describe nuclear matter in the spirit of the FTS1 theory, we
introduce terms cubic and higher in $\omega$ and $\phi$ fields to
be treated as perturbations around the BR background as
\be
\L_{n-body}=a \phi\omega^2 + b \phi^3 +c \omega^4 +d \phi^4 +e
\phi^2\omega^2 +\cdots\label{nbody} \ee where $a$ -- $e$  are
``natural" (possibly density-dependent) constants to be
determined. By inserting for the $\phi$ and $\omega$ fields the
solutions of the static mean field equations given by
${{\L}_{BR}}$,
\be
{m_s^\star}^2\bar\phi &=&h\sum_i\bar{N}_iN_i\label{meanphi}
\\
{m^\star_\omega}^2\bar\omega_0 &=&g_v\sum_i N_i^\dagger N_i \ee we
see that at the mean-field level, ${\L_{n-body}}$ generates three-
and higher-body forces with the exchanged masses density-dependent
{\it\`a la} BR. Note that at this point, the scaling factor $\Phi$
and the mean field value (\ref{meanphi}) are not necessarily
locked to each other by self-consistency.

As the first trial, we will consider the drastically simplified
model by dropping the n-body term (\ref{nbody}) and minimally
modifying the BR Lagrangian (\ref{toyBRP}). We shall do this by
letting as mentioned above the vector coupling ``run'' as a
function of density. For this,  we use the observation made in
\cite{LBLK96} that the nucleon flow probing higher density
requires that $g_v^\star/m_\omega^\star$ be independent of density
at low densities and decrease slightly at high densities. We shall
therefore take, to simulate this particular many-body correlation
effect, the vector coupling to scale as
\be
\frac{g_v^\star}{g_v}=\frac{1}{1+z\rho/\rho_0}\label{gvscaling}
\ee with $z$ equal to or slightly greater than $y$. The $h$ is
assumed not to scale although it is easy to take into account the
density dependence if necessary.

The scaling (\ref{gvscaling}) seems at odds with the prediction
made with the Skyrme model~\cite{mr85} where using the Skyrme
model with the quartic Skyrme term inversely proportional to the
coupling $e^2$, it was found that
\be
\frac{e}{e^\star}\sim \sqrt{\frac{g_A^\star}{g_A}}.\nonumber \ee
It is tempting to identify (via $SU(6)$ symmetry) $e$ with $g_v$
that we are discussing here since the Skyrme quartic term can
formally be obtained from a hidden gauge-symmetric Lagrangian by
integrating out the $\rho$ meson field. If this were correct, one
would predict that the vector coupling increases -- and not
decreases -- as density increases since we know that $g_A^\star$
is quenched in dense matter. This identification could be too
naive and incomplete in two respects, however. First of all, this
skyrmion formula is a large-$N_c$ relation and secondly the Skyrme
quartic term embodies {\it all} short-distance physics in one
dimension-four term in a derivative expansion. Thus the constant
$1/e$ must represent a lot more than just the vector-meson
($\rho$) degree of freedom. Furthermore we are concerned with the
$\omega$ degree of freedom which in a naive derivative expansion
would give a six-derivative term. The BR-scaled model we are
constructing should involve not only short-distance physics
presumably represented by the $1/e$ term (consistent with the
understanding that the quenching of $g_A$ is a short-distance
phenomena) but also longer-range correlations. Therefore the
qualitative difference should surprise no one.

The truncated Lagrangian that we shall consider then is
\be
\L_{BR} &=& \bar{N}(i\gamma_{\mu}(\del^\mu+ig_v^\star\omega^\mu
)-M^\star+h\phi )N \nonumber\\ & &-\frac 14 F_{\mu\nu}^2 +\frac 12
(\partial_\mu \phi)^2 +\frac{{m^\star_\omega}^2}{2}\omega^2
-\frac{{m^\star_s}^2}{2}\phi^2.\label{toyBRPP} \ee As suggested in
\cite{gr,BR96,tsp}, chiral in-medium Lagrangians can be brought to
a form equivalent to a Walecka-type model. The scalar field
appearing here transforms as a singlet, not as the fourth
component of $O(4)$ of the linear sigma model. As it stands, the
Lagrangian (\ref{toyBRPP}) does not look chirally invariant. This
is because we have dropped pion fields which play no role in the
ground state of nuclear matter. In considering fluctuations around
the ground state, they (and other pseudo-Goldstone fields such as
kaons) should be reinstated. The chiral singlet $\omega$ field and
$\phi$ field can be considered as auxiliary fields brought in from
a Lagrangian consisting of multi-Fermion field
operators~\cite{tsp} via a Hubbard-Stratonovich transformation.

Since we treat density-dependent parameters, we must be careful in
thermodynamic consistency. After showing the way to treat
density-dependent parameters in the next section, we display the
results of our model for nuclear matter properties. We will argue
in the next subsection that the energy density from
(\ref{edensity}) is independent of the way how the parameters move
as density increases.
\subsection{Thermodynamic consistency in medium}\label{thermo}

In this subsection we address the issue of thermodynamic
consistency of the Lagrangian (\ref{toyBRPP}) treated in the mean
field approximation. For instance, it is not obvious that the
presence of the density-dependent parameters in the Lagrangian
does not spoil the self-consistency of the model, in particular,
energy-momentum conservation in the medium and also certain
relations of Fermi-liquid structure of the matter. The purpose of
this section is to show that there is no inconsistency in doing a
field theory with BR scaling masses and other parameters. This
point has not been fully appreciated by workers in the field.
The Euler-Lagrange equations of motion are in the same as the ones
that arise from the field for the Lagrangian wherein the masses
and constants are not BR scaling. While this procedure gives
correct energy density, pressure and compression modulus, the
energy-momentum conservation is not automatically assured. In
fact, if one were to compute the pressure from the energy-density
${\E}$, one would find that it does not give $\frac 13 <T_{ii}>$
(where $T_{\mu\nu}$ is the conserved energy-momentum tensor and
the bra-ket means the quantity evaluated in the mean-field
approximation as defined before) unless one drops certain terms
without justification. This suggests that it is incorrect to take
the masses and coupling constants independent of fields in
deriving, by Noether theorem, the energy-momentum tensor. So the
question is: how do we treat the field dependence of the BR
scaling masses and constants?

One possible solution to this problem is as follows. In Section
\ref{scaling}, the density dependence of the Lagrangian arose as
the vacuum expectation value of the scalar field $\chi$ that
figures in the QCD trace anomaly. By vacuum we mean the state of
baryon number zero modified from that of true vacuum by the strong
influence of the baryons in the system. See later for more on this
point. It corresponded to the condensate of a quarkonium component
of the scalar $\chi$ with the gluonium component -- which lies
higher than the chiral scale -- integrated out. It was assumed to
scale in dense medium in a Skyrmion-type Lagrangian subject to
chiral symmetry. Now in the language of a chiral Lagrangian
consisting of the nucleonic matter field $N$ with other massive
fields integrated out, this scalar condensate would be some
function of the vacuum expectation value of $\bar{N}N$ or
$\bar{N}\gamma_0N$ coming from multi-Fermion field operators
mentioned above. How these four-Fermi and higher-Fermi field terms
can lead to BR scaling in the framework of chiral perturbation
theory was discussed in \cite{tsp}. We shall follow this strategy
in this paper leaving other possibilities (such as dependence on
the mean fields of the massive mesons) for later investigation.
For this, it is convenient to define
\be
\check\rho u^\mu \equiv \bar{N}\gamma^\mu N \ee with unit fluid
4-velocity \be u^\mu =\frac{1}{\sqrt{1-\vb^2}}(1,\vb )
=\frac{1}{\sqrt{\rho^2-\jb^2}}(\rho ,\jb ) \ee with the baryon
current density
\be
\jb = <\bar{N}\gammab N > \ee and the baryon number density \be
\rho =<N^\dagger N > =\sum_in_i.\label{rhon} \ee We will take
$n_i$ to be given by the Fermi distribution function,  $n_i=\theta
(k_F-|\Bk_i |)$ at $T=0$. We should replace $\rho$ in
(\ref{toyBRPP}) by $\check\rho$ for consistency of the model. The
definition of $\check\rho$ makes our Lagrangian Lorentz-invariant
which will later turn out to be useful in deriving relativistic
Landau formulas. With this, the Euler-Lagrange equation of motion
for the nucleon field is
\be
\frac{\delta\L}{\delta\bar{N}}&=&
\frac{\del\L}{\del\bar{N}}+\frac{\del\L}{\del\check\rho}\frac{\del\check\rho}
{\del\bar{N}} \nonumber\\ &=&[i\gamma^\mu (\del_\mu
+ig_v^\star\omega_\mu -iu_\mu\check\Sigma ) -M^\star +h\phi ]
N\nonumber\\ &=&0\label{fer} \ee with
\be
\check\Sigma &=&\frac{\del\L}{\del\check\rho}\\ &=&
m_\omega^\star\omega^2\frac{\del m_\omega^\star}{\del\check\rho}
-m_s^\star\phi^2\frac{\del m_s^\star}{\del\check\rho}
-\bar{N}\omega^\mu\gamma_\mu N\frac{\del
g_v^\star}{\del\check\rho} -\bar{N}N\frac{\del
M^\star}{\del\check\rho}. \nonumber\ee This additional term which
may be related to what is referred to in many-body theory as
``rearrangement terms" plays a crucial role in what
follows.\footnote{For a recent discussion on rearrangement terms
as well as thermodynamic consistency in the context of standard
many-body theory, see \cite{flw}. The notion of density-dependent
parameters in many-body problems is of course an old
one~\cite{kohnsham}.} The equations of motion for the bosonic
fields are
\be
(\del^\mu\del_\mu +m_s^{\star 2})\phi &=& h\bar{N}N \label{ph} \\
\del_\nu F_\omega^{\nu\mu}+m_\omega^{\star 2}\omega^\mu &=&
g_v^\star \bar{N} \gamma^\mu N. \ee

We start with the conserved canonical energy-momentum tensor
constructed a la Noether from the Lagrangian (\ref{toyBRPP}):
\be
T^{\mu\nu}&=&i\bar{N}\gamma^\mu\del^\nu N
+\del^\mu\phi\del^\nu\phi
-\del^\mu\omega_\lambda\del^\nu\omega^\lambda\nonumber\\ &
&-\frac12 [(\del\phi )^2 -m_s^{\star 2}\phi^2-(\del\omega
)^2+m_\omega^{\star 2}\omega^2 -2\check\Sigma\bar{N}\slash{u} N
]g^{\mu\nu}. \label{emtensor} \ee We shall compute thermodynamic
quantities from (\ref{emtensor}) using the mean field
approximation which amounts to taking
\be
N &=&\frac{1}{\sqrt{V}}\sum_i
a_i\sqrt{\frac{E_{\kappa_i}+m_N^\star}{2E_{\kappa_i}}}
\left(\begin{array}{c} \chi\\
\frac{\sigmab\cdot\kappab_i}{E_{\kappa_i}+m_N^\star}\chi
\end{array}\right)
\exp{(i\kappab_i\cdot\xb -i(g_v^\star\omega_0-u_0\Sigma+E_i) t)}
\nonumber\\ h\phi &=&C_h^2<\bar{N}N
>=C_h^2\sum_in_i\frac{m_N^\star } {\sqrt{\kappab^2_i+m_N^{\star
2}}}\label{phi}\\ g_v^\star\omega &=&C_v^2(\rho , \jb
)=C_v^2\sum_in_i\left(
1,\frac{\kappab_i}{\sqrt{\kappab^2_i+m_N^{\star 2}}}\right)
\nonumber\ee where $a_i$ is the annihilation operator of the
nucleon $i$, with
\be
n_i =<a^\dagger_i a_i>, \ee and \be \Sigma
&=&\langle\check\Sigma\rangle ,\\ \kappab_i &\equiv
&\Bk_i-C_v^2\jb +\ub\Sigma , \ee and
\be
E_{\kappa_i} &=&\sqrt{\kappab^2_i+m_N^{\star 2}}. \ee $\chi$ is
the spinor and $\sigmab$ is the Pauli matrix. We have defined
\be
C_v(\check\rho ) \equiv \frac{g_v^\star (\check\rho
)}{m_\omega^\star (\check\rho )} \ee and
\be
C_h(\check\rho ) \equiv \frac{h}{m_s^\star (\check\rho)}\equiv
\frac{1}{\C_h(\check\rho )}. \ee In this approximation, the energy
density is
\be
\E &=&<T^{00}>\label{edensity}\\ &=& <i\bar{N}\gamma^0\del^0N
+\frac12 m_s^{\star 2}\phi^2 -\frac12 m_\omega^{\star
2}\omega^2+\check\Sigma\bar{N}\slash{u} N > \nonumber\\ &=&
\frac12 C_v^2(\rho^2+\jb^2) +\frac12 \C_h^2(m_N^\star -M^{\star
})^2 +\sum_l n_l\sqrt{\kappab^2_l+m_N^{\star 2}}-
\Sigma\ub\cdot\jb .\nonumber \ee Note that the $\Sigma$-dependent
terms cancel out in the comoving frame ($\vb =0$), so that the
resulting energy-density is identical to what one would obtain
from the Lagrangian in the mean field with the density-dependent
parameters taken as field-independent quantities.\\ \indent Given
the energy density (\ref{edensity}), the pressure can be
calculated by (at $T=0$)
\be
p&=&-\frac{\del E}{\del V}=\rho^2\frac{\del\E
/\rho}{\del\rho}=\mu\rho -\E\nonumber\\ &=& \frac12 C_v^2(\rho
)\rho^2 -\Sigma_0 \rho -\frac12 \C_h^2(\rho )(m_N^\star -M^{\star
}(\rho ))^2\nonumber\\ & &-\frac{\gamma}{2\pi^2}\left(
E_F(\frac{m_N^{\star 2}}{8}k_F
-\frac{1}{12}k_F^3)-\frac{m_N^{\star 4}}{8}\ln [(k_F+
E_F)/m_N^\star ]\right)\label{P} \ee where $\mu$ is the chemical
potential -- the first derivative of the energy density with
respect to $\rho$ in the comoving frame ($\vb=0$):
\be
\mu\equiv \frac{\del}{\del\rho}\E|_{\vb=0} =C_v^2\rho
+E_F-\Sigma_0\label{mu} \ee with $E_F=\sqrt{k_F^2+m_N^{\star 2}}$
and $\Sigma_0=\langle\check\Sigma \rangle_{\vb =0}$. To check that
this is consistent, we calculate the pressure from the
energy-momentum tensor (\ref{emtensor}) in the mean field at
$T=0$:
\be
p_t&\equiv&\frac13 <T_{ii}>_{\vb =0}\\ &=& \frac13 \langle
i\bar{N}\gamma_i\del_iN -\frac{1}{2}(m_\omega^{\star
2}\omega^2-m_s^{\star 2}\phi^2 -2\check\Sigma N^\dagger N
)g_{ii}\rangle_{\vb=0} \nonumber\\ &=&\frac12 C_v^2(\rho )\rho^2
-\frac12 \C_h^2(\rho )(m_N^\star -M^{\star }(\rho ))^2-\Sigma_0
\rho\nonumber\\ &&-\frac{\gamma}{2\pi^2}\left(
E_F(\frac{m_N^{\star 2}}{8}k_F
-\frac{1}{12}k_F^3)-\frac{m_N^{\star 4}}{8}\ln [(k_F+
E_F)/m_N^\star ]\right). \nonumber\ee This agrees with (\ref{P}).
Thus our equation of state conserves energy and momentum.

We showed that a simple effective chiral Lagrangian with BR
scaling parameters is thermodynamically consistent, a point which
is important for studying nuclear matter under extreme conditions.
It is clear however that this does not require that the masses
appearing in the Lagrangian scale according to BR scaling only.
What is shown in this subsection is that masses and coupling
constants could depend on density without getting into
inconsistency with general constraints of chiral Lagrangian field
theory. This point is important for applying (\ref{toyBRPP}) to
the density regime $\rho\sim 3\rho_0$ appropriate for the CERES
dilepton experiments and also kaon production at GSI(Gesellschaft
f\"ur Schwerionenforschung) where deviation from the simple BR
scaling of \cite{BR91} might occur.
\subsection{Results}\label{result}

Based on the thermodynamic consistency of density-dependent
effective theories proven at the mean field level, we check
whether the model Lagrangian (\ref{toyBRPP}) can describe the
infinite nuclear matter properties successfully. The
characteristic properties we try to reproduce are compression
modulus, $m^\star_N$, and binding energy at normal nuclear matter
density, and the saturation density itself.
\begin{table}
\caption{Parameters for the Lagrangian (\ref{toyBRPP})  with
$y=0.28$, $m_s=700{\mbox{MeV}}$, $m_\omega =783 {\mbox{MeV}}$,
$M=939{\mbox{MeV}}$}\label{yeqz} \vskip .3cm
\begin{center}
\begin{tabular}{cccc}\hline\hline
SET&$h$& $g_v$&$z$\\ \hline S1&6.62&15.8&0.28 \\
S2&5.62&15.3&0.30\\ S3&5.30&15.2&0.31 \\ \hline\hline
\end{tabular}
\end{center}
\end{table}
\begin{table}
\caption{Nuclear matter properties predicted with the parameters
of Table \ref{yeqz}. The effective nucleon mass (later identified
with the Landau mass) is $m_N^\star=M^\star-h\phi_0$.}\label{ryez}
\vskip .3cm
\begin{center}
\begin{tabular}{cccccc}\hline\hline
SET&$E/A-M$(MeV)&$k_{eq}$(MeV)&K(MeV)& $m_N^\star /M$&$\Phi
(k_{eq})$\\ \hline S1&-16.0&257.3&296&0.619&0.79 \\
S2&-16.2&256.9&263&0.666&0.79 \\ S3&-16.1&258.2&259&0.675&0.78
\\ \hline\hline
\end{tabular}
\end{center}
\end{table}

In Table \ref{yeqz}, three sets of parameters are listed. We take
the measured free-space masses for the $\omega$ and the nucleon
and for the scalar $\phi$ for which the free-space mass cannot be
precisely given, we take $m_s=700$ MeV (consistent with what is
argued in \cite{BRPR96}) so that at nuclear matter density, it
comes close to what enters in the FTS1. The resulting fits to the
properties of nuclear matter are given in Table \ref{ryez} for the
parameters given in Table \ref{yeqz}. These results are
encouraging. Considering the simplicity of the model, the model --
in particular with the S2 and S3 set -- is remarkably close in
nuclear matter to the full FTS1. The compression modulus comes
down toward the low value that is currently favored.  In fact, the
somewhat higher value obtained here can be easily brought down to
about 200 MeV without modifying other quantities if one admits a
small admixture of the {\it residual} many-body terms
(\ref{nbody}), as we shall shortly show. The effective nucleon
Landau mass $m_N^\star/M\approx 0.67$ is in good agreement with
what was obtained in QCD sum rule calculations~\cite{furnstahljin}
and also in the next section (i.e., 0.69) by mapping BR scaling to
Landau-Migdal Fermi-liquid theory. We shall see below that this
has strong support from low-energy nuclear properties. What is
also noteworthy is that the ratio ${\R}\equiv
(g_v^\star/m_\omega^\star)^2$ forced upon us -- though not
predicted -- is independent of the density (set S1) or slightly
decreasing with density (sets S2 and S3), as required in the
nucleon flow data as found by Li, Brown, Lee and Ko~\cite{LBLK96}.
In FTS1 theory, it is the higher polynomial terms in $\omega$ and
$\phi$ defining the mean fields that are responsible for the
reduction in ${{\R}}$ needed in \cite{LBLK96}. In
Dirac-Brueckner-Hartree-Fock theory, it is found~\cite{brockmann}
that while ${{\R}}$ takes the free-space value ${{\R}}_0$ for
$\rho\approx \rho_0$, it decreases to ${{\R}}\approx 0.64{{\R}}_0$
at $\rho \approx 3\rho_0$ due to rescattering terms which in our
language would correspond to the many-body correlations.

\begin{figure}
\setlength{\epsfysize}{4.0in} \centerline{\epsffile{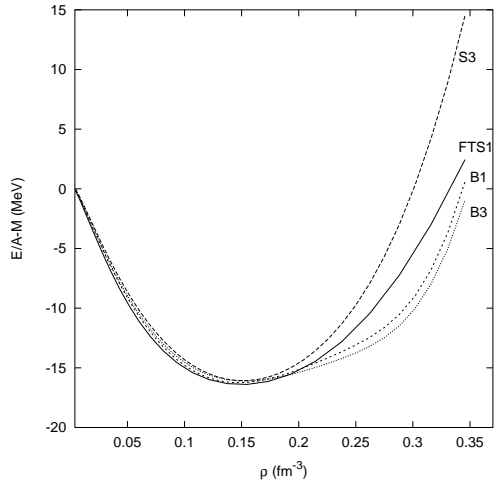}}
\caption{$E/A-M$ vs. $\rho$ for FTS1 theory (``T1'' parameter),
the ``S3'', ``B1'' and ``B3'' models defined in Table
\ref{fit}.}\label{good}
\end{figure}

The assumption that the many-body correlation terms in
(\ref{nbody}) can be entirely subsumed in the dropping vector
coupling may seem too drastic. Let us see what small residual
three-body and four-body terms in (\ref{nbody}) as many-body
correlations (over and above what is included in the running
vector coupling constant) can do to nuclear matter properties. For
convenience we rewrite (\ref{nbody}) by inserting dimensional
factors as
\be
\L_{n-body}&=&\frac{\eta_0}{2}m^2_\omega\frac{\phi}{f_\pi}\omega^2
-\frac{\kappa_3}{3!} m_s^2\frac{\phi^3}{f_\pi} \label{nbody2}\\ &
&+\frac{\zeta_0}{4!}g_v^2\omega^4
-\frac{\kappa_4}{4!}m_s^2\frac{\phi^4}{f_\pi^2}
+\frac{\eta_1}{2}m^2_\omega\frac{\phi^2}{f_\pi^2}\omega^2
\nonumber \ee and demand that the coefficients $\eta$, $\zeta$ and
$\kappa$ so defined be natural. The results of this analysis are
given in Table \ref{fit} and Fig.\ \ref{good} for various values
of the residual many-body terms and compared with those of the
truncated model (\ref{toyBRPP}) with S3 parameter set. The
coefficients are chosen somewhat arbitrarily to bring our points
home, with no attempt made for a systematic fit. (It would be easy
to fine-tune the parameters to make the model as close as one
wishes to FTS1 theory.) It should be noted that while the
equilibrium density or Fermi momentum $k_{eq}$, the effective
nucleon mass $m_N^\star$ and the binding energy $B$ stay more or
less unchanged, within the range of the parameters chosen, from
what is given by the BR-scaled model (\ref{toyBRPP}) with the S3
parameters, the compression modulus $K$ can be substantially
decreased by the residual many-body terms. Figure \ref{good} shows
that as expected, lowering of the compression modulus is
accompanied by softening of the equation of state at $\rho >
\rho_0$. While the equilibrium property other than the compression
modulus is insensitive to the many-body correlation terms, the
equation of state at larger density can be quite sensitive to
them. This is because for the generic parameters chosen, the
$m_N^\star$ can vanish at a given density above $\rho_0$ at which
the approximation is expected to break down and hence the
resulting value cannot be trusted. The B2 and B4 models do show
this instability {at $\rho\gsim 1.5\rho_0$}. (See Fig.\
\ref{comparison}.)
\begin{table}
\caption{Effect of many-body correlations on nuclear matter
properties using the Lagrangian (\ref{toyBRPP}) $+$
(\ref{nbody2}). We have fixed the free-space masses
 $m_s=700{\mbox{MeV}}, m_\omega =783 {\mbox{MeV}}, M=939{\mbox{MeV}}$ and
set $\eta_1=0$ for simplicity. The equilibrium density $k_{eq}$,
the compression modulus $K$, and the binding energy $B=M-E/A$ are
all given in units of MeV.} \label{fit} \vskip .3cm {\small
\begin{tabular}{ccccccccc cccc}\hline\hline
SET&$h$&$g_v$&$y$&$z$&$\eta_0$&$\zeta_0$&$\kappa_3$&$\kappa_4$&$k_{eq}$&
$\frac{m_N^*}{M}$&$K$&$B$\\ \hline S3&5.30&15.2&0.28&0.31& & & &
&258.2&0.675&259&16.1\\ B1&5.7&15.3&0.28&0.30& &
&0.5&-4.9&256.0&0.666&209&16.2 \\
B2&5.7&15.3&0.28&0.30&-0.055&0.18& & &257.3&0.661&201&16.1 \\
B3&5.6&15.27&0.28&0.30& & &0.31 &-4.1&259.1&0.659&185&16.1\\
B4&5.6&15.3 &0.28&0.31& & &0.9&-8.1 &256.4&0.669&191&16.1 \\
C1&5.7&15.3&0.28&0.30&-0.05&0.155& & &256.3&0.665&218&16.2\\
C2&5.8&15.3&0.28&0.30&-0.11&0.35& & &256.1&0.662&161&16.2 \\
\hline\hline
\end{tabular}
}
\end{table}

It is quite encouraging that the simple minimal model
(\ref{toyBRPP}) with BR scaling captures so much of the physics of
nuclear matter. Of course, by itself, there is no big deal in what
is obtained by the truncated model: It is not a prediction. What
is not trivial, however, is that once we have a Lagrangian of the
form (\ref{toyBRPP}) which defines the mean fields, then we are
able to control with some confidence the background around which
we can calculate fluctuations, which was the principal objective
we set at the beginning of this approach. The power of the simple
Lagrangian is that we can now treat fluctuations at {\it higher
densities} as one encounters in heavy-ion collisions,  not just at
an equilibrium point. The description of such fluctuations does
not suffer from the sensitivity with which the equation of state
depends at $\rho >\rho_0$ on the many-body correlation terms
(\ref{nbody}).

The simple Lagrangian (\ref{toyBRPP}) embodies the effective field
theory of QCD discussed by Furnstahl, Serot, and Tang~\cite{tang}
anchored on general considerations of chiral symmetry. This
Lagrangian should be viewed as an effective Lagrangian that
results from two successive renormalization group ``decimations",
one leading to a chiral liquid structure~\cite{lynn} at the chiral
symmetry scale and the other with respect to the Fermi
surface~\cite{shankar,froehlich}. The advantage of (\ref{toyBRPP})
is that it can, on the one hand, be connected to Landau
Fermi-liquid fixed point theory of nuclear matter and, on the
other hand, be extrapolated to the regime of hadronic matter
produced under extreme conditions as encountered in relativistic
heavy ion processes. It would, for instance, allow one, starting
from the ground state of nuclear matter, to treat on the same
footing the dilepton processes observed in CERES experiments as
explained in \cite{LKB} and kaon production at
SIS(Schwerionen-Synchrotron) energy and kaon condensation in dense
matter relevant to the formation of compact stars as discussed in
\cite{LLB}.
\subsection{Landau Fermi-liquid properties of the BR-scaled model}\label{liq}

The next issue we address is the connection between the mean-field
theory of the chiral Lagrangian (\ref{model}) and Landau's
Fermi-liquid fixed point theory as formulated in Section
\ref{integ}. As far as we know, this connection is the only means
available to implement chiral symmetry of QCD in dense matter
based on effective field theory. For this, we shall follow closely
Matsui's analysis of Walecka model in mean field~\cite{matsui}
exploiting the similarity of our model to the latter.
\subsubsection{Quasiparticle interactions}

The quasiparticle energy $\varepsilon_i$ and quasiparticle
interaction  $f_{ij}$ are, respectively, given by first and second
derivatives with respect to $n_i$:
\be
\varepsilon_i&=&\frac{\del\E}{\del n_i},\\
f_{ij}&=&\frac{\del\varepsilon_i}{\del n_j}. \ee A straightforward
calculation gives
\be
\varepsilon_i &=&C_v^2\rho +\sqrt{\kappab_i^2+m_N^{\star 2}}
+C_v\rho^2\frac{\del C_v}{\del n_i} -C_v^2\jb^2\frac{\del
C_v}{\del n_i} \nonumber\\ & &+\C_h(m_N^\star-M^\star
)^2\frac{\del \C_h}{\del n_i} -\C_h^2(m_N^\star-M^\star
)\frac{\del M^\star}{\del n_i} -\Sigma\ub\cdot\frac{\del\jb}{\del
n_i}\label{quasienergy} \ee and
\be
f_{ij}&=&\frac{\del\varepsilon_i}{\del
n_j}|_{\jb=\vb=0}\nonumber\\ &=&C_v^2+4C_v\rho\frac{\del
C_v}{\del\rho} +\frac{m_N^\star}{E_i}\frac{\del m_N^\star}{\del
n_j} +\rho^2[(\frac{\del
C_v}{\del\rho})^2+C_v\frac{\del^2C_v}{\del\rho^2}] \nonumber\\ &
&+(m_N^\star -M^\star )^2[(\frac{\del \C_h}{\del\rho})^2
+\C_h\frac{\del^2\C_h}{\del\rho^2}] +2\C_h\frac{\del
\C_h}{\del\rho}(m_N^\star -M^\star ) \frac{\del}{\del
n_j}(m_N^\star -M^\star )\nonumber\\ & &-2\C_h\frac{\del
\C_h}{\del\rho}(m_N^\star -M^\star ) \frac{\del
M^\star}{\del\rho}-\C_h^2\frac{\del M^\star}{\del\rho}
\frac{\del}{\del n_j}(m_N^\star -M^\star ) -\C_h^2(m_N^\star
-M^\star )\frac{\del^2 M^\star}{\del\rho^2}\nonumber\\ &
&-(C_v^2-\frac{\Sigma_0}{\rho} )
\frac{\Bk_i}{E_i}\cdot\frac{\del\jb}{\del n_j} \label{fij}\ee
with $E_i=\sqrt{\Bk_i^2+m_N^{\star 2}}$. Note that $C_v$, $\C_h$,
and $M^\star$ are functions of
\be
\langle\check\rho\rangle =u_0\rho-\ub\cdot\jb \ee in the mean
field approximation. In arriving at (\ref{fij}), we have used the
observation that in the limit $\jb\rightarrow 0$, we have
\be
\frac{\del u_0}{\del n_i} &\rightarrow& 0,\nonumber
\\
\frac{\del^2u_0}{\del n_i\del n_j} &\rightarrow& \frac{1}{\rho^2}
\frac{\del\jb}{\del n_i}\cdot\frac{\del\jb}{\del n_j},\nonumber
\\
\frac{\del\ub}{\del n_i} &\rightarrow&
\frac{1}{\rho}\frac{\del\jb}{\del n_i},\nonumber
\\
\frac{\del\langle\check\rho\rangle}{\del n_i} &\rightarrow&
1,\nonumber
\\
\frac{\del^2\langle\check\rho\rangle}{\del n_i\del n_j}
&\rightarrow& -\frac{1}{\rho} \frac{\del\jb}{\del
n_i}\cdot\frac{\del\jb}{\del n_j}\nonumber \ee and that if $f$ is
taken to be a function of the expectation value of $\check\rho$,
then as $\jb\rightarrow 0$, we have
\be
\frac{\del f}{\del n_i}&=&\frac{\del f}{\del
\langle\check\rho\rangle} \frac{\del\langle\check\rho\rangle}{\del
n_i}\nonumber\\ &\rightarrow &\frac{\del f}{\del \rho}\\
\frac{\del^2f}{\del n_i\del n_j}
&=&\frac{\del^2f}{\del\langle\check\rho\rangle^2}
\frac{\del\langle\check\rho\rangle}{\del n_i}\cdot
\frac{\del\langle\check\rho\rangle}{\del n_j} +\frac{\del
f}{\del\langle\check\rho\rangle}
\frac{\del^2\langle\check\rho\rangle}{\del n_i\del n_j}\nonumber\\
&\rightarrow&\frac{\del^2f}{\del\rho^2} -\frac{1}{\rho}\frac{\del
f}{\del\rho} \frac{\del\jb}{\del n_i}\cdot\frac{\del\jb}{\del
n_j}. \ee In the absence of the baryon current, $\jb=0$, the
quantities $\frac{\del m_N^\star}{\del n_j}$ and
$\frac{\del\jb}{\del n_j}$  simplify to
\be
\frac{\del m_N^*}{\del n_j}=\frac{ \frac{\del
M^\star}{\del\rho}-2C_h\frac{\del C_h}{\del\rho} \sum_l
n_l\frac{m_N^\star}{E_l} -C_h^2\frac{m_N^\star}{E_j}}
{1+C_h^2\sum_l n_l\frac{\Bk_l^2} {E_l^{3/2}}} \ee and
\be
\frac{\del\jb}{\del n_j}= \frac{\frac{\Bk_j}{E_j}}
{1+(C_v^2-\frac{\Sigma_0}{\rho})\sum_l n_l\frac{
\frac{2}{3}\Bk_l^2+m_N^{\star 2}}{E_l^{3/2}}}.\label{jeq} \ee
Writing in the standard way
\be
f_l=(2l+1)\int\frac{d\Omega}{4\pi}P_l(\frac{\Bk_i\cdot\Bk_j}{k_F^2})
f_{ij}(|\Bk_i|=|\Bk_j|=k_F), \ee we see that the last term in
(\ref{fij}) contributes to $f_1$ and the sum of the rest at the
Fermi surface (i.e. $|\Bk_j|=k_F$) to $f_0$. So
\be
F_0&\equiv&\frac{\lambda k_FE_F}{2\pi^2}f_0=\frac{3E_F}{k_F}\rho
f_0\nonumber\\ &=&\frac{3E_F}{k_F}\rho \big[
C_v^2+4C_v\rho\frac{\del C_v}{\del\rho}
+\frac{m_N^\star}{E_F}\frac{\del m_N^\star}{\del n_j} +\rho^2\{
(\frac{\del C_v}{\del\rho})^2+C_v\frac{\del^2C_v}{\del\rho^2}\}
\label{f0}\\ & &+(m_N^\star -M^\star )^2\left\{ (\frac{\del
\C_h}{\del\rho})^2 +\C_h\frac{\del^2\C_h}{\del\rho^2}\right\}
+2\C_h\frac{\del \C_h}{\del\rho}(m_N^\star -M^\star )
\frac{\del}{\del n_j}(m_N^\star -M^\star )\nonumber\\ &
&-2\C_h\frac{\del \C_h}{\del\rho}(m_N^\star -M^\star ) \frac{\del
M^\star}{\del\rho}-\C_h^2\frac{\del M^\star}{\del\rho}
\frac{\del}{\del n_j}(m_N^\star -M^\star ) -\C_h^2(m_N^\star
-M^\star )\frac{\del^2 M^\star}{\del\rho^2}\big]\nonumber \ee and
\be
F_1&\equiv&\frac{\lambda k_FE_F}{2\pi^2}f_1\nonumber\\
&=&-\frac{3(C_v^2-\frac{\Sigma_0}{\rho} )\rho}
{E_F+(C_v^2-\frac{\Sigma_0}{\rho} )\rho}. \label{f1} \ee
\subsubsection{Some relations for relativistic Fermi-liquid}

Here we bridge the model (\ref{toyBRPP}) to relativistic
Fermi-liquid theory. For it we will show that the thermodynamic
properties of any model like (\ref{toyBRPP}), which has
density-dependent parameters and is Walecka-type, can be described
in terms of relativistic Landau parameters derived from the mean
field approximation of the model in the same way as in Section
\ref{relation}.

First let us calculate the compression modulus K defined by
\be
K\equiv 9\rho\frac{\del^2\E (\jb=0)}{\del\rho^2}. \ee It comes out
to be
\be
K&=&\frac{3k_F^2}{E_F}+9\rho [C_v^2 +4C_v\rho\frac{\del
C_v}{\del\rho} +\frac{m_N^\star}{E_F}\frac{\del
m_N^\star}{\del\rho} +\rho^2\{(\frac{\del C_v}{\del\rho})^2
+C_v\frac{\del^2C_v}{\del\rho^2}\} \label{modul}\\ & &+(m_N^\star
-M^\star )^2\{(\frac{\del \C_h}{\del\rho})^2
+\C_h\frac{\del^2\C_h}{\del\rho^2}\} +2\C_h\frac{\del
\C_h}{\del\rho} (m_N^\star -M^\star
)\frac{\del}{\del\rho}(m_N^\star -M^\star ) \nonumber\\ &
&-2\C_h\frac{\del \C_h}{\del\rho}(m_N^\star -M^\star ) \frac{\del
M^\star}{\del\rho}-\C_h^2\frac{\del M^\star}{\del\rho}
\frac{\del}{\del\rho}(m_N^\star -M^\star ) -\C_h^2(m_N^\star
-M^\star )\frac{\del^2 M^\star}{\del\rho^2}] \nonumber\ee
Comparing (\ref{f0}) and ({\ref{modul}), we obtain Eq.
(\ref{commodul});
\be
K=\frac{3k_F^2}{E_F}(1+F_0).\label{K} \ee In our model
\be
k_F\left(\frac{\del
k_i}{\del\varepsilon_i}\right)_{k=k_F,\vb=0}=E_F\equiv
\sqrt{k_F^2+m_N^{\star 2}}.\label{energymass} \ee It is verified
that our model satisfies the relativistic Landau Fermi-liquid
formula for the compression modulus (\ref{commodul}).

As shown in Section \ref{relation}, the relativistic Landau liquid
satisfies the mass relation (\ref{mass});
\be
k_F\left(\frac{\del
k_i}{\del\varepsilon_i}\right)_{k=k_F,\vec{v}=0} =\mu
(1+F_1/3).\nn \ee One can see from Eqs.\ (\ref{mu}), (\ref{f1}),
and (\ref{energymass}) that (\ref{mass}) is satisfied exactly in
our model.

And lastly the relativistic relation for the first sound velocity
(\ref{firstsound}) is satisfied automatically since
(\ref{commodul}) and (\ref{mass}) are satisfied in our model. So
all the relativistic Landau Fermi-liquid relations in Section
\ref{relation} are satisfied with density-dependent Walecka-type
models like (\ref{toyBRPP}).
\subsubsection{Discussions}

The crucial question is really how to understand the scaling
masses and constants as one varies temperature and density as
considered in \cite{BR91}. If one takes the basic assumption that
the chiral Lagrangian in the mean field with BR scaling parameters
corresponds to Landau's Fermi-liquid fixed point theory, then one
should consider first fixing the Fermi momentum $k_F$ and let
renormalization group flow come to the fixed points of the
effective mass $m_L^\star$ for the nucleon and Landau parameters
$\Fc$~\cite{shankar}. In this case, the scaling quantities would
seem to be dependent upon $\Lambda/k_F$, not on the fields
entering into the effective Lagrangian. This paper however shows
that if one wants to approach the Fermi-liquid fixed point theory
starting from an effective chiral Lagrangian of QCD, it is
necessary to take into account the fact that the scaling arises
from the effect of multi-Fermi interactions figuring in chiral
Lagrangians as implied by chiral perturbation theory described in
\cite{tsp}. This is probably due to the fact that we are dealing
with two-stage ``decimations'' in the present problem -- with the
Fermi surface formed from a chiral Lagrangian as a nontopological
soliton (i.e., ``chiral liquid"~\cite{lynn}) -- in contrast to
condensed matter situations where one starts {\it ab initio} with
the Fermi surface without worrying about how the Fermi surface is
formed. Our result suggests that there will be a duality in
describing processes manifesting the scaling behavior. In other
words, the change of ``vacuum" by density exploited in \cite{BR91}
could equally be represented by a certain (possibly infinite) set
of interactions among hadrons -- e.g., four-Fermi and higher-Fermi
terms in chiral Lagrangians -- canonically taken into account in
many-body theories starting from the usual matter-free vacuum. A
notable evidence may be found in the two plausible explanations of
the low mass enhancement in CERES dilepton yields in terms of
scaling vector meson masses~\cite{LKB} and in terms of hadronic
interactions giving rise to increased widths~\cite{rcw97,wambach}.
Recently the relation between two descriptions are discussed by
Brown {\it et al}.\ \cite{BLRRW} and also by Kim {\it et al.}\
\cite{krbr99}. It is argued that the description based on the
reaction dynamics and on the meson spectral function should be
reliable at low density where the effective degrees of freedom are
hadrons and have no contradiction with the description on BR
scaling. There should however be a ``crossover" region at higher
density at which BR scaling will become more efficient or we
should go over to constituent quarks. How to relate the two
description in the ``crossover" density regime remains an open
problem.

In discussing the properties of dense matter, such as the BR
scaling of masses and coupling constants, e.g., $f_\pi^\star$ , we
have been using a Lagrangian which preserves Lorentz invariance.
This seems to be at odds with the fact that the medium breaks
Lorentz symmetry. One would expect for instance that the space and
time components of a current would be characterized by different
constants. Specifically such quantities as $g_A$, $f_\pi$ etc.
would be different if they were associated with the space
component or time component of the axial current. So a possible
question is: How is the medium-induced symmetry breaking
accommodated in the formalism which will be discussed in the next
section?

Section \ref{thermo} and this section provide the answer to this
question. Here the argument is given in an exact parallel to
Walecka mean field theory of nuclear matter. One writes an
effective Lagrangian with all symmetries of QCD which in the mean
field defines the parameters relevant to the state of matter with
density. The parameters that become constants (masses, coupling
constants etc.) at given density are actually functionals of
chiral invariant bilinears in the nucleon fields. When the scalar
field $\phi$ and the bilinear ${\psi}^\dagger\psi$, where $\psi$
is the nucleon field, develop a non-vanishing expectation value
Lorentz invariance is broken and the time and space components of
a nuclear current pick up different constants. This is how for
instance the Gamow-Teller constant $g_A$ measured in the space
component of the axial current is quenched in medium while the
axial charge measured in the axial charge transitions is enhanced
as described in the next chapter. If one were to calculate the
pion decay constant in medium, one would also find that the
quantity measured in the space component is different from the
time component. The way Lorentz-invariant Lagrangians figure in
nuclear physics is in some sense similar to what happens in
condensed matter physics.
\subsection{Mesons in medium}\label{mesons}

It should be recalled that we extracted the scaling parameter
$\Phi$ from the in-medium property of the vector mesons. Here we
will present evidences for the predicted scaling in the meson
masses. There are some preliminary experimental indications for
decrease in matter of the $\rho$ meson mass in recent nuclear
experiments~\cite{evidence1,evidence2} but we expect more
definitive results from future experiments at GSI and other
laboratories. In fact, this is currently a hot issue in connection
with the recent dilepton data coming from relativistic heavy ion
experiments at CERN(European Organization for Nuclear Research).

When heavy mesons such as the vector mesons $\rho$, $\omega$ and
the scalar $\sigma$ are reinstated in the chiral Lagrangian, then
the mass parameters of those particles in the Lagrangian, when
written in a chirally invariant way, are supposed to appear with
star and are assumed to scale according to Eq.\ (\ref{BRscaling}).
The question is: What is the physical role of these mass
parameters? If we assume that the mesons behave also like
quasiparticles, that is, like  weakly interacting particles with
the ``dropping masses," then physical observables will be
principally dictated by the tree diagrams of those particles
endowed with the scaling masses. In this case, the masses figuring
in the Lagrangian could be identified in some sense as
``effective" masses of the particles in the matter. This line of
reasoning was used in the work of Li, Ko and Brown~\cite{LKB} to
interpret the low mass enhancement of the CERES data~\cite{CERES}.
As discussed in Section \ref{model}, this treatment is consistent
with an effective Lagrangian which in the mean field approximation
yields the nuclear matter ground state as well as fluctuations
around the ground state. The parameters of the theory, as well as
their density dependence are determined by the properties of the
ground state. The work of this section shows that this scheme is
internally consistent. However we emphasize that the scaling we
have established is for the mesons that are highly off-shell and
it may not be applied to mesons that are near on-shell without
further corrections (e.g., widths etc.).

Suppose one probes the propagation of an $\omega$ meson in nuclear
medium as in HADES(High Acceptance Di-Electron Spectrometer) or
TJNAF(Thomas Jefferson National Accelerator Facility) experiments,
say through dilepton production. The $\omega$'s will decay
primarily outside of the nuclear medium, but let us suppose that
experimental conditions are chosen so that the leptons from the
$\omega$ decaying inside dense matter can be detected. See
\cite{frimansoyeur} for discussions on this issue. The question is
whether the dileptons will probe the BR-scaled mass or the
quantity given by (\ref{omegaeff}). The behavior of the $\omega$
mass would differ drastically in the two scenarios. A
straightforward application of FTS1 theory would suggest that at a
density $\rho\lsim \rho_0$, the $\omega$ mass as ``seen'' by the
dileptons will increase slightly instead of decrease. Since in
FTS1 theory, the vector coupling $g_v$ does not scale, this means
that $(g_v^\star/m_\omega^\star)$ will effectively decrease. On
the other hand if the vector coupling constant drops together with
the mass at increasing density as in the BR scaling model, the
situation could be quite different, particularly if dileptons are
produced at a density $\rho\sim 3\rho_0$ as in the CERES
experiments: The $\omega$ will then be expected to BR-scale up to
the phase transition. It has been recently suggested~\cite{kurt}
that at some high density, Lorentz symmetry can be spontaneously
broken giving rise to light $\omega$ mesons as ``almost Goldstone"
bosons when a small explicit Lorentz symmetry breaking term via
chemical potential is introduced. By introducing the term, they
assume a state which is chirally symmetric ($\la \bar{q}q\ra =0$)
but breaks Lorentz symmetry ($\la q^\dagger q\ra \neq 0$). The
assumed state is metastable at $\mu <\mu_c$ but becomes a global
minimum at $\mu >\mu_c$. At $\mu >\mu_c$, $\omega$ would become
light but not massless due to the explicit breaking. Such mesons
could be a source of copious dileptons at some density higher than
normal matter density. Thus measuring the $\omega$ mass shift
could be a key test of the BR scaling idea as opposed to the
FTS1-type interpretations. This interesting issue will be studied
in forthcoming experiments at GSI and TJNAF. It is interesting
that the dropping $\omega$ mass is also found in a recent QCD sum
rule calculation based on current correlation functions by Klingl,
Kaiser and Weise~\cite{KKW} who, however, do not see the dropping
of the $\rho$ mass because of the large broadening of $\rho$ peak.
If we can describe the $\omega$ meson in medium as a quasiparticle
an $\omega$-nuclear bound state is feasible even in light
nuclei~\cite{kkw98}. The process like
$^7$Li(d,$^3$He)$^6_\omega$He is expected to be seen in
GSI~\cite{hhg98} if such a bound state exists.
\section{Fermi-liquid theory vs.\ chiral Lagrangian}
\subsection{Electromagnetic current} \label{vector}

We will here give a brief derivation of the Landau-Migdal formula
for the convection current for a particle of momentum $\Bk$
sitting on top of the Fermi sea responding to a slowly varying
electromagnetic (EM) field. We will then analyze it in terms of
the specific degrees of freedom that contribute to the current.
This will be followed by a description in terms of a chiral
Lagrangian as discussed in \cite{FR96}. This procedure will
provide the link between the two approaches.
\subsubsection{Landau-Migdal formula for the convection current}\label{migdalan}

Following Landau's original reasoning adapted by Migdal to nuclear
systems, we start with the convection current given by
\be
\Jb=\sum_{\sigma,\tau}\int \frac{d^3 p}{(2\pi)^3}(\Bnabla_p\ve_p)
n_p\frac 12 (1+\tau_3)\label{landauJ} \ee where the sum goes over
the spin $\sigma$ and isospin $\tau$ which in spin- and
isospin-saturated systems may be written as a trace over the
$\sigma$ and $\tau$ operators. More precisely, this is a matrix
element of the current operator corresponding to the response to
the EM field of a nucleon (proton or neutron) sitting on top of
the Fermi sea. The sum over spin and isospin and the momentum
integral go over all occupied states up to the valence particle.
What we want is a current operator and it is deduced after the
calculation is completed. One can of course work directly with the
operator but the result is the same. We consider a variation of
the distribution function from that of an equilibrium state \bq
n_p = n_p^0 + \delta n_p, \eq where the superscript $0$ refers to
equilibrium. The variation of the distribution function induces a
variation of the quasiparticle energy \bq \ve_p = \ve_p^0
+\delta\ve_p. \eq In the equilibrium state the current is zero by
symmetry, so we have
\be
\Jb&=&\sum_{\sigma,\tau}\int \frac{d^3 p}{(2\pi)^3}
\left((\Bnabla_p\ve^0_p)\delta n_p+(\Bnabla_p\delta \ve_p) n^0_p
\right)\frac 12 (1+\tau_3),\nonumber\\ &=&\sum_{\sigma,\tau}\int
\frac{d^3 p}{(2\pi)^3} \left((\Bnabla_p\ve^0_p)\delta
n_p-(\Bnabla_p n^0_p) \delta \ve_p) \right)\frac 12 (1+\tau_3)
\label{J} \ee to linear order in the variation. We consider a
nucleon added at the Fermi surface of a system in its ground
state. Then
\be
\delta n_p=\frac{1}{V}\delta^3
(\pb-\bm{k})\frac{1\pm\tau_3}{2}\label{deltan} \ee and
\be
\Bnabla_p n^0_p=-\frac{\Bp}{k_F}\delta (p-k_F) \ee where $\bm{k}$
with $|\bm{k}|=k_F$ is the momentum of the quasiparticle. The
modification of the quasiparticle energies due to the additional
particle is given by
\be
\delta\ve_{p}=\sum_{\sigma^\prime,\tau^\prime} \int \frac{d^3
p^\prime}{(2\pi)^3} f_{p,p^\prime} \delta
n_{p^\prime\sigma^\prime\tau^\prime}\label{deltaepsilon}. \ee
\indent Combining  (\ref{qpint}), (\ref{J}), (\ref{deltan}) and
(\ref{deltaepsilon}) one finds that the first term of (\ref{J})
gives the operator
\be
\Jb^{(1)}=\frac{\bm{k}}{m_L^\star} \frac{1+\tau_3}{2},\label{qpJ}
\ee where $\bm{k}$ is taken to be at the Fermi surface. The second
term yields
\be
\delta \Jb=\delta \Jb_s +\delta \Jb_v=\frac{\bm{k}}{M}
\left(\frac{\tilde{F}_1 +\tilde{F}_1^\prime
\tau_3}{6}\right)\label{deltaJ}, \ee where
\be
\delta \Jb_s &=&
 \frac{\Bk}{m_L^\star}\frac 12
\frac{F_1}{3},\label{deltas}\\ \delta \Jb_v &=&
\frac{\bm{k}}{m_L^\star} \frac{\tau_3}{2}\frac{F^\prime_1}{3} =
\frac{\bm{k}}{m_L^\star} \frac{\tau_3}{2} \frac{F_1}{3}
+\frac{\bm{k}}{m_L^\star} \frac{\tau_3}{2}
\frac{F^\prime_1-F_1}{3}. \label{deltav} \ee For convenience,
let's define\footnote{The definition of $m_L^\star$ is in
(\ref{Landaumass})}
\be
\tilde{F}_l\equiv (M/m_L^\star)F_l\label{deftilde} \ee with
analogous definitions of $\tilde{F}_l^\prime$, etc.. It gives
another representation of (\ref{eff-mass})
\be
\frac{M}{m_L^\star}=1-\frac{\tilde{F}_1}{3}. \ee Putting
everything together we recover the well known result of
Migdal~\cite{migdal,BWBS} \bq \Jb=\frac{\bm{k}}{M}g_l =
\frac{\bm{k}}{M} \left(\frac{1+\tau_3}{2}+ \frac{1}{6}
(\tilde{F}^\prime_1-\tilde{F}_1)\tau_3\right),\label{Jtotal} \eq
where \bq g_l=\frac{1+\tau_3}{2}+\delta g_l\label{gl} \eq is the
orbital gyromagnetic ratio and \bq \delta g_l=\frac{1}{6}
(\tilde{F}^\prime_1-\tilde{F}_1)\tau_3.\label{deltagyro} \eq Thus,
the renormalization of $g_l$ is purely isovector. This is due to
Galilean invariance, which implies a cancellation in the isoscalar
channel.

We have derived Migdal's result using standard Fermi-liquid theory
arguments. This result can also be obtained~\cite{bentz} by using
the Ward identity, which follows from gauge invariance of the
electro-magnetic interaction. This is of course physically
equivalent to the above formulation. We shall now identify
specific hadronic contributions to the current (\ref{Jtotal}) in
two ways: the Fermi-liquid theory approach and the chiral
Lagrangian approach.
\subsubsection{Pionic contribution}

In Fermi-liquid theory approach, all we need to do is to compute
the Landau parameter $F_1$ from the pion exchange. The
one-pion-exchange contribution to the quasiparticle interaction is
\be
f^{\pi -exch.}_{\pb\sigma\tau,\pb^\prime\sigma^\prime\tau^\prime}
= \frac 13 \frac{f^2}{m_\pi^2}
\frac{\qb^2}{\qb^2+m_\pi^2}\left(S_{12} (\qbhat) +\frac 12
(3-\sigmab\cdot\sigmab^\prime)
\right)\frac{3-\taub\cdot\taub^\prime}{2}\label{vpi} \ee where
$\qb=\pb-\pb^\prime$ and $f=g_{\pi NN} (m_\pi/2M)\approx 1$. In a
relativistic formulation sketched in Appendix B, we can Fierz the
one-pion exchange. Done in this way, the Fierzed scalar channel is
canceled by a part of the vector channel and the remaining vector
channel makes a natural contribution to the pionic piece of $F_1$.
The one-pion-exchange contribution to the Landau parameter
relevant for the convection current is \bq \frac{F_1(\pi)}{3}=
-F_1^\prime(\pi) = -\frac{3f^2 m_L^\star}{8\pi^2 k_F}
I_1\label{F1pi} \eq where
\be
I_1=\int_{-1}^1 dx \frac{x}{1-x+\frac{m_\pi^2}{2k_F^2}}= -2 +
(1+\frac{m_\pi^2}{2k_F^2})\ln
(1+\frac{4k_F^2}{m_\pi^2}).\label{I1} \ee Note that $F_1(\pi )$
satisfies
\be
F_1(\pi )=-\frac{3m_L^\star}{k_F} \left. \frac{d\Sigma_\pi
(p)}{dp}\right|_{p=\kF} \label{piF1}\ee with one-pion-exchange
Fock contribution to the self-energy $\Sigma (p)$ and includes the
higher order contribution in $m_L^\star$. Thus, from Eq.\
(\ref{deltagyro}), the one-pion-exchange contribution to the
gyromagnetic ratio is \bq \delta
g_l^\pi=\frac{M}{\kF}\frac{f^2}{4\pi^2}I_1\tau_3. \label{gyro-pi}
\eq In the next subsubsection we include contributions also from
other degrees of freedom.

Let's obtain the convection current from a chiral Lagrangian and
compare it with the results given above. In absence of other meson
degrees of freedom, we can simply calculate Feynman diagrams given
by a chiral Lagrangian defined in matter-free space.
Nonperturbative effects due to the presence of heavy mesons
introduce a subtlety that will be treated below.
\begin{figure}
\centerline{\epsfig{file=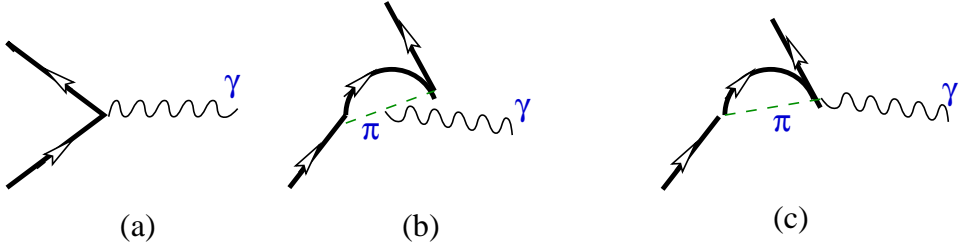,width=5in}} \caption{\small
\label{miyazawa}Feynman diagrams contributing to the EM convection
current in effective chiral Lagrangian field theory. Figure (a) is
the single-particle term and (b, c) the next-to-leading chiral
order pion-exchange current term. Figure (c) does not contribute
to the convection current; it renormalizes the spin gyromagnetic
ratio.}
\end{figure}

In the leading chiral order, there is the single-particle
contribution Fig.\ \ref{miyazawa}a which for a particle on the
Fermi surface with the momentum $\Bk$ is given by
\be
\Jb_{1-body}=\frac{\bm{k}}{M} \frac{1+\tau_3}{2}.\label{J1} \ee
Note that the nucleon mass appearing in (\ref{J1}) is the
free-space mass $M$ as it appears in the Lagrangian, not the
effective mass $m_L^\star$ that enters in the Fermi-liquid
approach, (\ref{qpJ}). To the next-to-leading order, we have two
soft-pion terms as discussed in \cite{tsp,KDR,chemrho,pmr}. We
should recall a well-known caveat here discussed already in
\cite{chemrho}. If one were to blindly calculate the convection
current coming from Fig.\ \ref{miyazawa}b, there would be a gauge
non-invariant term that is present because the hole line is
off-shell. Figure \ref{miyazawa}c contains also a gauge
non-invariant term which is exactly the same as in Fig.\
\ref{miyazawa}b but with an opposite sign, so in the sum of the
two graphs, the two cancel exactly so that only the
gauge-invariant term survives. Of course we now know that the
off-shell dependence is not physical and could be removed by field
redefinition {\it ab initio}. To the convection current we need,
only Fig.\ \ref{miyazawa}b contributes
\be
\Jb_{2-body}=\frac{\bm{k}}{\kF}\frac{f^2}{4\pi^2} I_1\tau_3 =
\frac{\bm{k}}{M}\frac{1}{6}
(\tilde{F}^\prime_1(\pi)-\tilde{F}_1(\pi))\tau_3. \label{J2} \ee
We should emphasize that the Landau parameters $\tilde{F}_1$ and
$\tilde{F}_1^\prime$ are entirely fixed by a chiral effective
Lagrangian for any density.

The sum of (\ref{J1}) and (\ref{J2}) agrees precisely with the
Fermi-liquid theory result (\ref{Jtotal}) and (\ref{F1pi}). This
formula first derived in \cite{br87} in connection with the
Landau-Migdal parameter is of course the same as the Miyazawa
formula \cite{miyazawa} derived nearly half a century ago. Note
the remarkable simplicity in the derivation starting from a chiral
Lagrangian. However, we should caution that there are some
non-trivial assumptions to go with the validity of the formula. As
we will see shortly, we will not have this luxury of simplicity
when other degrees of freedom enter.
\subsubsection{Vector meson contributions and BR scaling}

So far we have computed only the pion contribution to $g_l$. In
nuclear physics, more massive degrees of freedom such as the
vector mesons $\rho$ and $\omega$ of mass $700\sim 800$ MeV and
the scalar meson $\sigma$ of mass $600\sim 700$ MeV play an
important role. When integrated out from the chiral Lagrangian,
they give rise to effective four-Fermion interactions:
\be\label{four-Fermion} {\L}_4=\frac{C_\phi^2}{2}(\bar{N}N)^2
-\frac{C_\omega^2}{2} (\bar{N}\gamma_\mu N)^2
-\frac{C_\rho^2}{2}(\bar{N}\gamma_\mu\tau N)^2 +\cdots \ee where
the coefficients $C's$ can be identified with \be\label{couplings}
C_M^2=\frac{g_M^2}{{m_M}^2} \ \ \ {\rm with}\ \ \ M=\phi,\ \rho,\
\ \omega. \ee For the moment, we make no distinction as to whether
one is taking into account BR scaling or not. For the Fermi-liquid
approach, this is not relevant since the parameters are not
calculated. However with chiral Lagrangians, we will specify the
scaling which is essential. Such interaction terms are
``irrelevant" in the renormalization group flow sense but can make
crucial contributions by becoming ``marginal" in some particular
kinematic situation. A detailed discussion of this point can be
found in \cite{shankar}. The effective four-Fermion interactions
play a key role in stabilizing the Fermi-liquid state and leads to
the fixed points for the Landau parameters. (The other fixed-point
quantity, i.e.\ the effective mass, is put in by fiat to keep the
density fixed.) In the two-nucleon systems studied in \cite{PKMR},
they enter into the next-to-leading order term of the potential,
which is crucial in providing the cut-off independence found for
cut-off masses $\gsim m_\pi$.

Again it suffices to compute the Landau parameters coming from the
velocity-dependent part of heavy meson exchanges in the
Fermi-liquid theory approach. We treat the effective four-Fermion
interaction (\ref{four-Fermion}) in the Hartree approximation.
Then the only velocity-dependent contributions are due to the
current couplings mediated by $\omega$ and $\rho$ exchanges. The
corresponding contributions to the Landau parameters are
\bq\label{f1-omega} F_1(\omega)=
-C_\omega^2\frac{2\kF^3}{\pi^2M} \eq and \bq\label{f1p-rho}
F_1^\prime(\rho)=
-C_\rho^2\frac{2\kF^3}{\pi^2M}. \eq
The derivations of relativistic $F_1(\omega )$ and $F_1^\prime
(\rho)$ are shown in Appendix C.

Now the calculation of the convection current and the nucleon
effective mass with the interaction (\ref{four-Fermion}) in the
Landau method goes through the same way as in the case of the
pion. The net result is just Eq.\ (\ref{Jtotal}) including the
contribution of the contact interactions
(\ref{f1-omega},\ref{f1p-rho}), i.e.,
\be
\tilde{F}_1&=&\tilde{F}_1(\pi) +\tilde{F}_1(\omega),\\
\tilde{F}^\prime_1&=&\tilde{F}^\prime_1(\pi)
+\tilde{F}^\prime_1(\rho). \ee Similarly, the nucleon effective
mass is determined by (\ref{eff-mass}) with
\be
F_1=F_1 (\pi) +F_1 (\omega). \ee

In chiral Lagrangina approach the most efficient way to bring in
the vector mesons into the chiral Lagrangian is to implement BR
scaling in the parameters of the Lagrangian. We shall take the
masses of the relevant degrees of freedom to scale in the manner
of BR as (\ref{BRscaled}). Note again that $M^\star$ is a BR
scaling nucleon mass which will turn out to be different from the
Landau effective mass $m_L^\star$~\cite{FR96}. For our purpose, it
is more convenient to integrate out the vector and scalar fields
and employ the resulting four-Fermi interactions
(\ref{four-Fermion}). The coupling coefficients are modified
compared to Eq.\ (\ref{couplings}), because the meson masses are
replaced by effective ones:
\be
C_M^2=\frac{g_M^2}{{m_M^\star}^2} \ \ \ {\rm with}\ \ \ M=\phi,\
\rho,\ \ \omega. \ee The coupling constants may also
scale~\cite{sbmr} but we omit their density dependence for the
moment.

The first thing we need is the relation between the BR scaling
factor $\Phi$ which was proposed in \cite{BR91} to reflect the
quark condensate in the presence of matter and the contribution to
the Landau parameter $F_1$ from the isoscalar vector ($\omega$)
meson. For this we first calculate the Landau effective mass
$m_N^\star$ in the presence of the pion and $\omega$
fields~\cite{FR96}
\be
\frac{m_L^\star}{M}=1+\frac 13 (F_1 (\omega) +F_1
(\pi))=\left(1-\frac 13 (\tilde{F}_1 (\omega) +\tilde{F}_1
(\pi))\right)^{-1}.\label{mstar1} \ee Next we compute the nucleon
self-energy using the chiral Lagrangian. Given the single
quasiparticle energy
\be
\ve_p =\frac{p^2}{M^\star}+C_\omega^2\bra N^\dagger N\ket
+\Sigma_\pi (p), \ee we get the effective mass as in \cite{FR96}
\be
\frac{m_L^\star}{M}=\frac{k_F}{m_N} \left( \left.
\frac{d}{dp}\ve_p\right|_{p=\kF}\right)^{-1} =\left(\Phi^{-1}
-\frac 13 \tilde{F}_1 (\pi)\right)^{-1}\label{mstar2} \ee from
Eqs.\ (\ref{BRscaled}), (\ref{deftilde}), and (\ref{piF1}).
Comparing (\ref{mstar1}) and (\ref{mstar2}), we obtain the
important result
\be
\tilde{F}_1 (\omega)=3(1-\Phi^{-1}).\label{mainrelation} \ee This
is an intriguing relation. It shows that the BR factor, which was
originally proposed as a precursor manifestation of the chiral
phase transition characterized by the vanishing of the quark
condensate at the critical point~\cite{BR91}, is intimately
related (at least up to $\rho\approx \rho_0$) to the Landau
parameter $F_1$, which describes the quasiparticle interaction in
a particular channel.  We believe that the BR factor can be
computed by QCD sum rule methods or obtained from current algebra
relations such as the GMOR relation evaluated in medium. As was
shown in \cite{FR96}, Eq.\ (\ref{mainrelation}) implies that the
BR factor governs in some, perhaps, intricate way low-energy
nuclear dynamics. The equivalence discussed above between the
physics of the vacuum property $\Phi$ and that of the
quasiparticle interaction $F_1$ due to the massive vector-meson
degree of freedom suggests that the ``bottom-up" approach -- going
up in density with a Lagrangian whose parameters are fixed at zero
density -- and the ``top-down" approach -- extrapolating with a
Lagrangian whose parameters are fixed at some high density -- can
be made equivalent at some intermediate point. If this is so in
the hot and dense regime probed by relativistic heavy ion
collisions, then the CERES data should also be understandable in
terms of hadronic interactions without making reference to QCD
variables. Because of the complexity of hadronic descriptions, it
will be difficult to relate the two directly but the recent
alternative explanation of the CERES data in terms of ``melting of
the vector mesons" inside nuclear matter manifested in the
increased width of the mesons due to hadronic
interactions~\cite{wambach} may be an indication for a possible
``dual" description at low density between what is given in QCD
variables (e.g., quark condensates) and what is given in hadronic
variables (e.g., the Landau parameter), somewhat reminiscent of
the quark-hadron duality in heavy-light-quark systems
\cite{qhduality}. A possible mechanism that could make the link
between the two descriptions was suggested recently by Brown {\it
et al}.\ \cite{BLRRW} and by Kim {\it et al}.\ \cite{krbr99}.

In the presence of the BR scaling, a non-interacting nucleon in
the chiral Lagrangian propagates with a mass $M^\star$, not the
free-space mass $M$. Thus, the single-particle current Fig.\
\ref{miyazawa}a is {\it not} given by (\ref{J1}) but instead by
\be
\Jb_{1-body}=\frac{\bm{k}}{M^\star}
\frac{1+\tau_3}{2}.\label{onebody} \ee Now the current
(\ref{onebody}) on its own does not carry conserved charge as long
as $M^\star\neq M$. This means that two-body currents are
indispensable to restore charge conservation. Note that the
situation is quite different from the case of Fermi-liquid theory.
In the latter case, the quasiparticle propagates with the Landau
effective mass $m_L^\star$ and it is the gauge invariance that
restores $m_L^\star$ to $M$. In condensed matter physics, this is
related to a phenomenon that the cyclotron frequency depends on
the bare mass, not on Landau effective mass. It may be referred to
as  Kohn effect~\cite{kohn}. The bare mass in Kohn effect is
restored due to the quasiparticle interactions with Galilean
invariance in the same way as for the convection current in
Section \ref{migdalan}~\cite{hkm98}. This clearly indicates that
gauge invariance is more intricate when BR scaling is implemented.
Indeed if the notion of BR scaling and the associated chiral
Lagrangian are to make sense, we have to recover the charge
conservation from higher-order terms in the chiral Lagrangian.
This constitutes a strong constraint on the theory.

Let us now calculate the contributions from the pion and heavy
meson degrees of freedom. The pion contributes in the same way as
before, so we can carry over the previous result of Fig.\
\ref{miyazawa}b,
\be
\Jvec_{2-body}^{\pi}=\frac{\bm{k}}{M}\frac 16 (\tilde{F}_1^\prime
(\pi) -\tilde{F}_1 (\pi))\tau_3.\label{pi} \ee This is of the same
form as (\ref{J2}) obtained in the absence of BR scaling. It is in
fact identical to (\ref{J2}) if we assume that one-pion-exchange
graph does not scale in medium at least up to nuclear matter
density. This assumption is supported by observations in
pion-induced processes in heavy nuclei. This means that the
observation that the one-pion-exchange potential does not scale
implies that the constant $g_A^\star/f_\pi^\star$ remains
unscaling at least up to normal nuclear matter density with
non-scaling pion mass. In what follows, we will make this
assumption implicitly.

The contributions from the vector meson degrees of freedom are a
bit trickier. They are given by Fig.\ \ref{nbar}.
\begin{figure}
\centerline{\epsfig{file=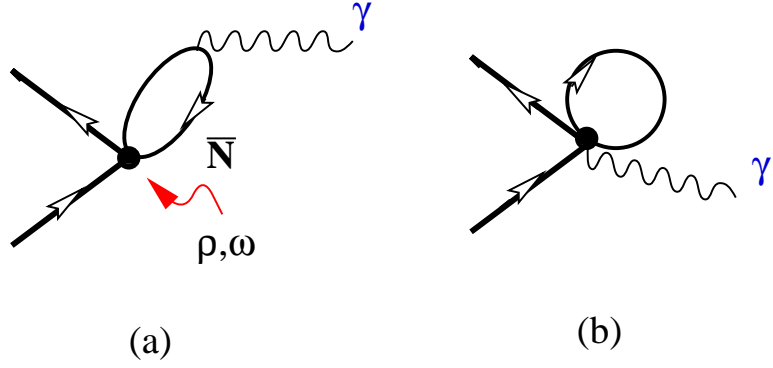,width=4in}} \caption{\small (a)
Feynman diagram contributing to the EM
 convection current from four-Fermi interactions corresponding to the
$\omega$ and $\rho$ channel (contact interaction indicated by the
blob) in effective chiral Lagrangian field theory. Th $\bar{N}$
denotes the anti-nucleon state that is given in the chiral
Lagrangian as a $1/M$ correction and the one without arrow is a
Pauli-blocked or occupied state. (b) The equivalent graph in
heavy-fermion formalism with the anti-nucleon line shrunk to a
point.}\label{nbar}
\end{figure}
Both the $\omega$ (isoscalar) and $\rho$ (isovector) channels
contribute through the antiparticle intermediate state as shown in
Fig.\ \ref{nbar}a. The antiparticle is explicitly indicated in the
figure. However in the heavy-fermion formalism, the backward-going
anti-nucleon line should be shrunk to a point as Fig.\ \ref{nbar}b,
leaving behind an explicit $1/M$ dependence folded with a factor
of nuclear density signaling the $1/M$ correction in the chiral
expansion. One can interpret Fig.\ \ref{nbar}a as saturating the
corresponding counter term although this has to be yet verified by
writing the full set of counter terms at the same order.  These
terms
with Fig.\ \ref{nbar}a
\be
\Jb_{2-body}^{\omega}&=&\frac{\Bk}{M}\frac 16 \tilde{F}_1
(\omega), \label{omega}\\ \Jb_{2-body}^{\rho}&=&\frac{\Bk}{M}\frac
16 \tilde{F}_1^\prime (\rho)\tau_3,\label{rho} \ee where
$\tilde{F}_1 (\omega)$ and $\tilde{F}_1^\prime(\rho)$ are given by
Eqs.\ (\ref{f1-omega},\ref{f1p-rho}) with $M$ replaced by
$M^\star$. Their relativistic forms are given in Appendix C. The
total current given by the sum of (\ref{onebody}), (\ref{pi}),
(\ref{omega}) and (\ref{rho}) precisely agrees with the
Fermi-liquid theory result (\ref{Jtotal}) when we take
\be
\tilde{F}_1&=&\tilde{F}_1 (\omega)+\tilde{F}_1 (\pi),\\
\tilde{F}_1^\prime &=&\tilde{F}_1^\prime (\rho)+\tilde{F}_1^\prime
(\pi). \ee \indent The way in which this precise agreement comes
about is nontrivial. What happens is that part of the $\omega$
channel restores the BR-scaled mass $M^\star$ back to the
free-space mass $M$ in the isoscalar current. (It has been known
since sometime that something similar happens in the standard
Walecka model (without pions and BR scaling) \cite{kurasawa}).
Thus, the leading single-particle operator combines with the
sub-leading four-Fermi interaction to restore the charge
conservation as required by the Ward identity. This is essentially
the ``back-flow mechanism" which is an important ingredient in
Fermi-liquid theory. We describe below the standard back-flow
mechanism as given in Sec.\ \ref{concept}, adapted to nuclear
systems with isospin degrees of freedom, and elucidate the
connection to the results obtained with the chiral Lagrangian in
this subsection.

The current so constructed is valid for a process occurring very
near the Fermi surface corresponding to the limit
$(\omega,\Bq)\rightarrow (0,\bf {0})$ where $q$ is the spatial
momentum transfer and $\omega$ is the energy transfer. In the
diagrams considered so far (Fig.\ \ref{miyazawa} and \ref{nbar})
the order of the limiting processes does not matter. However, the
particle-hole contribution, which we illustrate in Fig.\ \ref{pth}
with the pion contribution, does depend on the order in which
$q=|\Bq|$ and $\omega$ approach zero. Thus, in the limit
$q/\omega\rightarrow 0$, the particle-hole contributions vanish
whereas in the opposite case $\omega/q\rightarrow 0$, they do not.
\begin{figure}
\centerline{\epsfig{file=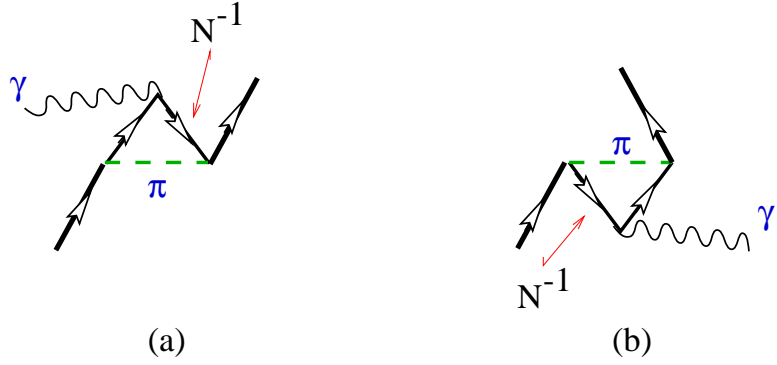,width=4in}} \caption{\small
Particle-hole contributions to the convection current. Here
backward-going nucleon line $N^{-1}$ denotes a hole. These graphs
vanish in the $q/\omega\rightarrow 0$ limit.}\label{pth}
\end{figure}
This can be seen by examining the particle-hole propagator
\be
\frac{n_k (1-n_{k+q})}{q_0 +\epsilon_k-\epsilon_{k+q}+i\delta} -
\frac{n_{k+q} (1-n_k)}{q_0+\epsilon_k-\epsilon_{k+q}-i\delta} \ee
where $(q_0, \Bq )$ is the four-momentum of the external (EM)
field. This vanishes if we set $q\rightarrow 0$ with $q_0$
non-zero but its real part is non-zero if we interchange the
limiting process since for $q_0=0$ we have
\be
\frac{\Bq\cdot\kbhat }{-\Bq\cdot\Bk /M}\delta (k_F-k). \ee Figure
\ref{pth} was computed by several authors (e.g., see \cite{bentz})
and is given in the limit $\omega /q\rightarrow 0$ by
\be
\Jb_{ph}=-\sum_{\tau^\prime} \la \taub
(1)\cdot\frac{1+\tau_3^\prime}{2}\taub (2)\ra
\int\frac{d^3p}{(2\pi )^3}\hat{\pb}\delta (k_F-|\pb |)f^\pi_s \ee
where $f_s^\pi $ is the spatial and spin part of the quasiparticle
interaction which is (\ref{vpi}) and (\ref{rfpi}) without isospin
part $(3-\taub\cdot\taub^\prime )/2$. The isospin factor is given
by the Fierz transformation:
\be
\sum_{\tau^\prime} \la\taub (1)\cdot\frac{1+\tau_3^\prime}{2}\taub
(2)\ra &=&\sum_{\tau^\prime} \la\frac34
-\frac14\taub\cdot\taub^\prime +\frac34 \tr [\tau_3^\prime
]-\frac14\taub\cdot\tr [\tau_3^\prime \taub^\prime ]\ra\nonumber\\
&=&\frac32-\frac12\tau_3.\label{isoint} \ee Note that the factor
$\frac32$ comes from $f_\pi$ and $\frac12\tau_3$ from
$f^\prime_\pi$. In the limit that we are concerned with (i.e.\
$T=0$ and $\omega /q\rightarrow 0$), the particle-hole
contribution to the current is
\be
\Jb_{ph} &=&-\frac{1}{3\pi^2}\kbhat k_F^2(f_1+f_1^\prime
\tau_3)\nonumber\\ &=&-\frac{\Bk}{M}\left( \frac{\tilde{F}_1 (\pi)
+\tilde{F}_1^\prime (\pi )\tau_3}{6}\right).\label{parthole} \ee
This holds in general regardless of what is being exchanged as
long as the exchanged particle has the right quantum numbers.
Contributions from heavy-meson exchanges are calculated in a
similar way. Adding the particle-hole contribution
(\ref{parthole}) to the Fermi-liquid result (\ref{Jtotal}) we
obtain the current of a dressed or localized quasiparticle
\be
\Jb_{locQP}=\frac{\Bk}{m_L^\star} \left(\frac{1+\tau_3}{2}\right).
\label{locQP} \ee Note that $\Jb_{ph}$ precisely cancels
$\delta\Jb$, Eq.\ (\ref{deltaJ}). The current $\Jb_{locQP}$ is the
total current carried by the wave packet of a localized
quasiparticle with group velocity ${\bf
{v}}_F=\frac{\Bk}{m_L^\star}$. However, the physical situation
corresponds to homogeneous (plane wave) quasiparticle excitations.
The current carried by a localized quasiparticle equals that of a
homogeneous quasiparticle excitation modified by the so called
back-flow current~\cite{pines}. The back-flow contribution
$(\Jb_{locQP}-\Jb_{LM})$ is just the particle-hole polarization
current in the $\omega/q\rightarrow 0$ limit, Eq.\
(\ref{parthole}).
\subsubsection{Phenomenological test}
It is not obvious that the effective nucleon mass computed in the
chiral Lagrangian approach is directly connected to a measurable
quantity although quasielastic electron scattering from nuclei
does probe some kind of effective nucleon mass and Walecka model
describes such a process in terms of an effective mass. To the
extent that the bulk of $m_N^\star$ is related to the condensate
through BR scaling as we can see in (\ref{mstar2}), the effective
mass in the chiral Lagrangian can be related to the quantity
calculated in the QCD sum rule approach for in-medium hadron
masses. In BR scaling, the parameter $\Phi$ is related to the
scaling of the vector meson ($\rho$) mass. There are several QCD
sum rule calculations for the $\rho$ meson in-medium mass starting
with \cite{SHLee}. The most recent one which closely agrees with
the GMOR formula in medium for the pion decay constant
$f_\pi^\star$ (see Eq.\ (\ref{BRscaling})) is the one by Jin and
Leinweber~\cite{jin}:
\be
\Phi(\rho_0)=0.78\pm 0.08.\label{Jinvalue} \ee We shall take this
value in what follows but one should be aware of the possibility
that this value is not quite firm. A caveat to this result was
recently discussed by Klingl, Kaiser, and Weise~\cite{KKW} who
show that the QCD sum rule can be saturated without the mass
shifting downward by increasing the vector meson width in medium.
For a discussion of the empirical constraints on the in-medium
widths of vector mesons, see Friman~\cite{friman97}.

Given this, we can compute $m_N^\star$ using (\ref{mstar2}) for
nuclear matter density since the pionic contribution $\tilde{F}_1
(\pi)$ is known. One finds~\cite{FR96}
\be
\frac{m_N^\star}{M}(\rho=\rho_0)\approx 0.70.\label{m*pred} \ee
This can be tested in an indirect way by looking at certain
magnetic response functions of nuclei as described below. An
additional evidence comes from QCD sum rule calculations. Again
there are caveats in the QCD sum rule calculation for the nucleon
mass even in free-space and certainly more so in medium.
Nevertheless the most recent result by Furnstahl, Jin, and
Leinweber~\cite{furnstahljin} is rather close to the prediction
(\ref{m*pred}):
\be
\left(\frac{m_N^\star
(\rho_0)}{M}\right)_{QCD}=0.69^{+0.14}_{-0.07}. \ee

If one writes the gyromagnetic ratio $g_l$ as
\be
g_l=\frac{1+\tau_3}{2} +\delta g_l \ee then the chiral Lagrangian
prediction is
\be
\delta g_l=\frac 16 (\tilde{F}_1^\prime -\tilde{F}_1)\tau_3 =\frac
49 \left[\Phi^{-1} -1- \frac 12 \tilde{F}_1 (\pi)\right]\tau_3.
\label{chptd} \ee In writing the second equality we have used
(\ref{F1pi}), (\ref{mainrelation}) and the nonet relation
$\tilde{F}^\prime (\rho)=\tilde{F} (\omega)/9$. At nuclear matter
density, we get, using (\ref{Jinvalue}),
\be
\delta g_l (\rho_0)\approx 0.23\tau_3. \ee This agrees with the
value extracted from the dipole sum rule in
$^{209}$Bi~\cite{schumacher},
\be
\delta g_l^{proton}=0.23\pm 0.03\label{experiment} \ee and agrees
roughly with magnetic moment data in heavy nuclei. Nuclear
magnetic moments are complicated due to conventional nuclear
effects. To make a meaningful comparison, one would have to
extract all ``trivial" nuclear effects and this operation brings
in inestimable uncertainties. It should be stressed that the
gyromagnetic ratio provides a test for the scaling nucleon mass at
$\rho\approx \rho_0$. It also gives a check of the relation
between the baryon property on the left-hand side of Eq.\
(\ref{mainrelation}) and the meson property on the right-hand
side. Instead of using (\ref{Jinvalue}) as an input to calculate
$\delta g_l$, we could take the experimental value
(\ref{experiment}) to determine, using (\ref{chptd}), the BR
scaling factor $\Phi$ at $\rho\approx \rho_0$. We would of course
get (\ref{Jinvalue}), a value which is consistent with what one
obtains in the QCD sum rule calculation and also in the in-medium
GMOR relation.

Recall that because of the pions which provide (perturbative)
non-local interactions to the Landau interaction, the Landau mass
for the nucleon scales differently from that of the vector mesons.
(See (\ref{BRscaling}) and (\ref{mstar2})). This difference is
manifested in the skyrmion description by the fact that the
coefficient of the Skyrme quartic term must also scale. In the
original discussion of the scaling based on the quark condensates
using the trace anomaly~\cite{BR91}, the Skyrme quartic term was
scale-invariant and hence the corresponding $g_A^\star$ was
non-scaling. So the scaling implied by (\ref{mstar2}) indicates
that the scaling of $g_A^\star$ is associated with the pionic
degrees of freedom. This is consistent with the description based
on the Landau-Migdal $g_0^\prime$ interaction between a nucleon
and a $\Delta$ resonance~\cite{rho,ohtawakamatsu,PJM} and also
with the QCD sum rule description of Drukarev and
Levin~\cite{drukarev} who attribute about 80\% of the quenching to
the $\Delta-N$ effect.

If we equate the skyrmion relation~\cite{FR96,mr85}
\be
\frac{m_N^\star}{M}=\sqrt{\frac{g_A^\star}{g_A}}\Phi \ee to
(\ref{mstar2}), we get
\be
\frac{g_A^\star}{g_A}=\left(1+\frac 13 F_1 (\pi)\right)^2
=\left(1-\frac 13 \tilde{F}_1
(\pi)\Phi\right)^{-2}.\label{quenching} \ee At nuclear matter
density, this predicts
\be
g_A^\star (\rho_0)\approx 1 \ee and
\be
\frac{g_A^\star}{g_A}\approx \frac{f_\pi^\star}{f_\pi}=\Phi. \ee
We will use this relation in deriving (\ref{CAAC}). It should be
emphasized that this relation, being unrelated to the vacuum
property, cannot hold beyond $\rho\approx \rho_0$. Indeed as
suggested by the scaling given in \cite{BR91}, $g_A^\star
(\rho)\approx \alpha g_A$ with $\alpha$ a constant independent of
density, for $\rho\gsim \rho_0$. It would be a good approximation
to set $g_A^\star$ equal to 1 for $\rho\gsim\rho_0$.

Since one expects that when chiral symmetry is restored, $g_A$
will approach 1, it may be thought that the evidence for
$g_A^\star\approx 1$ in nuclei is directly connected with chiral
restoration. This is not really the case. Neither in the skyrmion
picture nor in QCD sum rules is the quenching of $g_A$ simply
related to a precursor behavior of chiral restoration. This does
not however mean that the quenching of $g_A$ carries no
information on the chiral symmetry restoration. As suggested
recently by Chanfray, Delorme and Ericson~\cite{CDE}, if one were
to compute {\it all} pion-exchange-current graphs at one-loop
order that contribute to the in-medium $g_A$, the effect of
medium-induced change in the quark condesate would be largely
accounted for. In a way, this argument is akin to that for the
Cheshire-Cat (or dual) phenomenon we are advocating in the
description of the quark condensate in terms of quasiparticles.
Another issue that has generated lots of debate in the past and
yet remains confusing is the interpretation of an effective
constant $g_A^{eff}\approx 1$ actually observed in medium and
heavy nuclei. The debate has been whether the observed
``quenching" is due to ``core polarizations" or ``$\Delta$-hole
effect" (or other non-standard mechanisms). Our view is that in
the presence of BR scaling, both are involved. In light nuclei in
which the Gamow-Teller transition takes place in low density, the
tensor force is mainly operative and the core polarization (i.e.,
multiparticle-multihole configurations) mediated by this tensor
force is expected to do  most of the quenching, while the
$\Delta$-hole effect directly proportional to density is largely
suppressed. The typical diagrammatic representations for the
second order core polarization is shown in Fig.\ \ref{coreph}.
\begin{figure}
\centerline{\epsfig{file=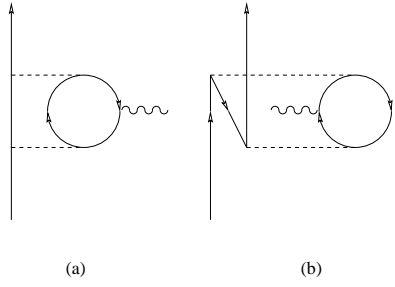,width=2in}}
\caption{\label{coreph} Examples of the second order core
polarization contribution to Gamow-Teller transition. Downward
nucleon line denote a hole. (a)2p1h (b)3p2h}
\end{figure}
In heavy nuclei, on the other hand, the tensor force is quenched
due to BR scaling, rendering the core polarization mechanism
ineffective while the increased density makes the $\Delta$-hole
effect dominant. Recently Park, Jung, and Min~\cite{PJM} applied
chiral perturbation theory to calculate $g_A^\star$ at normal
nuclear matter density $\rho_0$.
\begin{figure}
\centerline{\epsfig{file=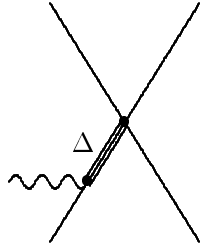,width=1in}} \caption{The
resonance-exchange graph in \cite{PJM} for four-Fermi contact
interaction contribution. Its Hartree contribution decreases axial
vector coupling constant in medium.}\label{SRC}
\end{figure}
\begin{figure}
\centerline{\epsfig{file=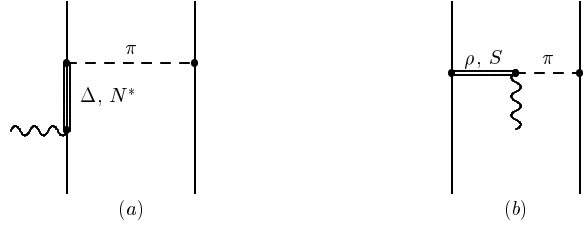,width=3in}} \caption{The
resonance-exchange graphs in \cite{PJM} for one-pion-exchange
contribution. Their Fock contributions enhance the axial vector
coupling constant with incorporating the short-range correlation
between nucleons. }\label{GTO}
\end{figure}
The resonance-exchange graphs that contribute are shown in Figs.\
\ref{SRC} and \ref{GTO} and the Landau-Migdal $g_0^\prime$ effect
contains both. The Hartree contribution from Fig.\ \ref{SRC},
i.e.\ $\Delta$-hole effect, makes $g_A^\star$ quenched and the
Fock contribution from Fig.\ \ref{GTO} enhances $g_A^\star$. The
magnitude of quenching is two or three times larger than the
enhancement. What is seen in nature, in our view, is the interplay
between these two.

The second form of (\ref{quenching}) shows that the quenching of
$g_A$ in matter is quite complex, both the pionic effect and the
vacuum condensate effect being confounded together. Again for the
reason given above, this relation cannot be extended beyond the
regime with $\rho\approx\rho_0$. We have no understanding of how
this formula and the $\Delta$-hole mechanism of
\cite{rho,ohtawakamatsu} are related. Our effort thus far has met
with no success. Understanding the connection would presumably
require the short-distance physics implied by both the
Landau-Migdal $g_0^\prime$ interaction and the Skyrme quartic term
(which is known to be more than just what results when the $\rho$
meson is integrated out of the chiral Lagrangian).
\subsection{Axial charge transition}\label{axi}

No one has yet derived the analogue to (\ref{Jtotal}) for the
axial current. Attempts using axial Ward identities in analogy to
the EM case have not met with success~\cite{bentzaxial}. The
difficulty has presumably to do with the role of the Goldstone
bosons in nuclear matter which is not well understood. In this
subsection, we analyze the expression for the axial charge
operator obtained by a straightforward application of the
Fermi-liquid theory arguments of Landau and Migdal and compare
this expression with that obtained directly from the chiral
Lagrangian using current algebra. For the vector current we found
precise agreement between the two approaches.
\subsubsection{Applying Landau quasiparticle argument}

The obvious thing to do is to simply mimic the steps used for the
vector current to deduce a Landau-Migdal expression for the axial
charge operator. We use both methods developed above and find that
they give the same result.

In free space, the axial charge operator nonrelativistically is
$\sim \Bsigma\cdot{\bf {v}}$ where ${\bf {v}}=\Bk/M$ is the
velocity. In the infinite momentum frame, it is the relativistic
invariant helicity ${\Bsigma}\cdot \nubhat$. It is thus tempting
to assume that near the Fermi surface, the axial charge operator
for a local quasiparticle in a wave packet moving with the group
velocity ${\bf v}_F=\Bk/m_L^\star$ is simply $\sim
{\Bsigma}\cdot{\bf v}_F$. This suggests that we take the axial
charge operator for a {\em localized} quasiparticle to have the
form
\be
 {\A0}^i_{locQP}= g_A\frac{\Bsigma\cdot\Bk}{m_L^\star}
\frac{\tau^i}{2}. \label{totalcharge} \ee As in the vector current
case, we take (\ref{totalcharge}) to be the $\omega/q\rightarrow
0$ limit of the axial charge operator. The next step is to compute
the particle-hole contribution to Fig.\ \ref{pth} (with the vector
current replaced by the axial current) in the $\omega/q\rightarrow
0$ limit. A simple calculation gives
\be
{\A0}^i_{ph}=-g_A\frac{\Bsigma\cdot\Bk}{m_L^\star}\frac{\tau^i}{2}
\Delta^\prime\label{phA0} \ee with
\be
\Delta^\prime=\frac{f^2k_F m_L^\star}{4m_\pi^2 \pi^2} (I_0-I_1)
\ee where $I_1$ was defined in (\ref{I1}) and
\be
I_0=\int_{-1}^1 dx\frac{1}{1-x+\frac{m_\pi^2}{2k_F^2}}=
\ln\left(1+\frac{4k_F^2}{m_\pi^2}\right). \ee In an exact parallel
to the procedure used for the vector current, we take the
difference
\be
{\A0}^i_{locQP}-{\A0}^i_{ph} \ee and identify it with the
corresponding ``Landau axial charge''(LAC):
\be
{\A0}^i_{LAC}=
g_A\frac{\Bsigma\cdot\Bk}{m_L^\star}\frac{\tau^i}{2}(1+\Delta^\prime).
\label{LAC} \ee

Let us now rederive (\ref{LAC}) with an argument analogous to that
proven to be powerful for the convection current. We shall do the
calculation using the pion exchange only but the argument goes
through when the contact interaction (\ref{four-Fermion}) is
included. We begin by assuming that the axial charge -- in analogy
to (\ref{landauJ}) for the convection current --  takes the form,
\be
{\A0}^i=g_A \sum_{\sigma\tau}\int \frac{d^3 p}{(2\pi)^3}
\Bsigma\cdot (\Bnabla_p\epsilon_p) n_p
\frac{\tau^i}{2}\label{a0me} \ee where $n_p$ and $\epsilon_p$ are
$2\times 2$ matrices with matrix elements
\be
[n_p (\Br,t)]_{\alpha\alpha^\prime}=n_p
(\Br,t)\delta_{\alpha\alpha^\prime} +\Bs_p
(\Br,t)\cdot\Bsigma_{\alpha\alpha^\prime}, \ee and
\be
[\epsilon_p (\Br,t)]_{\alpha\alpha^\prime} = \epsilon_p (\Br,t)
\delta_{\alpha\alpha^\prime} +{\Beta}_p (\Br,t)\cdot
\Bsigma_{\alpha\alpha^\prime} \ee with
\be
{\Bs}_p (\Br,t)=\frac 12 \sum_{\alpha\alpha^\prime}
\Bsigma_{\alpha\alpha^\prime} [n_p (\Br,t)]_{\alpha^\prime\alpha}.
\label{sigmap} \ee In general $n=4$ in the spin-isospin space. But
without loss of generality, we could confine ourselves to $n=2$ in
the spin space with the isospin operator explicited as in
Eq.(\ref{a0me}). Then upon linearizing, we obtain from
(\ref{a0me})
\be
{\A0}^i=g_A \sum_{\sigma\tau}\int \frac{d^3 p}{(2\pi)^3}
\left(\Bsigma\cdot(\Bnabla_p\epsilon^0_p)\delta n_{p\sigma\tau} -
\Bsigma\cdot (\Bnabla_p n^0_p)\delta\epsilon_{p\sigma\tau}\right)
\frac{\tau^i}{2}+\cdots\label{a0mes} \ee where
\be
\delta n_{p\sigma\tau}=\frac 1V \delta^3 (\Bp-\Bk )
\frac{1+\sigma_3}{2}\frac{\tau^i}{2} \label{deltanp} \ee and
\be
\delta\epsilon_{p\sigma\tau}=\sum_{\sigma^\prime,\tau^\prime} \int
\frac{d^3 p^\prime}{(2\pi)^3}
f_{p\sigma\tau,p^\prime\sigma^\prime\tau^\prime} \delta
n_{p^\prime\sigma^\prime\tau^\prime}.\label{deltaen} \ee in
analogy with (\ref{deltaepsilon}). The equation (\ref{a0mes}) is
justified if the density of polarized spins is much less than the
total density of particles (assumed to hold here). The first term
of (\ref{a0mes}) with (\ref{deltanp}) yields the quasiparticle
charge operator
\be
{\A0}^i_{QP}=g_A\frac{\Bsigma\cdot\Bk}{m_L^\star}\frac{\tau^i}{2}\label{A0}
\ee while the second term represents the polarization of the
medium, due to the pion-exchange interaction (\ref{vpi})
\be
\delta
{\A0}^i=g_A\frac{\Bsigma\cdot\Bk}{m_L^\star}\frac{\tau^i}{2}
\Delta^\prime.\label{deltaA0} \ee The sum of (\ref{A0}) and
(\ref{deltaA0}) agrees precisely with the Landau charge
(\ref{LAC}).

It is not difficult to take into account the full Landau-Migdal
interactions (\ref{qpint}) which includes the one-pion-exchange
interaction as well as other contributions to the quasiparticle
interaction. Thus, the general expression is obtained by making
the replacement \be\label{zero-minus} \Delta^\prime \rightarrow
\frac 13 G_1^\prime - \frac{10}{3}H_0^\prime +\frac 43 H_1^\prime
- \frac{2}{15}H_2^\prime \ee in (\ref{deltaA0}). This combination
of Fermi-liquid parameters corresponds to a $\ell = \ell ^\prime =
1, J=0$ distortion of the Fermi sea~\cite{BSJ}. We will see later
that the result obtained with the naive Landau argument may not be
the whole story, since the one-pion-exchange contribution
disagrees, though by a small amount, with the chiral Lagrangian
prediction derived below.
\subsubsection{Chiral Lagrangian prediction}

We now calculate the axial charge using our chiral Lagrangian that
reproduced the Landau-Migdal formula for the convection current.
Consider first only the pion-exchange contribution. In this case
we can take the unperturbed nucleon propagator to carry the free
space mass $M$. The single-particle transition operator
corresponding to Fig.\ \ref{miyazawa}a is given by
\be
{\A0}^i_{1-body}=g_A\frac{\Bsigma\cdot\Bk}{M}\frac{\tau^i}{2}.
\label{a01bod} \ee There is no contribution of the type of Fig.\
\ref{miyazawa}b because of the (G-)parity conservation. The only
contribution to the two-body current comes from Fig.\
\ref{miyazawa}c and is of the form~\cite{delorme}
\be
{\A0}^i_{2-body}=g_A\frac{\Bsigma\cdot\Bk}{M}\frac{\tau^i}{2}\Delta
\label{a02bod} \ee with
\be
\Delta=\frac{f^2 k_F M}{2g_A^2m_\pi^2\pi^2}\left(I_0-I_1-
\frac{m_\pi^2}{2k_F^2}I_1\right).\label{Delta} \ee The factor
$(1/g_A^2)$ in (\ref{Delta}) arose from replacing
$\frac{1}{f_\pi^2}$ by $\frac{g_{\pi NN}^2}{g_A^2 M^2}$ using the
Goldberger-Treiman relation.

Now consider what happens when  the vector degrees of freedom are
taken into account.  Within the approximation adopted, the only
thing that needs be done is to implement the BR scaling. The
direct intervention of the vector mesons $\rho$ and $\omega$ in
the axial-charge operator is suppressed by the chiral counting, so
they will be ignored here. This means that in the single-particle
charge operator, all that one has to do is to replace $M$ by
$M^\star=M\Phi$ in (\ref{a01bod}):
\be
{\A0}^i_{1-body}=g_A\frac{\Bsigma\cdot\Bk}{M\Phi}\frac{\tau^i}{2}
\label{a01bod*} \ee and that in the two-body charge operator
(\ref{a02bod}), $f_\pi$ should be replaced by $f_\pi\Phi$ and $M$
by $M\Phi$:
\be
{\A0}^i_{2-body}=g_A\frac{\Bsigma\cdot\Bk}{M\Phi}\frac{\tau^i}{2}
\Delta.\label{a02bod*} \ee In the two-body operator, there is a
factor $(g_A/f_\pi)$ coming from the $\pi NN$ vertex which as
mentioned before, is assumed to be non-scaling at least up to
nuclear matter density~\cite{gebhenley}, in consistency with the
observation that the pion-exchange current
 does not scale in medium.

The total predicted by the chiral Lagrangian (modulo higher-order
corrections) is then
\be
g_A\frac{\Bsigma\cdot\Bk}{M\Phi}\frac{\tau^i}{2} (1+\Delta). \ee
which differs from the charge operator obtained by the Landau
method, (\ref{LAC}).
\subsubsection{Comparison between vector and axial current}

An immediate question (to which we have no convincing answer) is
whether or not the difference between the two approaches -- the
Fermi-liquid vs.\ the chiral Lagrangian -- is genuine or a defect
in either or both of the approaches. One possible cause of the
difference could be that {\it both} the assumed localized
quasiparticle charge, Eq.(\ref{totalcharge}), {\it and} the
effective axial charge, Eq.(\ref{a0me}), are incomplete. We have
looked for possible additional terms that could contribute but we
have been unable to find them. So while not ruling out this
possibility, we turn to the possibility that the difference is
genuine.

It is a well-known fact that the conservation of the vector
current assures that the EM charge or the weak vector charge is
$g_V=1$ but the conservation of the axial vector charge does not
constrain the value of the axial charge $g_A$ , that is, $g_A$ can
be anything. This is because the axial symmetry is spontaneously
broken. In the Wigner phase in which the axial symmetry would be
restored, one would expect that $g_A=1$. It therefore seems that
the Goldstone structure of the ``vacuum" of the nuclear matter is
at the origin of the difference.

To see whether there can be basic differences, let us look at the
effect of the pion field. The cancellation between the two-body
current $\Jb_{2-body}^\pi$ (\ref{pi}) and $\Jb_{ph}^\pi$
(\ref{parthole}) leaving only a term that changes $M^\star$ to
$m_L^\star$ in the one-body operator with a BR scaling mass,
Eq.(\ref{onebody}), in the EM case can be understood as follows.
Both terms involve the two-body interaction mediated by a
pion-exchange. It is obvious how this is so in the latter.  To see
it in the former, we note that it involves the insertion of an EM
current in the propagator of the pion. Thus the sum of the two
terms corresponds to the insertion of an EM current in all
internal hadronic lines of the one-pion exchange self-energy graph
of the nucleon.  The two-body pionic current -- that together with
the single-particle current preserves gauge invariance -- is in
turn related to the one-pion-exchange potential $V_\pi$. Therefore
what is calculated is essentially an effect of a nuclear force.
Now the point is that the density-dependent part of the sum (that
is, the ones containing one hole line) -- apart from a term that
changes $M^\star$ to $m_L^\star$ in (\ref{onebody}) -- vanishes in
the $\omega/q\rightarrow 0$ limit. In contrast, the cancellation
between (\ref{deltaA0}) and (\ref{phA0}) in the case of the axial
charge, has no corresponding interpretation. While the
one-pion-exchange interaction is involved in the particle-hole
term (\ref{phA0}),  (\ref{deltaA0}) cannot be interpreted as an
insertion of the axial vector current into the pion propagator
since such an insertion is forbidden by parity. In other words,
Eq.(\ref{deltaA0}) does not have a corresponding Feynman graph
which can be linked to a potential. We interpret this as
indicating that there is no corresponding Landau formula for the
axial charge in the same sense as in the vector current case.

In a chiral Lagrangian formalism, each term is associated with a
Feynman diagram. As mentioned, there is no contribution to the
convection current from a diagram of the type Fig.\
\ref{miyazawa}c (apart from a gauge non-invariant off-shell term
which cancels the counter part in Fig.\ \ref{miyazawa}b). Instead
this diagram renormalizes the spin gyromagnetic ratio. In
contrast, the corresponding diagram for the axial current does
contribute to the axial charge (\ref{a02bod}). As first shown in
\cite{KDR}, the contribution from Fig.\ref{miyazawa}c for both the
vector current and the axial-vector current is current algebra in
origin and constrained by chiral symmetry. Furthermore it does not
have a simple connection to nuclear force. While the convection
current is completely constrained by gauge invariance of the EM
field, and hence chiral invariance has little to say, both the EM
spin current and the axial charge are principally dictated by the
chiral symmetry. This again suggests that the Landau approach to
the axial charge cannot give the complete answer even at the level
of quasiparticle description. There is however a caveat here: in
the Landau approach, the nonlocal pionic and local four-Fermion
interactions (\ref{four-Fermion}) enter together in an intricate
way as we saw in the EM case. Perhaps this is also the case in the
axial charge, with an added subtlety due to the presence of
Goldstone pions. It is possible that the difference is due to the
contribution of the four-Fermion interaction term to
(\ref{zero-minus}) which cancels out in the limit
$\omega/q\rightarrow 0$ but contributes in the
$q/\omega\rightarrow 0$ limit. This term cannot be given a simple
interpretation in terms of chiral Lagrangians. Amusingly the
difference between the results (see below) turns out to be small.
\subsubsection{Numerical comparison}

To compare the two results, we rewrite the sum of (\ref{A0}) and
(\ref{deltaA0}), i.e., ``Landau axial charge" (LAC), using
(\ref{eff-mass}) and (\ref{F1pi})
\be
 {\A0}^i_{LAC}=g_A\frac{\Bsigma\cdot\Bk}{M\Phi}\frac{\tau^i}{2}(1+
\tilde{\Delta}) \label{LACp} \ee where
\be
\tilde{\Delta}=\frac{f^2 k_F M\Phi}{4\pi^2 m_\pi^2} \left(I_0-I_1
+ \frac{3m_\pi^2}{2k_F^2}I_1\Phi^{-1}\right)\label{delta"} \ee and
the sum of (\ref{a01bod*}) and (\ref{a02bod*}), i.e., the
``current-algebra axial charge"(CAAC), as
\be
{\A0}^i_{CAAC}=g_A\frac{\Bsigma\cdot\Bk}{M\Phi}\frac{\tau^i}{2}
(1+\Delta)\label{CAAC} \ee where
\be
\Delta=\frac{f^2 k_F M}{2g_A^2m_\pi^2\pi^2}\left(I_0-I_1-
\frac{m_\pi^2}{2k_F^2}I_1\right). \ee We shall compare
$\tilde{\Delta}$ and $\Delta$ for two densities $\rho=\frac
12\rho_0$ ($k_F=1.50 m_\pi$) and $\rho=\rho_0$ ($k_F=1.89m_\pi$)
where $\rho_0$ is the normal nuclear matter density $0.16/{\rm
fm}^3$.

For numerical estimates, we take
\be
\Phi (\rho)=\left(1+0.28\frac{\rho}{\rho_0}\right)^{-1} \ee which
gives $\Phi(\rho_0)=0.78$ found in QCD sum rule
calculations~\cite{sbmr}. Somewhat surprisingly, the resulting
values for $\tilde{\Delta}$ and $\Delta$ are close; they agree
within 10\%. For instance at $\rho\approx \rho_0/2$,
$\tilde{\Delta}\approx 0.48$ while $\Delta\approx 0.43$ and at
$\rho\approx \rho_0$, $\tilde{\Delta}\approx 0.56$ while
$\Delta\approx 0.61$. Whether this close agreement is coincidental
or has a deep origin is not known.
\subsubsection{Test: axial charge transition in heavy nuclei}
\indent\indent The axial charge transition in heavy nuclei
\be
A(J^+)\leftrightarrow B(J^-) \ee with change of one unit of
isospin $\Delta T=1$ provides a test of the axial charge operator
(\ref{CAAC}) or (\ref{LACp}). To check this, consider the
Warburton ratio $\epsilon_{MEC}$~\cite{warburton}
\be
\epsilon_{MEC}=M_{exp}/M_{sp} \ee where $M_{exp}$ is the measured
matrix element for the axial charge transition and $M_{sp}$ is the
theoretical single-particle matrix element. There are theoretical
uncertainties in defining the latter, so the ratio is not an
unambiguous object but what is significant is Warburton's
observation that in heavy nuclei, $\epsilon_{MEC}$ can be as large
as 2:
\be
\epsilon_{MEC}^{HeavyNuclei}=1.9\sim 2.0. \ee More recent
measurements -- and their analyses -- in different
nuclei~\cite{otherexp} quantitatively confirm this result of
Warburton.

To compare our theoretical prediction with the Warburton analysis,
we calculate the same ratio using (\ref{CAAC})
\be
\epsilon_{MEC}^{CAAC}=\Phi^{-1} (1+\Delta).\label{epsilonth} \ee
The formula(\ref{CAAC}) differs from what was obtained in
\cite{kuboderarho} in that here the non-scaling in medium of the
pion mass and the ratio $g_A/f_\pi$ is taken into account. We
believe that the scaling used in \cite{kuboderarho} (which
amounted to having $\Delta/\Phi$ in place of $\Delta$ in
(\ref{epsilonth})) is not correct.\\ \indent The enhancement
corresponding to the ``Landau formula" (\ref{LACp}) is obtained by
replacing $\Delta$ by $\tilde{\Delta}$ in (\ref{epsilonth}). Using
the value for $\Phi$ and $\Delta$ at nuclear matter density, we
find
\be
\epsilon_{MEC}^{th}\approx 2.1\ \ \ (2.0) \ee in good agreement
with the experimental results of \cite{warburton} and
\cite{otherexp}. Here the value in parenthesis is obtained with
the Landau formula (\ref{LACp}). The difference between the two
formulas (i.e., current algebra vs. Landau) is indeed small. This
is a check of the scaling of $f_\pi$ in combination with the
scaling of the Gamow-Teller constant $g_A$ in medium.
\section{Summary}\label{conc}

An attempt is made and some success is obtained in this review to
relate an effective chiral Lagrangian to an effective field theory
for nuclear matter. The aim is to bridge between what we know at
normal nuclear density and what can be expected under the extreme
condition, relevant in neutron stars and in relativistic heavy ion
collisions. Furnstahl, Serot and Tang's effective chiral model
Lagrangian FTS1~\cite{tang}, which describes successfully the
phenomenology of finite nuclei and infinite nuclear matter, is
taken to imply that an effective chiral Lagrangian calculated in
high chiral orders corresponds to Lynn's chiral soliton with the
chiral liquid structure~\cite{lynn} in mean field. This provides
the ground state around which quantum fluctuations can be
calculated. Note that FTS1 is simply one of the available theories
that are consistent with the symmetries of QCD and successful
phenomenologically. We do not imply that FTS1 is the best one can
construct as an effective theory of nuclear matter.

The scalar sector in FTS1 develops a large anomalous dimension,
which is interpreted as a signal of a strong coupling situation.
It is suggested that the strong coupling theory
can be transformed into a weak coupling theory if the chiral
Lagrangian is rewritten in terms of the parameters given by BR
scaling. A simple model, whose mass parameters are BR-scaled, is
constructed and is shown to describe ground state properties of
nuclear matter very well with fits comparable to the full FTS1
theory. The simple BR-scaled Lagrangian gives the background at
any arbitrary density around which fluctuations can be calculated.
Tree diagrams yield the dominant contributions. It is shown that
we can make simple Walecka-type models, including our simple
model. The models are thermodynamically consistent and the
dependence of parameters on density are represented by the
interactions of hadrons. One can also map the density-dependent
model Lagrangian into relativistic Landau Fermi-liquid theory.
Thus a quasiparticle picture of a strongly correlated system at
densities away from the normal nuclear matter density is obtained.

The BR scaling parameter $\Phi$ has been identified with a Landau
Fermi-liquid parameter by means of nuclear responses to the EM
convection current. The Landau effective mass of the nucleon
$m_L^\star$ is given in terms of $\Phi$ and pion cloud, i.e.\ the
Goldstone boson of the broken chiral symmetry, through the Landau
parameter $\tilde{F}_1^\pi$. The relation between the exchange
current correction to the orbital gyromagnetic ratio $\delta g_l$
and $m_L^\star$ provides the crucial link between $\Phi$ and
$F_1^\omega$ which comes from the massive degree of freedom in the
isoscalar vector channel dominated by the $\omega$ meson. The
axial charge transition in heavy nuclei provides a relation
between $\Phi$ and the in-medium pion decay constant $f^\star_\pi
/f_\pi$. These relations are found to be satisfied very accurately
and to connect physics of relativistic heavy ion collision data,
e. g., dilepton data of CERES and nucleon and kaon flow data of
FOPI(4 pi multiparticle detector) and KaoS(Kaon Spectrometer),
etc.\ to low energy spectroscopic properties, e.g., $m_L^\star$,
$\delta g_l$, etc.\ in heavy nuclei via BR scaling.
\section{Open issues}
While as an exploration our results are satisfying, there are
several crucial links that remain conjectural in the work and
require a lot more work. We mention some issues for future
studies.
\begin{itemize}
\item
We have not yet established in a convincing way that a
nontopological soliton coming from a high-order effective chiral
Lagrangian accurately describes nuclear matter that we know of.
The first obstacle here is that a realistic effective Lagrangian
that contains sufficiently high-order loop corrections including
non-analytic terms has not yet been constructed. Lynn's argument
for the existence of such a soliton solution and identification
with a drop of nuclear matter is based on a highly truncated
Lagrangian (ignoring non-analytic terms). We are simply assuming
that the FTS1 Lagrangian is a sufficiently realistic version (in
terms of explicit vector and scalar degrees of freedom that are
integrated out by Lynn) of Lynn's effective Lagrangian. To prove
that this assumption is valid is an open problem.
\item
We do not understand clearly the role and the origin of the
anomalous dimension $d_{an}\approx 5/3$ for the quarkonium scalar
field in FTS1. It is an interesting problem how the scalar in FTS1
comes to include higher order interactions in its anomalous
scaling dimension through decimations. And our argument for
interpreting the FTS1 with such large anomalous dimension as a
strong-coupling theory which can be reinterpreted in terms of a
weak-coupling theory expressed with BR scaling is heuristic at
best and needs to be sharpened, although our results strongly
indicate that it is correct.
\item
There is also the practical question as to how far in density the
predictive power of the BR-scaled effective Lagrangian can be
pushed. In our simple numerical calculation, we used a
parameterization for the scaling function $\Phi (\rho)$ of the
simple geometric form which can be valid, if at all, up to the
normal matter density as seems to be supported by QCD sum rule and
dynamical model calculations. At higher densities, the form used
has no reason to be accurate. By using the empirical information
coming from nucleon and kaon flows, one could infer its structure
up to, say,
 $\rho\sim 3 \rho_0$ and if our
argument for kaon condensation is correct -- and hence kaon
condensation takes place at $\rho\lsim 3\rho_0$, then this will be
good enough to make a prediction for the critical density for kaon
condensation. In calculating compact-star properties in supernovae
explosions, however, the equation of state for densities
considerably higher than the normal matter density, say, $\rho
\gsim 5\rho_0$ is required. It is unlikely that this high density
can be accessed within the presently employed approximations. Not
only will the structure of the scaling function $\Phi$ be more
complicated but also the correlation terms that are small
perturbations at normal density may no longer be so at higher
densities, as pointed out by Pandharipande, Pethick, and
Thorsson~\cite{panda} who approach the effect of correlations from
the high-density limit. In particular, the notion of the scaling
function $\Phi$ will have to be modified in such a way that it
will become a non-linear function of the fields that figure in the
process. This would alter the structure of the Lagrangian field
theory. Furthermore there may be a phase transition (such as
spontaneously broken Lorentz symmetry, Georgi vector limit, chiral
phase transition or meson condensation) lurking nearby in which
case the present theory would have already broken down. These
caveats will have to be carefully examined before one can
extrapolate the notion of BR scaling to a high-density regime as
required for a reliable calculation of the compact-star structure.
How the scaling parameters extrapolate beyond normal nuclear
matter density is not predicted by theory and should be deduced
from lattice measurements and heavy-ion experiments that are to
come. Corrections to BR scaling as massive mesons approach
on-shell need be taken into account. The fit to the available
CERES data indicates however that the extrapolation to higher
density -- perhaps up to the chiral phase transition -- is at
least approximately correct under the conditions that prevail in
nucleus-nucleus collisions at SPS energies. How this could come
about was discussed in \cite{BLRRW}.
\item
In addition the behavior of $g_v^\star$ at $\rho >\rho_0$ also
deviates from our simple form. It is expected to drop more
rapidly~\cite{LBLK96}. Indeed a recent calculation \cite{WRW} of
kaon attraction to ${{\O} (Q^2)}$ in chiral perturbation theory
that is highly constrained by the ensemble of on-shell
kaon-nucleon data and that includes both Pauli and short-range
correlations for many-body effects is found to give at most about
120 MeV attraction at nuclear matter density. Thus the crucial
input here is the strength of the $K^-$-nuclear interaction in
dense medium. The attraction decreases from the analysis of the
$K$-mesic atom by Friedman, Gal, and Batty~\cite{friedman}
indicating the 200 MeV attraction. If the attraction
 came down to
$100\sim 120$ MeV as found in \cite{WRW}, this would give a strong
constraint on the constants that enter in the four-Fermi
interactions in the chiral Lagrangian. This would presumably
account for the need for a dropping vector coupling $g_v^\star$
required for $\rho\gsim \rho_0$. Moreover Kim and Lee~\cite{kl99}
found recently by renormalization group analysis that the coupling
constant $g_{\rho NN}$ drops as density increases. This crucial
information is also expected to come from on-going heavy-ion
experiments.
\item
Although we show that BR scaling parameters can be written in
terms of Landau parameter via nonrelativistic EM current, it
remains to formulate the relativistic mapping along the line
developed in Section \ref{liq} where thermodynamic properties of a
simple BR-scaled chiral model Lagrangian in the mean field were
shown to be consistent with relativistic Landau formula derived in
Section \ref{relation}. This work is needed to go to higher
density region. Such a work is in progress. Furthermore
quasiquarks must become relevant degrees of freedom at high
density. Thus quasihadron liquid is shifted to quasiquark liquid
as density increases. The investigation of the change in the shift
may give a good way to connect the low density physics to the
higher one.
\item
As seen in Section \ref{axi} it is not clear how we deal with
Goldstone bosons in the scheme of Landau-Migdal approach. The
Landau axial charge and current algebra axial charge are not the
same, though they give similar numerical values. We do not know
even whether it is possible or not that the Landau-Migdal approach
can treat Goldstone boson properly. The study of it will show the
way to treat the Landau Fermi-liquid theory and its scope.
\item
Finally one could ask more theoretical questions as to in what way
our effective Lagrangian approach is connected to standard chiral
effective theory, which does not concern scale symmetry and its
anomaly, proper and if the theory is to be fully predictive, how
one can proceed to calculate the corrections to the tree-diagram
results we have obtained. The first issue, a rigorous derivation
of BR scaling starting from an effective chiral action via
multiple scale decimations required for the problem is yet to be
formulated but the main ingredients, both theoretical and
phenomenological, seem to be available. The second issue is of
course closely tied with what the appropriate expansion parameter
is in the theory. These matters are addressed in the paper but
they are somewhat scattered all over the place and it might be
helpful to summarize them here. The answers to these questions are
not straightforward since there are two stages of ``decimation" in
the construction of our effective Lagrangian: The first is the
elimination of high-energy degrees of freedom for the effective
Lagrangian that gives rise to a soliton (i.e., chiral liquid) and
here the relevant scale is the chiral symmetry breaking scale
$\sim$ 1 GeV and the second is that given a chiral liquid which we
argued can be identified as the Fermi-liquid fixed point, the
decimation involved here is for the excitations of scale $\Lambda$
above (and below) the Fermi surface for which the expansion is
made in $1/N$. As discussed in Section \ref{rgliq}, $1/N$ $\sim$
$\Lambda/k_F$ where $\Lambda$ is the cutoff in the Fermi system.
In bringing in a BR-scaled chiral Lagrangian, we are relying on
chiral symmetry considerations {\it applied} to a system with a
density defined by nuclear matter. Thus the link to QCD proper of
the effective theory we use for describing fluctuations around the
nuclear matter ground state must be tenuous at best.
As recently re-emphasized by Weinberg~\cite{wein97}, low-energy
effective theories need not be in one-to-one correspondence with a
fundamental theory meaning that one low-energy effective theory
could arise through decimation from several different
``fundamental" theories. This applies not only to theories with
global symmetry but also to those with local gauge symmetry. In
the present case, this aspect is more relevant since there is a
change in degrees of freedom between the nonperturbative regime in
which we are working and the perturbative regime in which QCD
proper is operative.
\end{itemize}

\section*{Acknowledgements}
I am very grateful to
Professor D.-P. Min and Professor M. Rho for their guidance and
encouragement throughout my
graduate years. Collaboration with Professors G.E. Brown and B.
Friman has been a great pleasure to me. I would like to thank C.-H. Lee and
R. Rapp for useful comments and discussions. This work was supported 
in part by the U.S. Department of Energy under DE-FG02-88ER40388, by the
Korea Science and Engineering Foundation through Center for
Theoretical Physics of Seoul National University, and by 
the Korea Ministry of Education under BSRI-98-2418.

\section*{Appendix}

\addcontentsline{toc}{section}{Appendix}
\appendix
\setcounter{equation}{0} \setcounter{figure}{0}
\renewcommand{\theequation}{\mbox{A.\arabic{equation}}}
\renewcommand{\thefigure}{\mbox{A.\arabic{figure}}}
\section*{A: Effect of many-body correlations on EOS}\label{appenda}
\addcontentsline{toc}{section}{A: Effect of many-body correlations
on EOS}

In this appendix, we briefly discuss the sensitivity of the EOS to
the correlation parameters of (\ref{nbody2}) at a density $\rho\
>\ \rho_0$. This is shown in Fig.\ \ref{comparison}. While the
parameter sets B1, B2, B3 and B4 give more or less the same
equilibrium density and binding energy (see Table \ref{fit}), the
parameter set B2 has an instability and B4 a local minimum at
$\sim$ 2 times the normal matter density whereas the sets B1 and
B3 give a stable state at all density, possibly up to meson
condensations and/or chiral phase transition. It is not clear what
this means for describing fluctuations at a density above $\rho_0$
but it indicates that given data at ordinary nuclear matter
density, it will not be feasible to extrapolate in a unique way to
higher densities unless one has constraints from experimental data
at the corresponding density. In our discussion, we relied on the
data from KaoS and FOPI collaborations to avoid the fine-tuning of
the parameters.
\begin{figure}
\setlength{\epsfysize}{6.in} \centerline{\epsffile{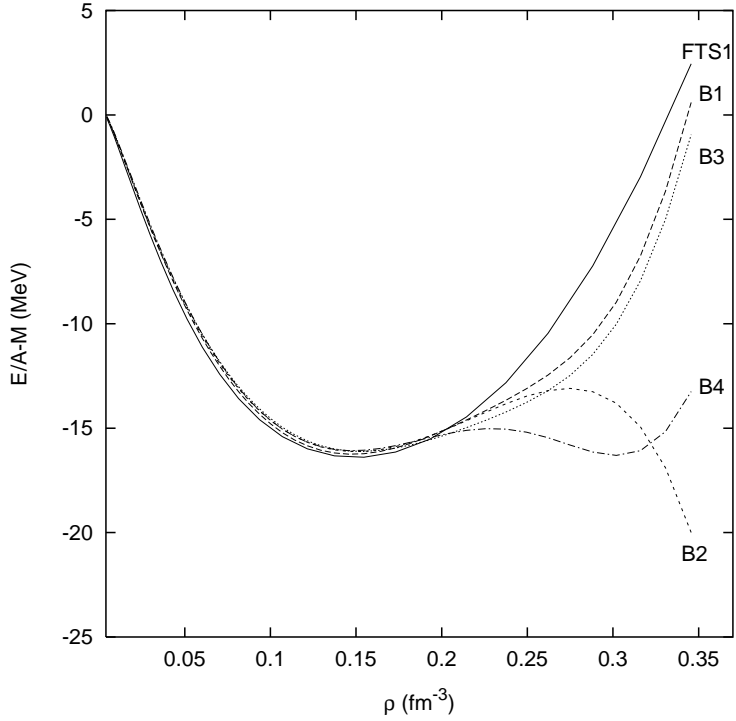}}
\caption{$E/A-M$ vs. $\rho$ for the B1, B2, B3 and B4 models given
in Table \ref{fit} compared with FTS1 theory. } \label{comparison}
\end{figure}
\setcounter{equation}{0} \setcounter{figure}{0}
\renewcommand{\theequation}{\mbox{B.\arabic{equation}}}
\renewcommand{\thefigure}{\mbox{B.\arabic{figure}}}
\section*{B: Relativistic calculation of $F^\pi_1$}\label{appndab}
\addcontentsline{toc}{section}{B: Relativistic calculation of
$F^\pi_1$}

In the text, the Landau parameter $F_1^\pi$ (or $f^\pi$) was
calculated nonrelativistically via the Fock term of Fig.\
\ref{f-pi}. Here we calculate it relativistically by
Fierz-transforming the one-pion-exchange graph and taking the
Hartree term. This procedure is important for implementing
relativity in the connection between Fermi-liquid theory and
chiral Lagrangian theory along the line discussed by Baym and
Chin~\cite{baymchin}.
\begin{figure}
\centerline{\epsfig{file=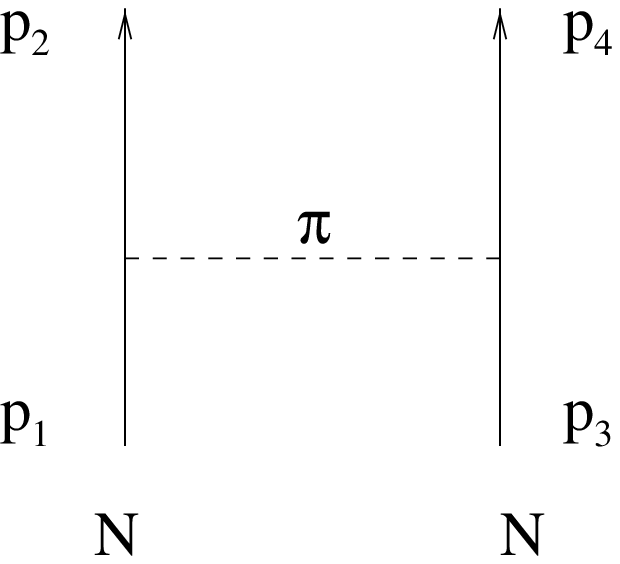,width=2.5in}} \caption{\small
The-one-pion-exchange diagram that gives rise to $F_1^\pi$.
}\label{f-pi}
\end{figure}

The one-pion-exchange potential in Fig.\ \ref{f-pi} is
\be
V_{\pi}= -g_{\pi NN}^2(\taub_{21}\cdot\taub_{43})
\frac{\bar{u}_2\gamma^5u_1\bar{u}_4\gamma^5
u_3}{(p_2-p_1)^2-m_\pi^2}. \ee The Dirac spinors are normalized by
\be
u^\dagger (p,s)u(p,s^\prime )=\delta_{ss^\prime}. \ee By a Fierz
transformation, we have
\be
\taub_{21}\cdot\taub_{43} =\frac12
(3\delta_{41}\delta_{23}-\taub_{41}\cdot\taub_{32})
\label{fierzt}\ee and
\be
\bar{u}_2\gamma^5u_1\bar{u}_4\gamma^5u_3
&=&\frac{1}{4}[\bar{u}_4u_1\bar{u}_2u_3- \bar{u}_4\gamma^\mu
u_1\bar{u}_2\gamma_\mu u_3\label{fierzf}\\ & &+
\bar{u}_4\sigma^{\mu\nu}u_1\bar{u}_2\sigma_{\mu\nu}u_3+
\bar{u}_4\gamma^\mu\gamma^5u_1\bar{u}_2\gamma_\mu\gamma^5u_3+
\bar{u}_4\gamma^5u_1\bar{u}_2\gamma^5u_3].\nonumber \ee
Remembering a minus sign for the fermion exchange, we obtain the
corresponding pionic contribution to the quasiparticle interaction
at the Fermi surface, $f^\pi = -V_\pi (\pb_1=\pb_4=\pb,
\pb_2=\pb_3=\pb^\prime ,\pb^2 =\pb^{\prime 2}=k_F^2)$ (see
(\ref{qpint})). Decomposing $f^\pi$ as
\be
f^\pi =\frac{3-\taub\cdot\taub^\prime }{2}
(f_S+f_V+f_T+f_A+f_P)\label{fss} \ee where $S$, $V$, $T$, $A$ and
$P$ represent scalar, vector, tensor, axial vector and
pseudoscalar channel respectively, we find
\be
f_S&=&-\frac{M^4f^2}{E_F^2m_\pi^2}\frac{1}{q^2+m_\pi^2}
\nonumber\\ f_V&=&\frac{M^4f^2}{E_F^2m_\pi^2}\frac{1}{q^2+m_\pi^2}
\left( 1+\frac{q^2}{2M^2} \right) \nonumber\\
f_T&=&-\frac{M^4f^2}{E_F^2m_\pi^2}\frac{1}{q^2+m_\pi^2} \left(
\sigmab\cdot\sigmab^\prime (1+\frac{q^2}{2M^2})
+\frac{2\sigmab^\prime\cdot\pb\sigmab\cdot\pb^\prime
-\sigmab\cdot\pb\sigmab^\prime\cdot\pb
-\sigmab\cdot\pb^\prime\sigmab^\prime\cdot\pb^\prime}{2M^2}
\right)\nonumber\\
f_A&=&\frac{M^4f^2}{E_F^2m_\pi^2}\frac{1}{q^2+m_\pi^2} \left(
\sigmab\cdot\sigmab^\prime
-\frac{2\sigmab\cdot\pb\sigmab^\prime\cdot\pb^\prime
-\sigmab\cdot\pb\sigmab^\prime\cdot\pb
-\sigmab\cdot\pb^\prime\sigmab^\prime\cdot\pb^\prime}{2M^2}
\right)\nonumber\\ f_P&=&0.\label{decomp} \ee with $E_F
=\sqrt{k_F^2+M^2}$ and $q=|\pb -\pb^\prime |$. Thus we obtain
\be
f^\pi &=&\frac{f^2}{m_\pi^2}\frac{M^2}{E_F^2}\frac{1}{q^2+m_\pi^2}
\left( \sigmab\cdot\qb\sigmab^\prime\cdot\qb
-\frac{q^2(1-\sigmab\cdot\sigmab^\prime )}{2}\right)
\frac{3-\taub\cdot\taub^\prime }{2}\label{rfpi}\\
&=&\frac13\frac{f^2}{m_\pi^2}\frac{M^2}{E_F^2}\frac{q^2}{q^2+m_\pi^2}
\left( 3\frac{\sigmab\cdot\qb\sigmab^\prime\cdot\qb}{q^2}
-\sigmab\cdot\sigmab^\prime +\frac12
(3-\sigmab\cdot\sigmab^\prime)\right)
\frac{3-\taub\cdot\taub^\prime }{2}.\nonumber \ee In the
nonrelativistic limit, $E_F \sim M$ and we recover (\ref{vpi}).
The factor $M/E_F$ comes since there is one particle in the unit
volume which decreases relativistically as the speed increases.
Note that only $f_S$ and $f_V$ in (\ref{decomp}) are
spin-independent and contribute to $F_1^\pi$. The $f_S$ is
completely canceled by the leading term of $f_V$ with the
remainder giving $F_1^\pi$. In this way of deriving the Landau
parameter $F_1$, it is the vector channel that plays the essential
role.

\setcounter{equation}{0} \setcounter{figure}{0}
\renewcommand{\theequation}{\mbox{C.\arabic{equation}}}
\renewcommand{\thefigure}{\mbox{C.\arabic{figure}}}
\section*{C: Relativistic calculation of $F_1(\omega )$ and
$\Jb_{2-body}^\omega$ }\label{appendc}
\addcontentsline{toc}{section}{C: Relativistic calculation of
$F_1(\omega )$ and $\Jb_{2-body}^\omega$ }

Here we compute the contribution of vector meson channel to Landau
parameter $F_1$ and EM current relativistically. The way to
compute the contribution of $\rho$-meson channel is almost the
same as $\omega$ channel. So we treat here $\omega$-meson channel
only.

The blob in Fig.\ \ref{nbar}a corresponding to four-Fermi
interactions can be expanded as Fig.\ \ref{rpa} in random phase
approximation.
\begin{figure}
\centerline{\epsfig{file=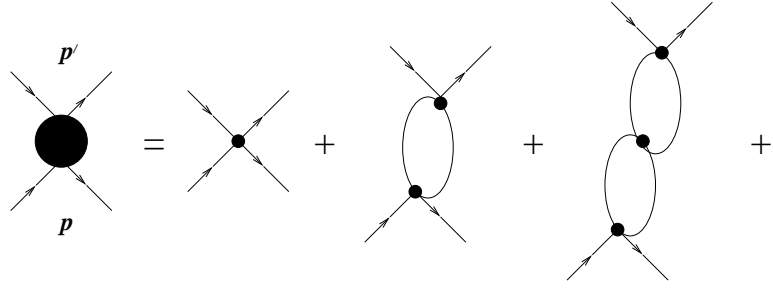,width=4in}}
\caption{Quasiparticle interactions in vector meson channel
represented by four-Fermi interaction. The large blob corresponds
to the blob in Fig.\ \ref{nbar}. }\label{rpa}
\end{figure}
One bubble is represented by
\be
\Pi^{\mu\nu}=\lim_{\omega_q\to 0}\lim_{q\to 0}
2\int\frac{d^4p}{(2\pi )^4} \tr [\gamma^\mu S(p)\gamma^\nu
S(p+q)]\label{bub1} \ee where $S(p)$ is the Fermion propagator.
The factor 2 come from isospin contribution. In the presence of
Fermi sea, we can divide $S(p)$ into the four parts;
\be
S(p)&=&\frac{1}{2\omega_p}[ (\omega_p\gamma_0-\pb\cdot\gammab
+M^\star )
(\frac{1-n_p}{p_0-\omega_p+i\delta}+\frac{n_p}{p_0-\omega_p-i\delta})
\nonumber\\
& &+\frac{\omega_p\gamma_0+\pb\cdot\gammab
-M^\star}{p_0+\omega_p-i\delta}] \ee with $\omega_p=\sqrt{M^{\star
2}+p^2}$ and $n_p =\theta (k_F-p)$ at $T=0$. The first term is the
free particle propagator in vacuum. The second is the particle
propagator for Pauli-blocked state in medium. The third is for
hole and the fourth is for antiparticle. Since vacuum
contribution, i.e.\ antiparticle-particle in vacuum contribution,
is canceled by counter terms and particle-hole contribution
vanishes in our limit $q/\omega_q\to 0$, the antiparticle-particle
in Pauli-blocked state contribution remains. Then (\ref{bub1})
becomes
\be
\Pi^{\mu\nu}=&&\int\frac{d^3p}{(2\pi )^3}
\frac{n_p}{4\omega_p^3}\tr [\gamma^\mu
(\omega_p\gamma_0-\pb\cdot\gammab +M^\star )\gamma^\nu
(\omega_p\gamma_0+\pb\cdot\gammab-M^\star )\nn\\ &+&\gamma^\mu
(\omega_p\gamma_0+\pb\cdot\gammab-M^\star )\gamma^\nu
(\omega_p\gamma_0 -\pb\cdot\gammab +M^\star ) ]. \ee Because of
rotational invariance and zero energy-momentum transfer, only
$\Pi^{ii}$ does not vanish.
\be
\Pi^{ii}=\frac43\int^{k_F}\frac{d^3p}{(2\pi )^3} \frac{3M^{\star
2}+2p^2}{\omega_p^3}=\frac{\rho}{E_F} \ee with $E_F=\sqrt{M^{\star
2}+k_F^2}$.

The quasiparticle interaction in $\omega$-meson channel in Fig.\
\ref{rpa} gives
\be
f^\omega_{pp^\prime}&=& C_\omega^2\left[\bar{u}(p)\gamma^\mu u(p)
\bar{u}(p^\prime )\gamma_\mu u(p^\prime )- \bar{u}(p)\gamma_i
u(p)\left( \sum_{i=1}^{\infty}(-
C_\omega^2\frac{\rho}{E_F})^i\right) \bar{u}(p^\prime )\gamma^i
u(p^\prime )\right]\nn\\ &=&C_\omega^2\left( \bar{u}(p)\gamma^\mu
u(p) \bar{u}(p^\prime )\gamma_\mu u(p^\prime )+
\bar{u}(p)\gamma_iu(p)\frac{C_\omega^2\rho /E_F}{1+C_\omega^2\rho
/E_F} \bar{u}(p^\prime )\gamma^i u(p^\prime )\right) \nn\\
&=&C_\omega^2-C_\omega^2\frac{\pb\cdot\pb^\prime}{\mu E_F} \ee
with chemical potential $\mu =E_F+C_\omega^2\rho$. Thus
\be
F_1(\omega )=-C_\omega^2\frac{2k_F^3}{\pi^2\mu}. \ee And the EM
current in Fig.\ \ref{nbar}a is
\be
\Jb_{2-body}^\omega&=&\bar{u}(k)\gammab u(k) \frac{C_\omega^2\rho
/E_F}{1+C_\omega^2\rho /E_F}\nn\\
&=&\frac{\Bk}{\mu}\frac{\tilde{F}_1(\omega )}{6}. \ee


\end{document}